\documentclass[twocolumn,pre,floatfix,amsmath,amssymb,amsfonts,showpacs]{revtex4}
\pdfoutput=1
\usepackage{graphicx,color}
\usepackage{subfigure}
\usepackage{url}
\usepackage{longtable}

\newcommand{\calA}{{\cal A}}
\newcommand{\calASL}{{\calA}_{Sk}}
\newcommand{\calAhL}{{\calA}_{hk}}
\newcommand{\calAwL}{{\calA}_{wk}}
\newcommand{\calAwSL}{{\calA}_{wSk}}
\newcommand{\calAswSxL}{{\calA}_{swSxk}}

\newcommand{\calAhxL}{{\calA}_{{h_x}k}}
\newcommand{\calB}{{\cal B}}

\newcommand{\calBzz}{{\calB}_{zz}}

\newcommand{\Aoo}{{A_{00}}}
\newcommand{\All}{{A_{11}}}
\newcommand{\Aol}{{A_{01}}}
\newcommand{\Alo}{{A_{10}}}
\newcommand{\Azz}{{A_{zz}}}
\newcommand{\calAoo}{{\calA_{00}}}
\newcommand{\calAll}{{\calA_{11}}}

\newcommand{\calAlo}{{\calA_{10}}}
\newcommand{\calAzz}{{\calA_{zz}}}

\makeatletter

\newcommand{\Rmnum}[1]{\expandafter\@slowromancap\romannumeral #1@}
\makeatother

\begin{document}

\title{Avalanche Spatial Structure and Multivariable Scaling Functions; Sizes, Heights, Widths, and Views through Windows}
\date{\today}

\author{Yan-Jiun Chen, Stefanos Papanikolaou, James P. Sethna}
\affiliation{Laboratory of Atomic and Solid State Physics (LASSP), Clark Hall,
Cornell University, Ithaca, NY 14853-2501, USA}
\author{Stefano Zapperi}
\affiliation{IENI-CNR,Via R. Cozzi 53, 20125 Milano, Italy and ISI Foundation, viale S. Severio
65, Torino, Italy}
\author{Gianfranco Durin}
\affiliation{Istituto Nazionale di Ricerca Metrologica, strada delle Cacce
91, 10135, Torino, Italy, and ISI Foundation, viale S. Severio 65, Torino,
Italy}

\date{\today}

\begin{abstract}

We introduce a systematic method for extracting multivariable universal scaling functions and critical exponents from data.  We exemplify our insights by analyzing simulations of avalanches in an interface using simulations from a driven quenched Kardar-Parisi-Zhang (qKPZ) equation.  We fully characterize the spatial structure of these avalanches- we report universal scaling functions for size, height and width distributions, and also local front heights.    Furthermore, we resolve a problem that arises in many imaging experiments of crackling noise and avalanche dynamics, where the observed distributions are strongly distorted by a limited field of view.  Through artificially windowed data, we show these distributions and their multivariable scaling functions may be written in terms of two control parameters, the window size and the characteristic length scale of the dynamics.  For the entire system and the windowed distributions we develop accurate parameterizations for the universal scaling functions, including corrections to scaling and systematic error bars, facilitated by a novel software environment {\it SloppyScaling}. 

\end{abstract}


\maketitle

\section{Introduction}

Systems that have crackling noise and avalanches exhibit scale invariance and power laws, which point to the notion of underlying universality~\cite{sethna2001crackling}.  These systems include many of the best-studied examples of non-equilibrium critical phenomena, and much progress has been made in a renormalization group context~\cite{le2011distribution, kardar97interfaces}.  The renormalization group implies that the long length and time behavior near critical points is governed by universal exponents and scaling functions.  However, the predictive power of these theoretical studies has hiterto been underutilized; the primary focus of experiments and numerical simulations has been on precise estimates of critical exponents, rather than on the universal joint predictions of properties involving several control parameters and/or measured quantities.

A wide variety of materials and natural systems have been studied in the context of non-equilibrium critical phenomena.  Many of these systems exhibit avalanches which have power law size distributions.  These include Barkhausen noise in ferromagnets~\cite{DurinZapperi04,ryu07nature,MagniDurin09,schwarz04PRL,Christian06PRB}, fluid imbibition into porous media, flux-line depinning~\cite{barabasi95fractal, kardar97interfaces, TangLeschhorn92,Buldyrev1992PRA, buldyrev1992anomalous, leschhorn96PRE}, and martensitic transformations~\cite{martensite1994,martensite2004}, to name a few.  In the first three of these systems, avalanches are the result of the jerky motion of an interface (domain wall, fluid front, flux-line) in a disordered environment, and can be described by the same family of front-propagation models.

In this manuscript, we study the spatial structures of avalanches in a front-propagation model in two dimensions, developing tools and methods needed for systematic study and extraction of these multiparameter universal scaling functions. To illustrate the utility and importance of these functions, we apply them systematically to a practical experimental problem -- the size distributions of avalanches seen through a viewing window.  This problem illustrates (a)~the complexity and sophistication of the different emergent size distributions, (b)~the relationships amongst the probability distributions of heights, widths, and sizes and their utility in generating predictions for windowed avalanches, and
(c)~the use of functional forms and least-squares fits to analyze and report on these multiparameter scaling functions. 

Imaging experiments have been used in recent years to study a wide variety of systems exhibiting crackling noise or similar dynamics.  Barkhausen noise is measured making use of the magneto-optical Kerr effect~\cite{Puppin00, ryu07nature, MagniDurin09}, allowing one to examine the domain wall motion in 2D thin films in real-time.  In experiments on superconducting vortices~\cite{welling2005fluxavalanches}, a magneto-optical (MO) setup is also used.  In experiments of fracture~\cite{2001slowcrack}, fluid imbibition~\cite{planet2009imbibition} and granular systems~\cite{aegerter2003rice,aegerter2004rice}, the dynamics are also followed with a camera.

These visualization experiments provide an unusual opportunity: we now can study the universal properties of the spatial morphology -- various distributions of heights, widths, angles, local heights, etc of either the avalanches or the fronts.  However, the measurements of these properties are often distorted by a limited field of view.  We hereby take this problem and develop the scaling theory for the universal functions needed to characterize the results of a generic imaging experiment -- the distribution of avalanche sizes seen through windows.

The limited field of view in experiments distort the size distributions of avalanches, and cause difficulties in characterizing the critical exponents.  Naturally, there is a bias towards small avalanches; large ones are cut off by the boundaries of the window.  It can also distort the size distribution if pieces of large avalanches cut off by window boundaries are counted as small ones. Experiments have taken a variety of approaches to deal with such windowing effects: ``Laser reflectometry'' ~\cite{Puppin00} on Barkhausen noise uses the magneto-optical Kerr effect, but lacking spatial resolution, lumps fragments and avalanches together; meanwhile, other optical Kerr experiments have shown~\cite{MagniDurin09} that the effective size exponent $\tau$ for this lumped distribution depends strongly on the window width.  Work by Kim {\it et al.}~\cite{KimPRL03} report quite striking distributions but do not specify whether their data includes avalanches that touch the boundary.  In experiments on superconducting vortices, or magnetic flux avalanches~\cite{welling2005fluxavalanches} avalanches exceeding a certain size are discarded.  In fluid imbibition~\cite{planet2009imbibition}, the edges of the system are purposely left out to avoid any distortion produced by side walls.   In granular systems, where avalanche dynamics in piles of rice are studied with real-time reconstruction~\cite{aegerter2003rice,aegerter2004rice}, and fracture experiments~\cite{2001slowcrack}, where the dynamics are followed with a high-speed camera, boundary effects are not considered but may also be important.        

We will show comprehensively how to analyze all of the size data lying within a window, and how to use the different classes of avalanches to get independent measures of various critical exponents.  Indeed, window-width finite-size effects need not be avoided, but properly treated may provide additional measures of the critical exponents and the spatial structure.

Characterizing spatial structures of avalanches must go far beyond the traditional focus on critical exponents.   Many experiments report power laws, however through this study we emphasize that one can make predictions about both power laws {\it and} scaling shapes from data, as has been demonstrated in a previous study of avalanche temporal shapes~\cite{BeyondScaling}.  Indeed,
traditional scaling collapse methods fail for functions of more than
two variables.  To optimally extract the behavior and estimate errors,
we need to do simultaneous analysis of many different properties.  We thus
introduce a software environment, {\em SloppyScaling}~\cite{SloppyScaling}, which
facilitates the exploration and development of simultaneous fits of
multiple data sets with parameterized forms of universal scaling functions.   With this approach we are taking the first steps towards the use of scaling methods as a practical engineering tool.

\section{Summary of Key Results}
\label{sec:summary}

Since our theme is multifaceted, readers may be interested in focusing on different aspects of this work.  In this section we present an overview of key results in this paper and a summary of their corresponding sections.

\begin{figure*}[t]
\begin{center}
\includegraphics[width=16cm]{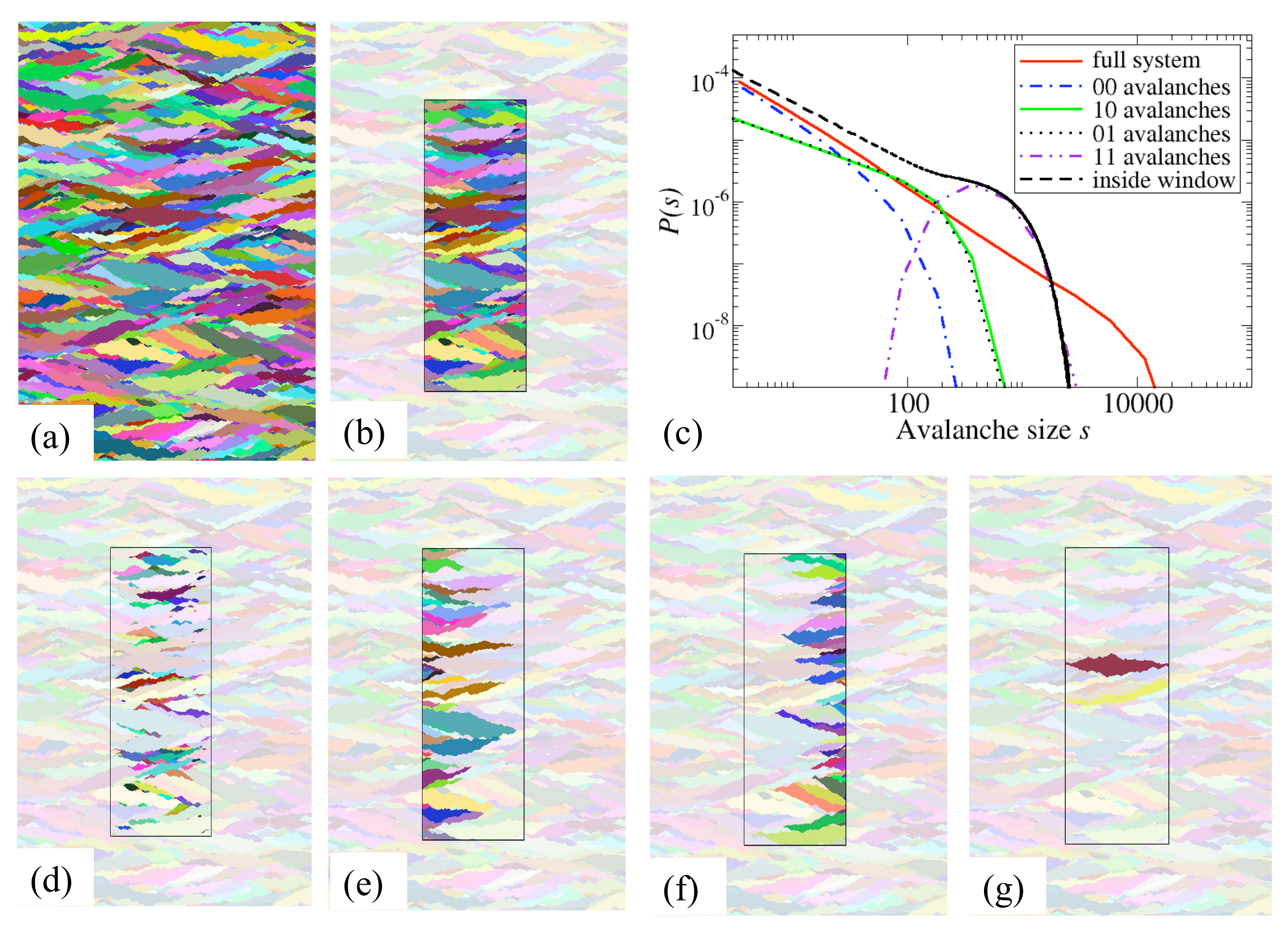}
\end{center}
\caption{{\bf Windowed distributions} (color online)
(a)~The full system of avalanches.  This figure shows a qKPZ simulation with avalanches.   (b)~A limited field of view:  we can only see part of the system.  The avalanches inside the artificial viewing window are brightly colored, and those outside are washed out.  Notice that the avalanches within the window are cut off at top and bottom, and (more importantly for short, wide avalanches with roughness exponent $\zeta<1$) on the two sides. (c)~The size distributions for the different types of avalanches:  (d)~internal 00 avalanches, (e)~split 10 avalanches, (f)~split 01 avalanches, and  (g)~spanning 11 avalanches
}
\label{fig:window_dist}
\end{figure*}

{\bf 1. Universal spatial structures of avalanches in directed percolation depinning.}   We provide a substantive analysis of the universal spatial morphology of avalanches in the quenched KPZ (qKPZ) model in 1+1 dimensions (the model is discussed in section ~\ref{sec:model}).   Figure~\ref{fig:window_dist}(a) shows avalanches in a typical simulation of this model.  Analogous to magnetic systems, we have added a ''demagnetization factor`` $k$ that parameterizes the restoring force, which allows us to access many metastable configurations of the front near the depinning transition, and controls the typical width of an avalanche, $L_k \sim k^{-\nu_k}$.  Figure~\ref{fig:ScaleInvariant} shows examples of these resulting avalanches from simulations of various $k$.  In section~\ref{sec:FullSystem}, we thoroughly examine the spatial structure of the avalanches, including sizes $s$ (the total area covered between pinned fronts, which would correspond to the total magnetization change in magnets, or the avalanche size of a rice pile), and also widths $w$ and heights $h$ (which measure the length of an avalanche in directions perpendicular and parallel to the direction of the motion of the front; this corresponds to studying the shapes of the magnetic domains or flux lines).  We examine and fit the distributions of these sizes, heights and widths in section~\ref{sec:FullSystem}.

{\bf 2. Avalanches in Windows.}  As mentioned in the Introduction, in many imaging experiments the limited field of view distorts the avalanche size distribution. This is illustrated in Fig.~\ref{fig:window_dist}(b) and (c). 
In Fig.~\ref{fig:window_dist}(b) most of the avalanches are
cut off by the left and right boundaries- if one were to count the area of these avalanches, we would count large ones as smaller ones. The resulting size distribution inside the window
(dashed line of Figure~\ref{fig:window_dist}(c)) has a very different power law
and shape compared to the full system distribution (solid line of Figure~\ref{fig:window_dist}(c)).

In section~\ref{sec:Window}, we show in detail how avalanches which cross
boundaries exhibit distinctly different size distributions and critical
exponents.  For simplicity, we consider a strip geometry where only the left/right boundaries are relevant (since avalanches in our system are flat and wide, few touch the top or bottom).  We can therefore separate avalanches into four different categories, 00, 01, 10, and 11, depending on whether avalanches touch (1) or do not touch (0) the left or right boundaries: internal avalanches (00) (Figure~\ref{fig:window_dist}(d)), split avalanches (10 and 01)(Figure~\ref{fig:window_dist}(e) and (f)), and spanning avalanches (11) (Figure~\ref{fig:window_dist}(g)).  The internal (00) avalanches share the same power law as that of the full distribution, with a cutoff controlled by both the system size $L_k$ and the window size $W$.  The split (10 or 01) avalanches have a modified power law- a smaller exponent as larger avalanches are counted as smaller halves. (See Figure~\ref{fig:window_dist})(c)).  The spanning avalanches also exhibit a smaller exponent, although this is not obvious in the shape of its scaling function.  We can see that it has both an outer cutoff due to $L_k$ and an  inner cutoff due to the window size $W$, since avalanches must be large enough
to span the window (purple dash-dotted line in Figure~\ref{fig:window_dist}(c)).  The internal and spanning avalanches also have distinct universal scaling functions with a cutoff controlled predominately by the window size for windows comparable to or smaller than the size of $L_k$.  In Section ~\ref{sec:Window} and Section ~\ref{sec:shapes} we give a thorough analysis of these modified power laws, the different scaling shapes, and the results of fits to data. 

Having established a sophisticated method of analyzing both experimental and simulation data, we can utilize this analysis to enhance the collection of data in visualization experiments.  Section~\ref{sec:Experiments} has some suggestions for how to collect data and simultaneously analyze the scaling behavior of different magnifications, and extract multiple exponents.  

{\bf 3. Functional forms}  A main emphasis of our work is that we fit an entire functional form instead of power laws~\cite{olga1999disorder}, this includes the shape of the scaling function, and analytic and singular corrections to scaling.  The benefit of approaching a scaling problem this way is that it allows us to account for both universal and non-universal effects in a consistent way.  Writing down the functional form that is given by the data for a certain universality class will also be useful for identifying and characterizing other systems that are thought to belong to the same universality class.

We have found that, to analyze the windowed distributions, we need to first thoroughly examine the spatial structure of avalanches for the full system.  In particular, in order to analyze the avalanche pieces left inside the window, we need to define height and width distributions and also joint distributions of sizes and widths. Section~\ref{sec:FullSystem} discusses these and also the results of fits for such distributions for the qKPZ model we are studying.  To focus on the scaling region, and minimize lattice effects, we will discuss these distributions in terms of {\it fractional area distributions}, $A(S) \propto S P(S)$, the average fraction of the system that a given size takes up.

A remarkable result is that size, height, and width distributions can be fitted
with a nearly identical functional form. For example the size distribution is:
\begin{align}
\label{eq:firstEq}
    A(S|L_k) &= (S/L_k^{1+\zeta})^{2-\tau} \calASL(S/L_k^{1+\zeta})/S \\
    	         &= S_k^{2-\tau} \exp((U_S S_k^{1/2} -Z_S S_k^{\delta_S}))/S. \nonumber
\end{align}
Here we have the shorthand $S_k = S/L_k^{1+\zeta}$.   The width $A(w|L_k)$ and height $A(h|L_k)$ distributions are similar, with the form of $\frac{1}{X} Y^{\alpha} \cal{A}(Y)$, where $X=\{S,h,w\}$ and $Y = \{ S/L_k^{1+\zeta}, h/L_k^\zeta, w/L_k\}$.  $\cal{A}(Y)$ is identical in form to the one quoted above in Eq~\ref{eq:firstEq}.

Fitting the size, height and width distributions at once, we can extract multiple exponents- not only the commonly measured size distribution exponent $\tau$, but also the exponent $\nu_k$ which, as mentioned in point 1, relates the typical width of an avalanche to $k$, and also the roughness exponent $\zeta$.  The roughness exponent $\zeta$, which measures the fluctuations of the interface, is typically quoted for front propagation models as a means of characterizing the universality class.  (In our analysis, we have found a range for the roughness exponent $\zeta$ (from $0.62$ to $0.72$) that differs from the literature value $0.63$~\cite{leschhorn96PRE};  we discuss this in Appendix~\ref{sec:zeta}.)   One must note that the functional forms we choose are a practical tool to summarize existing information.  While they may be inspired by analytical calculations, and chosen to be consistent with known asymptotics, they should be trusted only in the ranges over which they have been measured.
 
{\bf 4. Multivariable scaling problems.}  The size distribution in Eq.~\ref{eq:firstEq} has a scaling form with one scaling variable.  However, in this paper we will consider many scaling forms with more than one variable, such as a joint size and width distribution (\ref{sec:FullSystem}), or the windowed distributions (section~\ref{sec:Window} and ~\ref{sec:shapes}).  In these cases, two or more scaling variables are important for describing the shape of the distribution (as seen in Figure~\ref{fig:window_dist}(c)).  For example, the general form for the $11$ windowed distributions is:
\begin{align}
\All(s|W \, L_k)= \frac{1}{s} \left(\frac{s_k}{W_k}\right)^{(2-\tau)(1+\zeta)/\zeta} \calAll(s_k,W_k).
\end{align}
Here the scaling functions become distributions with two scaling variables, the rescaled size $s_k$ and rescaled window width $W_k$.  The traditional ``scaling collapse'' methods become problematic when multiple scaling variables are simultaneously important; this has hitherto retarded the effective study and use of these powerful universal joint distributions. 

We present a systematic method for analyzing scaling problems with multiple control variables.   In our problem, the two control parameters are the demagnetization factor $k$, and the window width $W$.  We will show that the interplay between $k$ and $W$ is important for determining the shape of the avalanche size distributions.  In particular, we can write scaling functions for the distributions of the avalanche pieces in terms of these two scaling variables $s_k = s/L_k^{1+\zeta}$ and $W_k = W/L_k$ (as seen in Figure~\ref{fig:window_dist}(c)).  For example, the scaling function for the 11 distribution is:
\begin{align}
\calAll(s_k, W_k) = \exp(-(T_{11}+U_{11} s_k^{1/2}+Z_{11} s_k^{\delta_{11}}  \cr
+D_{11} \left(\frac{s_k}{W_k}\right)^{m1}+  C_{11} \left(\frac{s_k}{W_k^{n_{11}}}\right)^{-m2})
\end{align}
Section~\ref{sec:shapes} discusses in detail the functions $\calAoo(s_k, W_k)$, $\calAlo(s_k, W_k)$, and $\calAll(s_k, W_k)$ that we consider.  One may also find the results of fits (figures and tables of parameters) in section~\ref{sec:Window}.

{\bf 5. SloppyScaling} The analysis in this paper is done in the software environment {\it SloppyScaling} and includes a Bayesian analysis of systematic error bars, which are explained in Appendix~\ref{subsec:systematic}.  {\it SloppyScaling} allows us to fit data without collapses, which as mentioned above is problematic when there is more than one scaling variable involved.  Included in the software setup are automatic fits of data to theory functions with nonlinear least-squares and ease of visualizing results.  This software may be applied to many different multivariable scaling problems, making full use of universality and the predictions of the renormalization group.  All of the fits in this paper and their corresponding figures (including axis labels) were generated directly and automatically using {\it SloppyScaling}.

\section{Model}
\label{sec:model}

\begin{figure}[tb]
\begin{center}
\includegraphics[width=8cm]{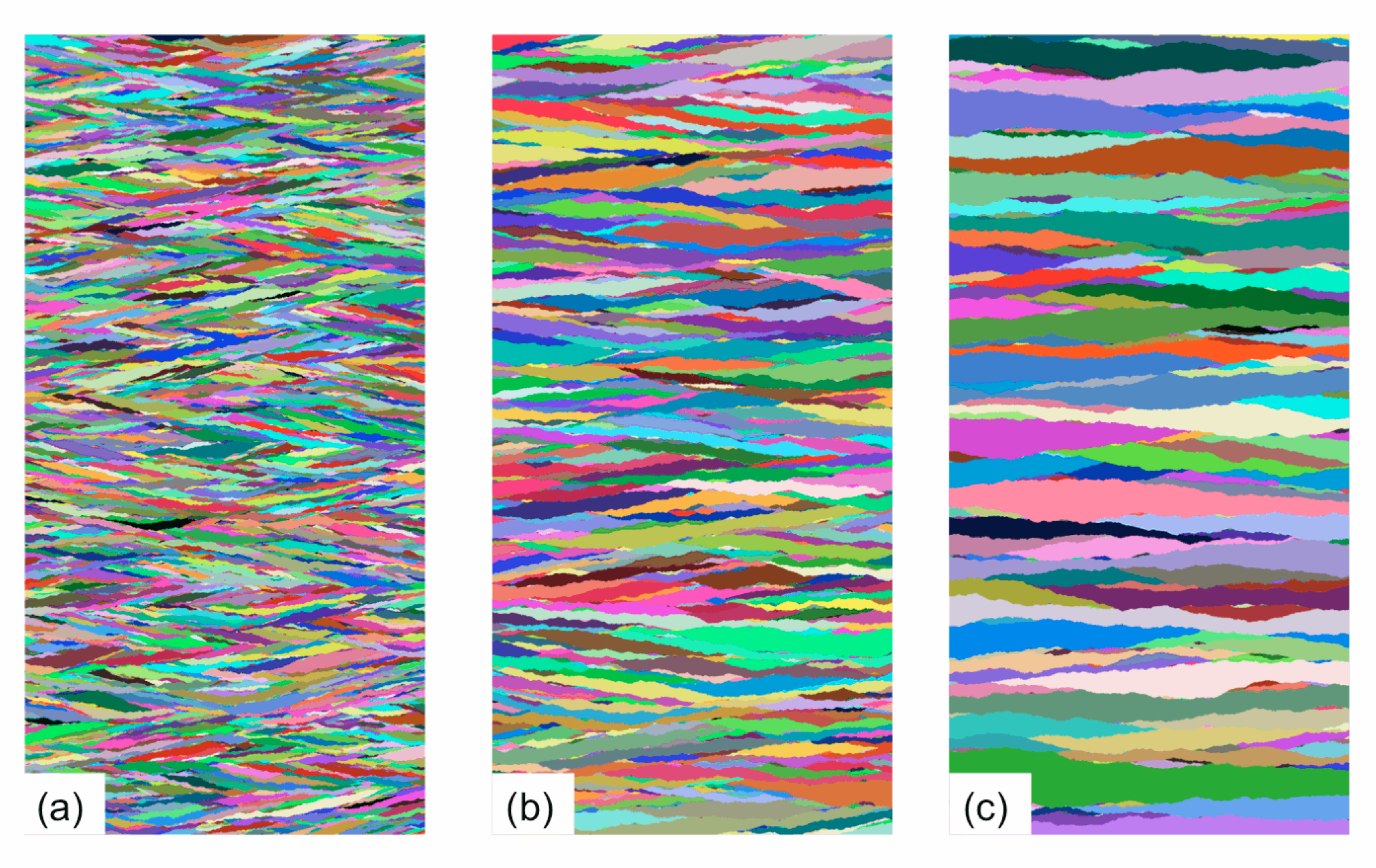}
\end{center}
\caption{{\bf Scale Invariance in Crackling Noise.} (color online) Three simulations of crackling noise, with different ``demagnetization fields", $k$.  (a)~$k=10^{-5}$ (b)~$k=10^{-6}$ (c)~$
k=10^{-7}$  ($k$ is what controls the typical avalanche sizes in a given system, giving a characteristic width $L_k$.)  Larger demagnetization fields stop avalanches more strongly, hence large $k$ corresponds to smaller avalanches.  The colored regions represent avalanches. The fronts are moving from bottom to top. Notice that the two simulations are statistically similar to one another apart from a rescaling of heights and widths. Note that most of the area is covered by the largest avalanches.}
\label{fig:ScaleInvariant}
\end{figure}

We use a model for imbibition fronts to produce avalanches, it is extensively studied, with well-established critical exponents.

We simulate crackling noise using a quenched KPZ model in 1+1 dimensions (see Appendix~\ref{subsec:numerics} for details on the implementation)~\cite{TangLeschhorn92}, with dynamics given by:
\begin{equation}
\frac{\partial h (x, t)}{\partial t } = F -k \langle h \rangle + \gamma \nabla^2 h + \lambda (\nabla h)^2 + \eta(x,h)
\label{eqn:qKPZ}
\end{equation}
where $h(x,t)$ is the height of the front, $F$ the driving force increasing quasistatically, linear and non-linear terms for the KPZ model controlled by the parameters $\gamma$ and $\lambda$ respectively, and $\eta$ Gaussian quenched disorder.  In the spirit of magnetic avalanche models~\cite{colaiori2008exactly}, we have added a term analogous to a demagnetization field $ - k \langle h \rangle$, which allows us to have pinned fronts at many metastable configurations.  Models like these have been simulated~\cite{barabasi95fractal} mostly near the depinning transition (and with $k=0$).  With $k \neq 0$, we define the area between each pinned front as an avalanche of size $S$.  Avalanches produced by this model are thought to belong to the directed percolation depinning (DPD) universality class ~\cite{leschhorn94PRE,leschhorn96PRE}.  The avalanches are self-affine, long and wide, with $\zeta<1$.   The roughness exponent $\zeta$ characterizes the ruggedness of the front, and also governs the scaling of avalanche heights $h$ with widths $w$, $h \propto w^\zeta$~\cite{barabasi95fractal,RossoKrauth01}.  In our model, $k$ controls the typical size of the avalanche (Figure~\ref{fig:ScaleInvariant}), the larger the $k$ the smaller the typical size of an avalanche.  We define the characteristic width of an avalanche in the full, unwindowed system to be $L_k = k^{-\nu_k}$.  We will write all the scaling forms in terms of the length $L_k$ rather than the demagnetizing factor
$k$ directly to emphasize the analogies to finite-size scaling, as we are also studying finite {\it window} sizes compared to the size of $L_k$.

\begin{figure*}[t]
\begin{center}
\subfigure[Size distribution (probability)]{\label{fig:ps-a}\includegraphics[scale=0.35]{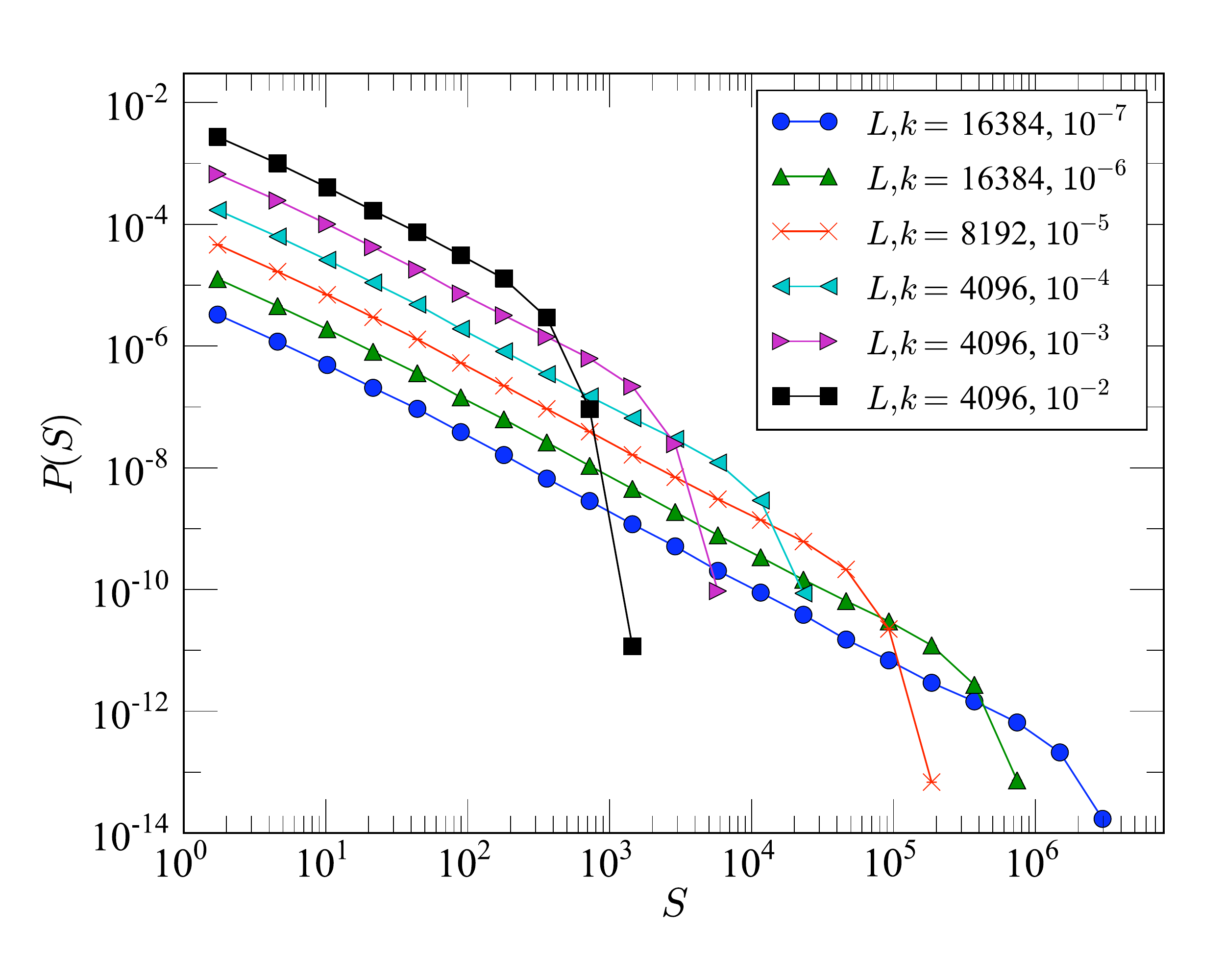}}
\subfigure[Area-weighted size distribution]{\label{fig:ps-b}\includegraphics[scale=0.35]{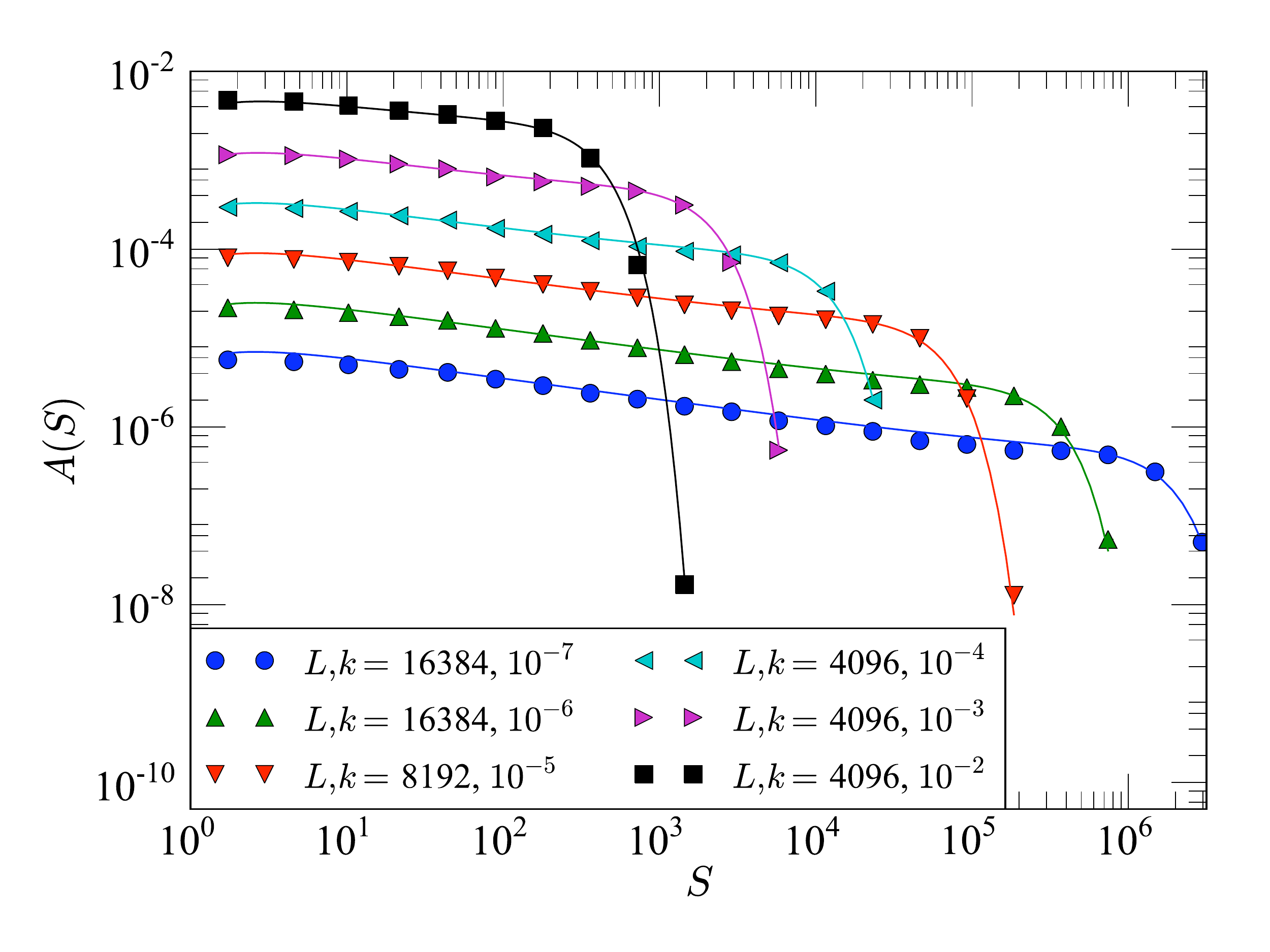}}
\caption{{\bf  $A(S)$ vs $P(S)$.}  (color online) Here one can see the difference between the more traditional $P(S)$ and the area-weighted $A(S)$.  (a)~$P(S)$: most of the area under the curve is from small avalanches, where non-universal lattice effects are important. (b)~$A(S)$: the normalization is dominated by large avalanches, avoiding the lattice effects, so that we can focus instead on the dependence on the large scale cutoffs- $L_k$ and, in later sections the window size.  The data here is from qKPZ simulations of different $k$ with different simulation size $L$.  The lines in (b) are a result of a joint fit with the maximum height and width distributions.  The fitting function and fitting parameters are shown in Table~\ref{table:shw-fit}.}
\label{fig:ps}
\end{center}
\end{figure*}

\section{Avalanche Shapes and Distributions of the Full System}
\label{sec:FullSystem}

In this section we will introduce various avalanche spatial distributions and their scaling forms.  These forms will allow us to motivate the windowed distributions, and also serve as an example where traditional collapses may lead to questionable results.  

\subsection{Area-weighted size distributions}

Traditionally, to describe an avalanche size distribution, we write the
probability distribution as a power law times a universal scaling function. For
example:
\begin{equation}
P(S|L_k) = S^{-\tau} {\cal P}(S/L_k^{1+\zeta}).
\label{eq:P(S)}
\end{equation}
$L_k$ is the characteristic width of an avalanche, and $h \sim L_k^\zeta$ the
typical largest height, and therefore the appropriate scaling variable to describe
the area of an avalanche is $S_k = S/L_k^{1+\zeta}$.

When studying the spatial properties as in our case, the probability distribution $P(S)$ (Eq.~\ref{eq:P(S)}) is not
the best choice, as its normalization is highly affected by non-universal
effects at the lattice spacing (Figure{~\ref{fig:ps}}). In fact, for $\tau > 1$,
the normalization integral
\begin{align}
N^{-1} &= \int_{a^2}^\infty P(S|L_k)\, dS \sim \int_{a^2}^{L_k^{1+\zeta}}
S^{-\tau}\, dS  \cr
&\sim a^{2(1-\tau)}-L_k^{(1-\tau)(1+\zeta)}
\end{align}
diverges at its lower (ultraviolet) limit $a\rightarrow 0$, but not for $L_k
\rightarrow \infty$. Although we could study
scaling functions that include a lattice cutoff, it is more interesting to focus
on the large (infrared) avalanche cutoff, which depends on
$L_k$ in a universal way. To this end, it is more appropriate to
make use of the first moment of $P(S)$, and to
consider the {\it area-weighted size distribution} $A(S)$
\begin{equation}
A(S) \sim S \cdot P(S).
\end{equation}
$A(S)dS$ has a natural physical interpretation: it is the fraction of the
full system area covered by avalanches with sizes between $S$ and $S+dS$. Its
scaling form is thus:
\begin{align}
\label{eq:AofSandkappa}
  A(S|L_k) &= L_k^{(\tau-2)(1+\zeta)} S^{1-\tau} \calASL(S/L_k^{1+\zeta}) \cr
      &= (S/L_k^{1+\zeta})^{2-\tau} \calASL(S/L_k^{1+\zeta})/S \cr
      &= S_k^{2-\tau} \calASL(S_k)/S.
\end{align}
We use the subscripts $Sk$ to distinguish the scaling of the size distributions
governed by $k$ from those governed by other control variables. The power of
$L_k$ we pull out of the scaling function is needed to normalize $A(S)$ to one,
since $A(S)$ is sensitive to the long-distance cutoff. In particular, 
\begin{widetext}
\begin{align}
N^{-1} &= \int_{a^2}^\infty A(S|L_k) dS 
= \int_{a^2}^\infty S_k^{2-\tau} \calASL(S_k)/S dS 
= \int_{a^2/L_k^{1+\zeta}}^\infty S_k^{2-\tau} \calASL(S_k)/S_k dS_k 
= \int_{a^2/L_k^{1+\zeta}}^\infty S_k^{1-\tau} \calASL(S_k) dS_k \cr
&= \int_{0}^\infty S_k^{1-\tau} \calASL(S_k) dS_k -
\int_0^{a^2/L_k^{1+\zeta}}S_k^{1-\tau} \calASL(S_k) dS_k
\approx 1 - \calASL(0)(a^2/L_k^{1+\zeta})^{2-\tau} \approx 1 \nonumber
\end{align}
\end{widetext}
where the last integral converges for $\tau < 2$ \footnote{All of the front
propagation models have $1<\tau<2$, but we
would need to use the second moment $S^2$ for the 3D nucleated
RFIM~\cite{SethnaDahmen04}, which has $\tau\sim 2.06$.} and
becomes small as $L_k$ becomes large. Notice that the normalization of a power
law must either diverge at the lower or
at the upper limit. By studying avalanches weighted by their first moment, the
normalization depends explicitly on $L_k$, the infrared cutoff. This is the
regime we are mainly interested in, since we would like to study the finite size
effects imposed by both $k$ and the window size $W$.  Alternatively, as done
in~\cite{Rosso07}, one may also define a scale $S_m = \frac{\langle S^2 \rangle}{2 \langle S\rangle}$,
redefine the sizes as $S/S_m$ and a corresponding $p(S/S_m)$ which has
normalization $\int_0^\infty \, ds sp(s)=1$. Here $p(s)$ is universal, but is
also not a probability distribution in the conventional sense. This definition
has an effect equivalent to what we do here: to make the function universal and
insensitive to non-universal lattice effects on normalization. Namely, our
definitions are related in the following way: $A(S) = \frac{S}{S^2_m} p(S/S_m)$
and $\calASL(S/S_m) = (S/S_m)^2 p(S/S_m)$. Here we use $S_m \equiv S_k
= S/L_k^{1+\zeta}$, which is consistent with their definition of $S_m$ up to a
constant factor. We prefer to focus on the more directly interpretable
fractional area distribution $A(S)$.

Furthermore, notice that we have been unorthodox in writing the scaling
form~(\ref{eq:AofSandkappa}) for $A(S|L_k)$ with a power of both $L_k$ and of
$S$ outside the scaling function. Normally one factors out a single variable
from the scaling function. For example, one could in principle write 
\begin{equation}
\label{eq:BSL}
A(S|L_k) = \frac{1}{S} {\cal B}_{Sk}(S_k).
\end{equation}
In this form, $A(S|L_k) \, dS$ is invariant under rescaling, and also clearly
preserves normalization. However, 
$\calB(S_k) = S_k^{2-\tau} \calA(S_k)$ vanishes as $S_k\to 0$, so this form
of the scaling function disguises the power law behavior of the avalanche size
distribution. By choosing the form $\calASL$ which is defined to be finite and
non-zero as $S\to 0$, we make manifest both the avalanche size dependence and
the system size dependence.

\subsection{Maximum height and width distributions}
\label{subsec:heightwidth}

In addition to the size distributions, we can also study the avalanche height
and width distributions. We define height along the direction of front
propagation, measuring the maximum height of an avalanche, and width
perpendicular to heights, measuring the maximum width of an avalanche. 

How do the height and width distributions scale? An avalanche of height $h$
has size $S \approx h \, w \sim h \, h^{1/\zeta} = h^{(1+\zeta)/\zeta}$,
so the system area $A(h|L_k) dh$ covered by avalanches with heights between $h$
and $h+dh$ scales as
\begin{align}
A(h|L_k) dh 
  &\sim A(S|L_k) dS \cr
  &= L_k^{(\tau-2)(1+\zeta)} S^{1-\tau} \calASL(S/L_k^{1+\zeta}) \, dS \cr
  &= L_k^{(\tau-2)(1+\zeta)} 
		h^{(1-\tau)(1+\zeta)/\zeta} \calAhL(h/L_k^\zeta) \, dS \cr
\end{align}
and since $dS/dh \sim h^{1/\zeta}$, 
\begin{align}
\label{eq:A_h}
A(h|L_k) &= L_k^{(\tau-2)(1+\zeta)} 
		h^{(2-\tau)(1+\zeta)/\zeta - 1} \calAhL(h/L_k^\zeta) \cr
    &= (h/L_k^\zeta)^{(2-\tau)(1+\zeta)/\zeta} \calAhL(h/L_k^\zeta)/h \cr
&= h_k^{(2-\tau)(1+\zeta)/\zeta} \calAhL(h_k)/h
\end{align}
where $h_k = h/L_k^\zeta$ is the appropriate scaling variable to describe the
avalanche height.

Similarly, we can write the scaling form for the width distributions as:
\begin{align}
\label{eq:A_w}
A(w|L_k) &= (w/L_k)^{(2-\tau)(1+\zeta)} \calAwL(w/L_k)/w \cr
&= w_k^{(2-\tau)(1+\zeta)} \calAwL(w_k)/w
\end{align}
where $w_k = w/L_k$ is the scaling variable for widths. 

Notice the pattern in all these formulas: $A(X|L)$ is a function with $\calA(Y)$ of
scaling variables (combinations of $X$ and $L$ invariant under the
renormalization group rescaling) multiplying a power of the scaling variable,
divided by the independent variable $X$, making $A(X|L) dX$ invariant under
rescaling, allowing it to be universal.

\begin{figure*}[htp]
\begin{center}
\subfigure[ Maximum
heights]{\includegraphics[height=6.5cm]{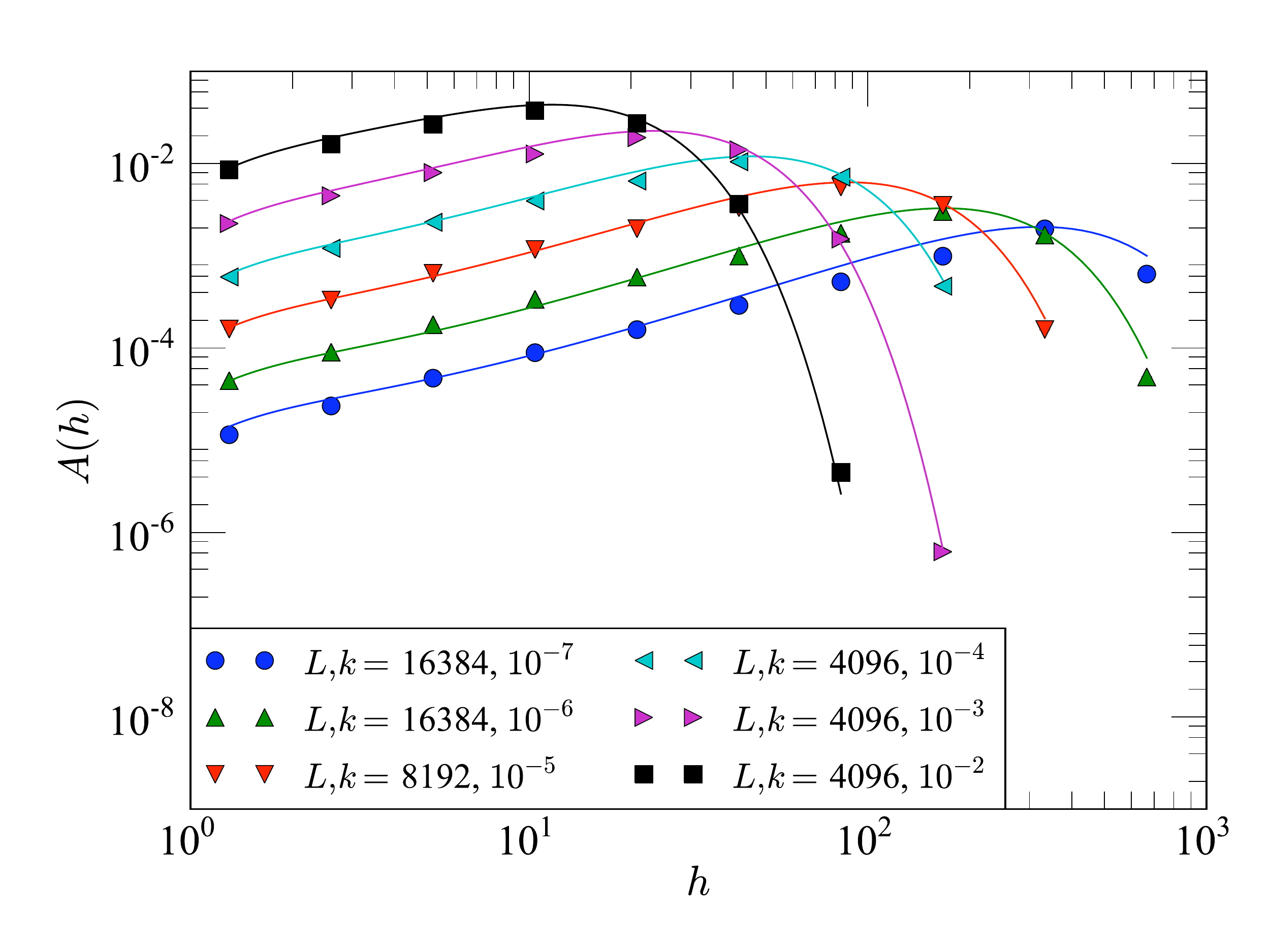}\label{fig:Ah}} \quad
\subfigure[ Maximum
widths]{\includegraphics[height=6.5cm]{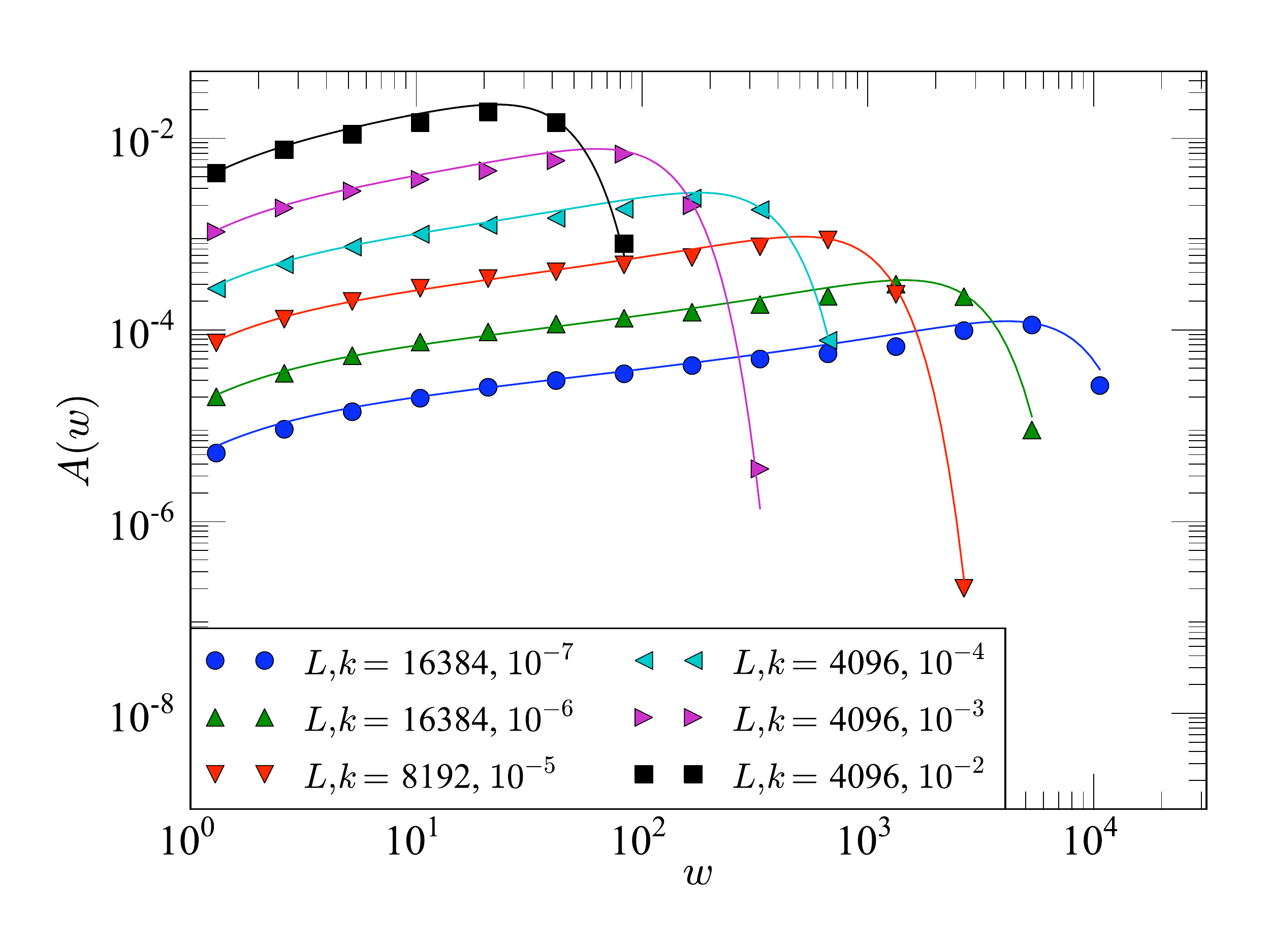}\label{fig:Aw}}
\caption{{\bf Area-weighted avalanche distributions for maximum heights and
widths} (color online) Area-weighted avalanche distributions of (a) maximum heights, and (b)
maximum widths for qKPZ simulations at different $k$, and simulation sizes $L$ (dots are binned data). The critical exponents were jointly fit with the size distributions $A(S|L_k)$ of
Fig.~\ref{fig:ps}, using the scaling forms of Eqs.~\ref{eq:A_h},~\ref{eq:A_w},
and~\ref{eq:Usf}. The best fit values for the critical exponents, parameters for
universal scaling functions, and non-universal corrections are given in
Table~\ref{table:shw-fit}.}
\label{fig:A_h_w}
\end{center}
\end{figure*}

\subsection{Joint fit of size, height and width distributions}
\label{sec:joint_fit}

In Figs.~\ref{fig:ps-b},~\ref{fig:Ah},~\ref{fig:Aw} we show the size, height
and width distributions for our qKPZ simulations at
various $k$, and at different simulation sizes $L$. The curves are theoretical fits
using a functional form of the type
\begin{align}
\label{eq:Usf}
\calA(Y) = &\exp(U_x Y^{1/2} - Z_x Y^{\delta_x})\exp(A^x_1/x+ A^x_2/x^2).
\end{align}
where $x = \{S, h, w\}$, and $Y = \{S_k, h_k, w_k\}$, and ${U_x, Z_x,
\delta_x, A_1^x, A_2^x}$ are (5x3) fitting parameters (results are listed in Table~\ref{table:shw-fit}). The first exponential term is the universal scaling function, while the second
accounts for non-universal analytic corrections at small $x$ due to lattice
effects~\cite{SethnaDahmen04}.  (See Appendix~\ref{subsec:corrections}).

One may ask why we choose this particular scaling form. When fitting data to a
function, there are many parameterizations we could use to describe the data.
This form is motivated from a functional renormalization group expansion by Le
Doussal and Wiese for static avalanche size distributions in a linear model~\cite{DoussalWiese09}. Our model differs in that
there is a nonlinear KPZ term leading to anisotropy, so that our avalanches
belong to a different universality class, the DPD universality class. 

Le Doussal and Wiese find the avalanche size distribution for the linear model,
for all static universality classes (random-bond, random-field, and
random-periodic), to leading order in $d = 4-\epsilon$ (where $d$ is the
dimension of the interface), to be:
\begin{equation}
P(S) \propto S^{-\tau} \exp(C(S/S_m)^{1/2}-\frac{B}{4}(S/S_m)^\delta)
\label{eq:FRG}
\end{equation}

Here their scaling form includes the large scale cutoff $S_m$. Le Doussal and
Wiese claim that their results for both static and dynamic avalanches agree up to one loop for
systems with $\zeta < 1$. Static avalanches are separated by equilibrium
configurations (or ground states), and dynamic avalanches are connected through
a sequence of metastable states. Our avalanches result from a moving interface
near the depinning transition, so they belong to a dynamic universality class.
We thus use Eq.~\ref{eq:FRG} as inspiration for the scaling form of 
Eq.~\ref{eq:Usf}. However, one may note that there is no theoretical basis that 
it should work, since it is from a distinctly different universality class.

One cannot determine the values of $\nu_k$ and $\zeta$ independently, if we fit the size, height, or width distributions with Eqns.~\ref{eq:AofSandkappa}, ~\ref{eq:A_h}, and ~\ref{eq:A_w} separately.  For example in the size distribution we can only determine the combination $\nu_k (1+\zeta)$.    We determine the three critical exponents $\tau$, $\nu_k$, and $\zeta$, by jointly fitting the size, height and width distributions.

The results of our fits are reported in Table~\ref{table:shw-fit} and shown in Figures~\ref{fig:ps-b} and ~\ref{fig:A_h_w}. In particular, we find $\zeta = 0.62 \pm 0.02$ which is close to the highly-precise value of
$\zeta = 0.63$ found in the literature~\cite{RossoKrauth01,
buldyrev1992anomalous, tang1992pinning, sneppen1992PRL, Amaral1995PRE}.
We note that the parameters $\delta_x$, which in principle control the
aymptotic decay of the scaling function, are estimated here from a fit to the entire distribution. The quoted errors do not represent a
confidence on the asymptotic decay - merely a confidence in the predictions
over the range where data has been fit.

\begin{widetext}
\begin{table*}[b]
\centering
\begin{tabular}{c | c | c | c}
\hline
~~~~~parameter~~~~&~~~~~~~best fit~~~~~~~&standard errors&systematic errors\\
~~~~~~~~~~~~~~~~~&~~~~~~~&in linear approx.&\\
\hline
\hline
\multicolumn{4}{l}{Universal Exponents} \\
\hline
\multicolumn{4}{l}{Shared Exponents}\\
\hline
$\tau$ & 1.2414 & $\pm$ 0.0006 & $\pm$ 0.04 \\
$\nu_k$ & 0.4513 & $\pm$ 0.0001 &$\pm$ 0.008\\
$\zeta$ & 0.6155 & $\pm$ 0.0004 &$\pm$ 0.02 \\
\hline
\multicolumn{4}{l}{$\calASL(S_k) = \exp(U_S S_k^{1/2} - Z_S S_k^{\delta_S})$}
\\
\hline
$U_{S}$ & 0.173 & $\pm$ 0.003 &$\pm$ 0.2 \\
$Z_{S}$ & 0.0099 & $\pm$ 0.0002 &$\pm$ 0.01 \\
$\delta_{S}$ & 1.832 & $\pm$ 0.004 &$\pm$ 0.3 \\
\hline
\multicolumn{4}{l}{$\calAhL(h_k) = \exp(U_h h_k^{1/2} - Z_h h_k^{\delta_h})$}
\\
\hline
$U_{h}$ & 0.94 & $\pm$ 0.01 &$\pm$ 0.9 \\
$Z_{h}$ & 0.307 & $\pm$ 0.004 &$\pm$ 0.3 \\
$\delta_{h}$ & 1.255 & $\pm$ 0.003&$\pm$ 0.2 \\
\hline
\multicolumn{4}{l}{$\calAwL(w_k) = \exp(U_w w_k^{1/2} - Z_w w_k^{\delta_w})$}
\\
\hline
$U_{w}$ & 0.401 & $\pm$ 0.005 &$\pm$ 0.6 \\
$Z_{w}$ & 0.0291 & $\pm$ 0.0004 &$\pm$ 0.2 \\
$\delta_{w}$ & 2.202 & $\pm$ 0.005 &$\pm$ 0.9 \\
\hline
\hline
\multicolumn{4}{l}{Non-Universal Exponents}\\
\hline
\multicolumn{4}{l}{$ \exp(A_1^{x}/x + A_2^{x}/x^2)$}\\
\hline
$A_1^s$ & -0.36 & $\pm$ 0.03 &$\pm$ 2 \\
$A_2^s$ & -0.35 & $\pm$ 0.06 &$\pm$ 4 \\
$A_1^h$ & 2.30 & $\pm$ 0.04 &$\pm$2\\
$A_2^h$ & -1.90 & $\pm$ 0.04 &$\pm$ 2\\
$A_1^w$ & -0.99 & $\pm$ 0.03 &$\pm$ 2 \\
$A_2^w$ & -0.06 & $\pm$ 0.03 &$\pm$ 1
\end{tabular}

\caption{{\bf Best Fit exponents and parameters} Here are the results of our
joint fit for the size $A(S|L_k)$, width $A(w|L_k)$, and height $A(h|L_k)$
distributions. The corresponding universal scaling forms which were fit are
quoted in the table alongside the parameter results; on the bottom of the table
are multiplicative corrections for each distribution, with $x$ equal to either
$S$, $w$, and $h$. Here systematic error bars which account for errors in the
theory (see Appendix~\ref{subsec:systematic} for explanation) are given. The traditional standard
error bars are typically $\sim 64$ times smaller than the systematic error bars
quoted; however, they are a gross underestimate of the actual errors expected
since our theory is both highly nonlinear and sloppy~\cite{FrederiksenPRL04}. 
We quote each parameter to the significant figure indicated by its standard
error, since the parameters are strongly correlated, truncating each parameter
to its significant figure would yield a poor fit.}
\label{table:shw-fit}
\end{table*}
\end{widetext}

\begin{figure}[htp]
\begin{center}
\includegraphics[scale=0.35]{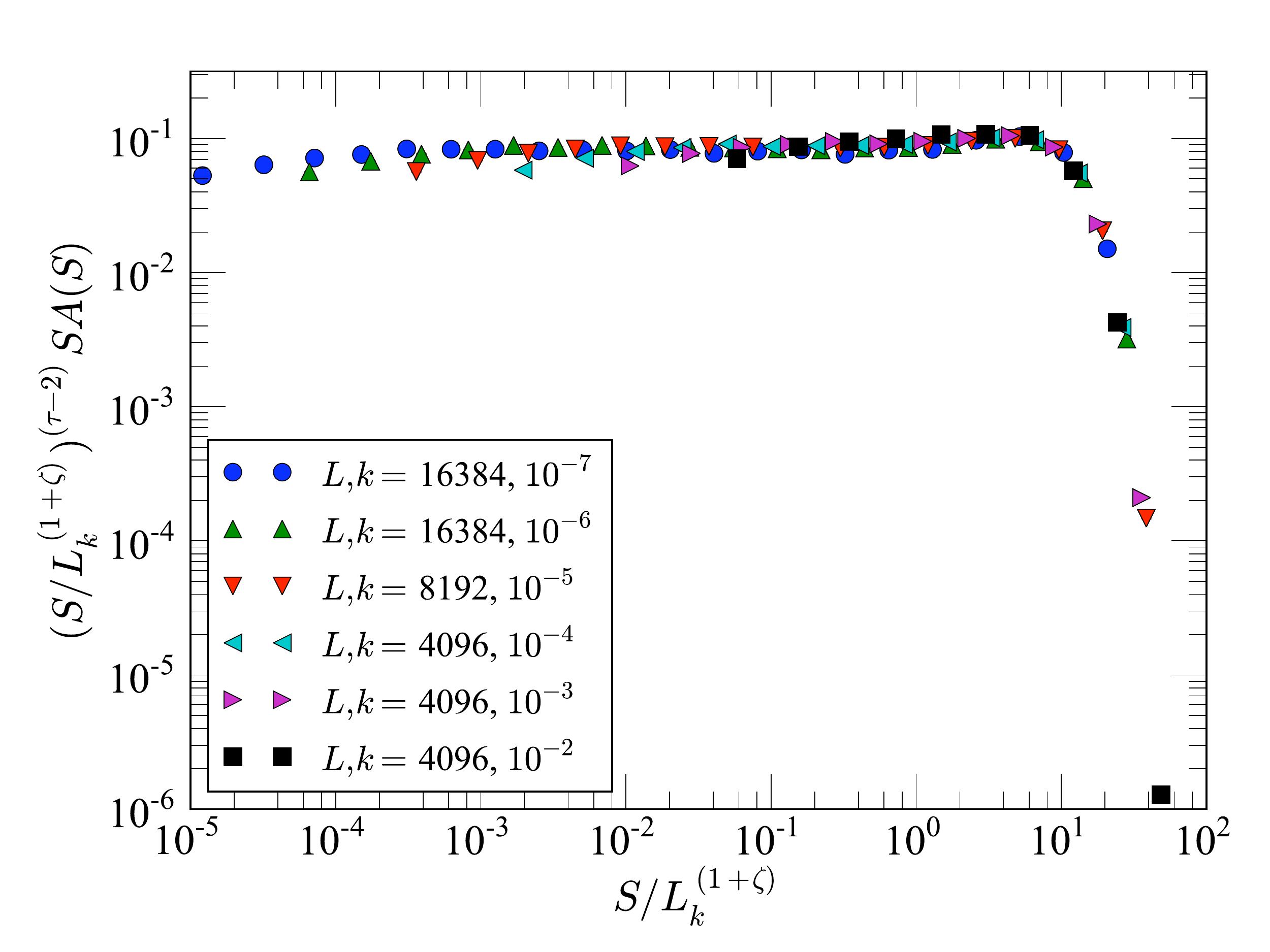}
\caption{{\bf  Size distribution $A(S|L_k)$ scaling collapse} (color online) We collapse the size distribution data using the universal exponents of $\tau=1.24$, $\nu_k=0.45$, and $\zeta=0.62$, the best fit values of the joint fit between $A(S|L_k)$, $A(h|L_k)$ and $A(w|L_k)$.}
\label{fig:Ascollapse}
\end{center}
\end{figure}

\begin{figure}[htp]
\begin{center}
\includegraphics[scale=0.35]{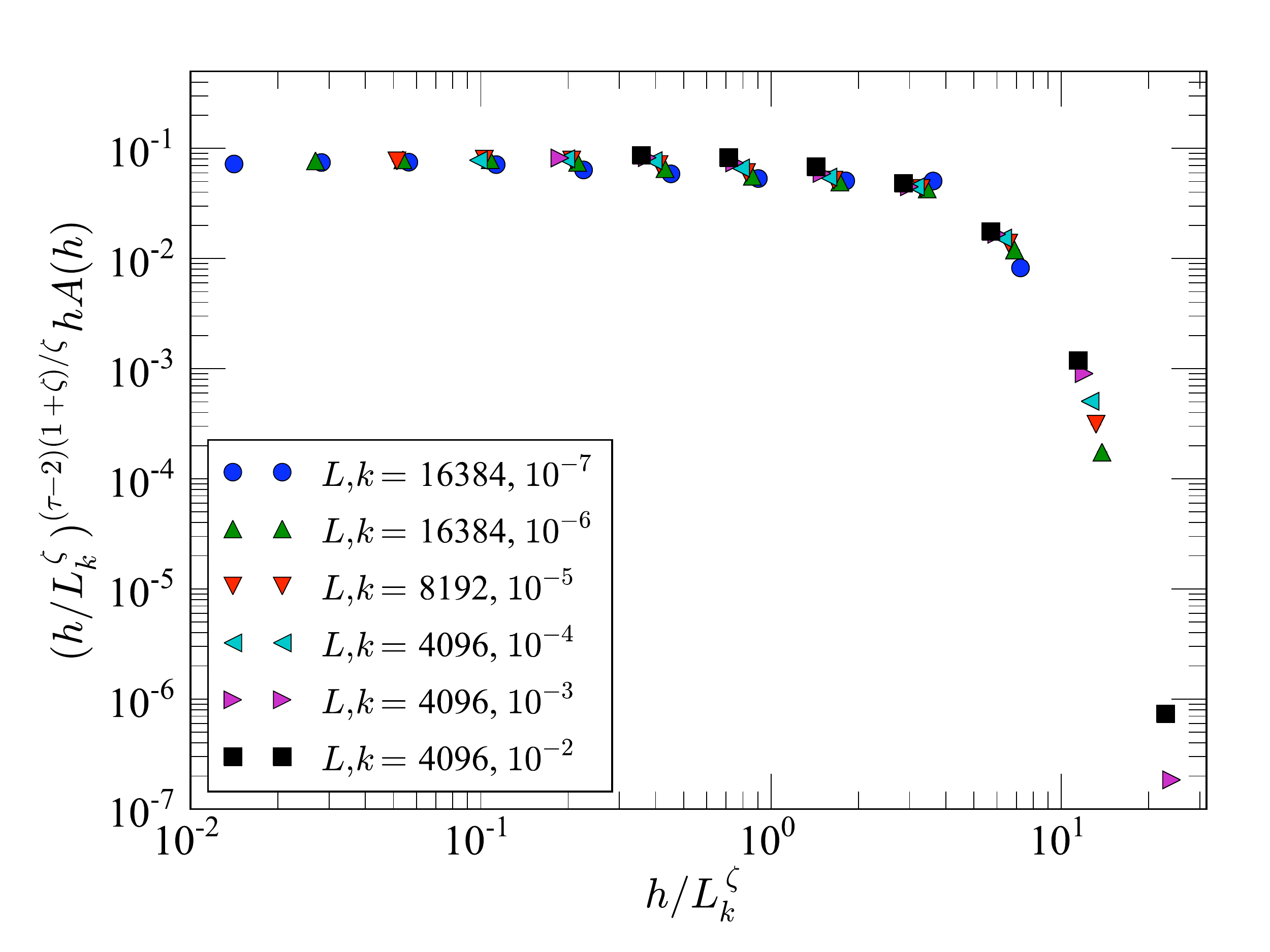}
\caption{{\bf  Height distribution $A(h|L_k)$ scaling collapse} (color online) We collapse the simulation data using the universal exponents of $\tau=1.24$, $\nu_k=0.45$, and $\zeta=0.62$, the best fit values of the joint fit between $A(S|L_k)$, $A(h|L_k)$ and $A(w|L_k)$.}
\label{fig:Ahcollapse}
\end{center}
\end{figure}

\begin{figure}[htp]
\begin{center}
\includegraphics[scale=0.35]{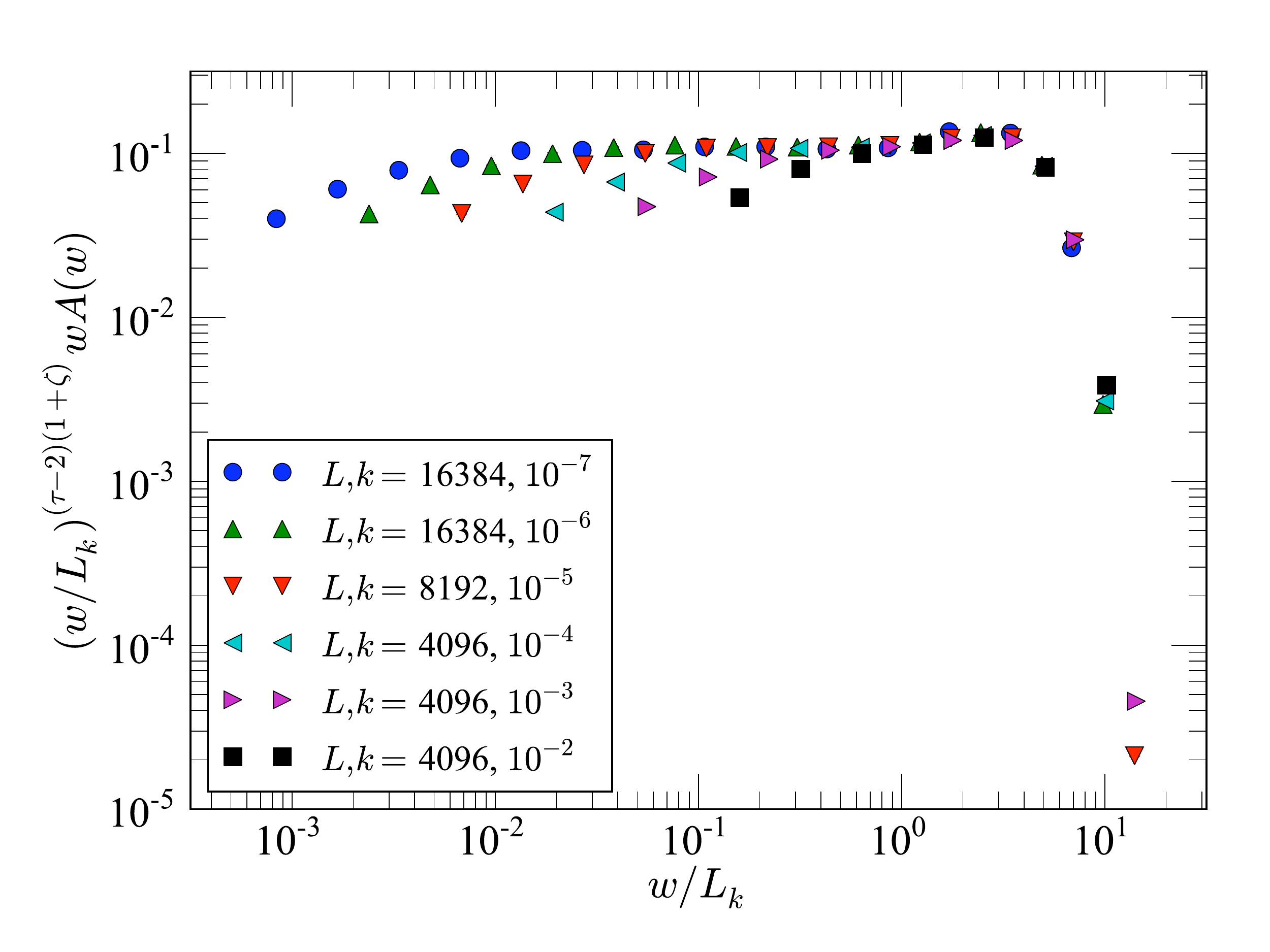}
\caption{{\bf  Width distribution $A(w|L_k)$ scaling collapse} (color online) We collapse the width distribution data using the universal exponents of $\tau=1.24$, $\nu_k=0.45$, and $\zeta=0.62$, the best fit values of the joint fit between $A(S)$, $A(h)$ and $A(w)$.}
\label{fig:Awcollapse}
\end{center}
\end{figure}

Finally, the respective scaling collapses for the size, height, and width distributions are shown in Figures~\ref{fig:Ascollapse},~\ref{fig:Ahcollapse}, and~\ref{fig:Awcollapse}.  Although scaling collapses are very useful in verifying critical behavior, we argue that they may be problematic for the purpose of determining critical exponents, and one should fit and make use of functional forms.  In Appendix~\ref{subsec:corrections} we will show how scaling collapses are unable to incorporate the effects of corrections to scaling, and how these corrections may cause a drift in the critical exponents.

\subsection{Local height distributions}
\label{subsec:localheights}

\begin{figure}[t]
\begin{center}
\includegraphics[scale=0.35]{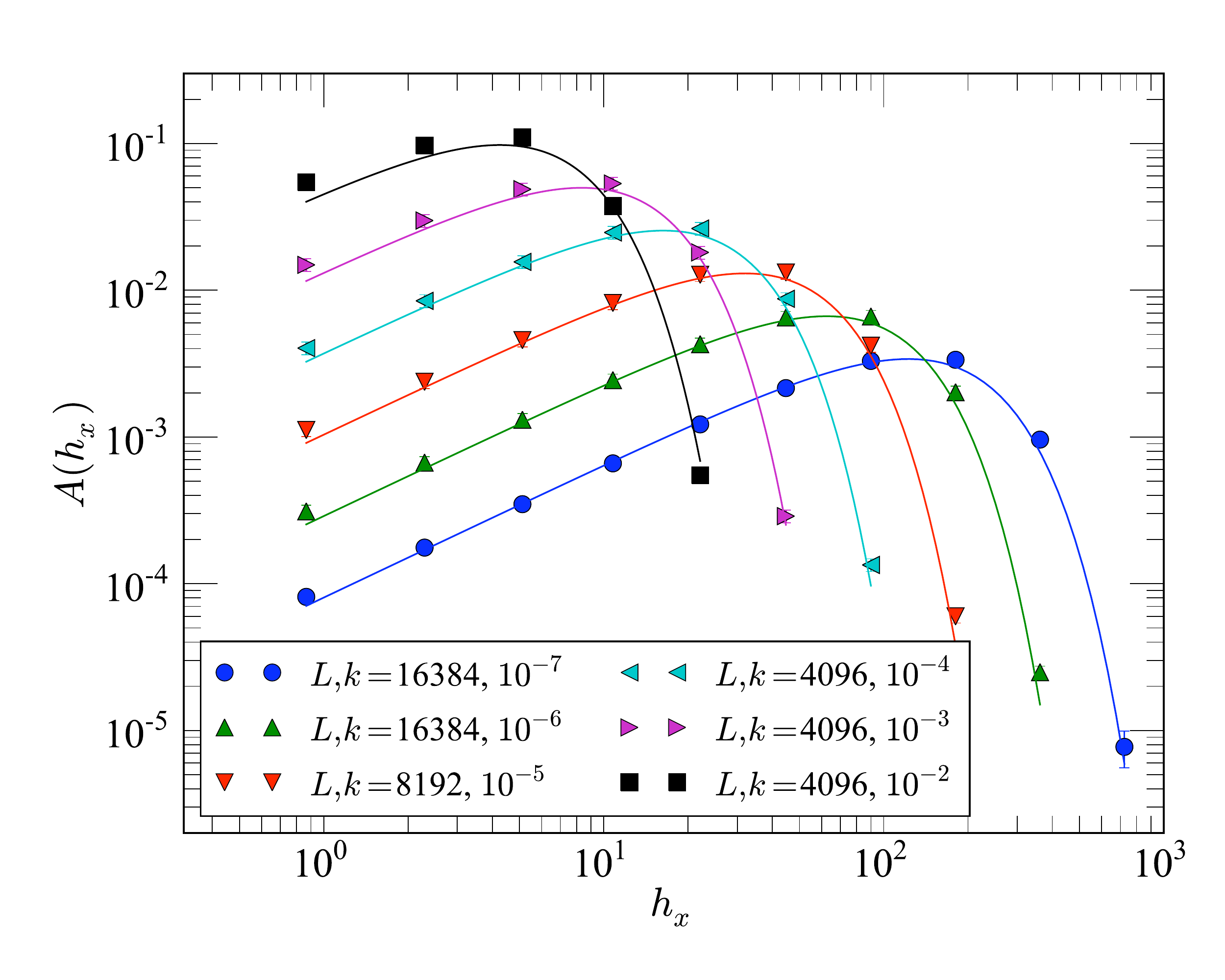}
\caption{{\bf Area-weighted local height distributions} (color online) Here are the area-weighted local height distributions, the fraction of area taken up by a cross sectional height $h_x$.  The fits shown in the figure were with the form of Eq.~\ref{eq:AhxL}, where ${\cal A}_{h_xk}$ is the scaling function of the fit to $11$ spanning avalanches  (Eq:~\ref{eq:A11function}) which cross both window boundaries, taking $W=0$.  Details for this function are explained in section~\ref{sec:Window}.}
\label{fig:AhxL}
\end{center}
\end{figure}

In our analysis of the next section, we will make use of another scaling function of the same form: not the maximum height of an avalanche, but the {\em distribution} of heights given by random cross sections of avalanches. Let $A(h_x|L_k) \, dh_x$ be the fraction of the system area consisting of points $(x,y)$ where the enclosing avalanche has vertical  cross-sectional height at $x$ in the range $(h_x, h_x + dh_x)$, then,
\begin{align}
\label{eq:AhxL}
A(h_x|L_k) \sim (h_x/L_k^\zeta)^{(2-\tau)(1+\zeta)/\zeta} {\cal A}_{h_x k} (h_x/L_k^\zeta)/h_x.
\end{align}
This distribution gives a different measure of the typical shape of an avalanche.  The local height distribution is shown in figure~\ref{fig:AhxL}.  Here the curves show a fit we have generated with a scaling function of the fit to $11$ spanning avalanches (Eq~\ref{eq:A11function}) which cross both window boundaries, taking the limit of $W=0$.  Our best measure of the local heights is equivalent to a $11$ distribution with window size $1$.  Details for how this works are provided in the section~\ref{subsec:11}.

\subsection{Joint distributions and multivariable scaling functions}
\label{subsec:joint}

Once we have distributions for the measures of size $S$, width $w$ and height $h$, we can also explore the forms of joint distributions. The area $A(w,S|L_k) \, dS\, dw$ of avalanches in the range of size $(S, S+dS)$ and widths $(w, w+dw)$ will go to zero strongly if the size $S$ becomes either much larger than or much smaller than the typical size $w^{1+\zeta}$ of an avalanche of width $w$ -- so we may factor out any combination of powers of $w$ and $S$ without changing the singularity.  It still makes sense, though, to factor out the $L_k$-dependence. If we choose to factor out powers of $S$, therefore, we find
\begin{align}
\label{eqn:Awsjoint}
&A(w,S|L_k) \\
 &= \left(\frac{S}{L_k^{1+\zeta}}\right)^{2-\tau} 
	\calAwSL\left(\frac{w}{S^{1/(1+\zeta)}}, \frac{w}{L_k}\right)
		 \frac{1}{S\, S^{1/(1+\zeta)}} \cr
  &= L_k^{(\tau-2)(1+\zeta)} S^{1-\tau-1/(1+\zeta)}
	\calAwSL\left(\frac{w}{S^{1/(1+\zeta)}}, \frac{w}{L_k}\right). \nonumber
\end{align}
where again we have a power of a dimensionless scaling variable, times a scaling function, divided by $S w \sim S S^{1/(1+\zeta)}$ since $A(w,S|L_k)$ is multiplied by $dS\, dw$ in its invariant form.

The last joint distribution that will be useful is related to right-most pieces of an avalanche. Consider the right-most piece of width $x$ of an avalanche of total width $w$ and total size $S$; let this segment have size $s$.  (This will be the size measured by a window that cuts the avalanche at the left-hand window boundary at $x$.) Let $A(s,w,S|x,L)$ be the fraction of the system covered by such avalanche pieces. Then, in the same logic as before, this five-variable distribution can be written as a power law times a universal function of four variables:
\begin{widetext}
\begin{align}
\label{eq:pieces}
A(s,w,S|x,L) = L_k^{(\tau-2)(1+\zeta)} s^{-\tau-1/(1+\zeta)}
	\calAswSxL\left(\frac{x}{s^{1/(1+\zeta)}},\frac{S}{s},
			\frac{w}{S^{1/(1+\zeta)}}, \frac{w}{L_k}\right).
\end{align}
\end{widetext}

One can clearly work out scaling forms for joint distributions of several variables and other combinations. The ones we have discussed here will be needed in our analysis of windowing effects. 

\section{Window Effects}
\label{sec:Window}

Now that we have laid the groundwork for exploring the shapes of avalanches, we focus on analyzing avalanches inside a viewing window.  In this section, we focus on how to define the right power laws and scaling; we also give results for fits, extracting critical exponents.  In the next section, we go into more detail about the scaling shapes of these distributions - the universal scaling {\it functions} for avalanche sizes viewed through windows.  

In imaging experiments one often runs into the problem of not being able to see the whole system, distorting the avalanche size distribution.  In particular, for Barkhausen noise, typical magnetic avalanches span many decades in size, far beyond the spatial resolution of optical microscopes.  The natural solution is to take measurements at a variety of magnifications.  Even though at the weakest magnifications the window size $W > L_k$ and most avalanches avoid the window boundaries, the effects of the boundaries will always dominate at the highest magnifications.  The analysis in this section not only provides a method to correct for finite-size-like window effects on exponents, but allows us to actively make use of all the data for a range of magnifications.

We show in detail in this section and the next how the avalanches which cross different boundaries exhibit distinctly different size distributions and critical exponents (Figure~{\ref{fig:window_dist}}).   As described in section~\ref{sec:summary}, we consider the avalanches measured in an infinite strip of width $W$ (Figure~\ref{fig:window_dist}(b)), for a system with characteristic length $L_k$.  We separate avalanches into different categories: internal avalanches (00), split avalanches (10 and 01), and spanning avalanches (11).   Let us call  $\Aoo(s|W,L_k) \, ds$ the area fraction covered by such avalanches with sizes in the range $s, s +ds$.  (For $\Aoo$, the segment size equals the total size.)      The split (10 or 01) avalanches will have area fraction $\Alo(s|W,L_k) \, ds$ for each $s$. The distribution $\Aol$  (Figure~\ref{fig:window_dist}(f)) of avalanches touching the right boundary naturally equals $\Alo$ on average.  $\All(s|W,L_k)\, ds$ is the fraction of the strip spanned by 11 spanning avalanches.  We mentioned in section~\ref{sec:summary} that the 00 avalanches have a power law that matches the full system, whereas the 10, 01, 11 avalanches all have modified power laws with a smaller exponent $\tau$.

Besides different power law scaling, the universal scaling functions for these different avalanche distributions are also distinct.  In particular, the cutoff dependence on window size is different for internal avalanches and split avalanches, while the spanning avalanches have both an outer and an inner cutoff due to the window size (since avalanches must be large enough to span the window).  We present the fits of these universal scaling functions in this section and discuss their shapes in more detail in the next.  

We know that all avalanches in the window are of one of the 00, 10, 01, 11 types, so
\begin{equation}
\int \, ds \Aoo(s|W,L_k) + 2 \Alo(s|W,L_k) + \All(s|W,L_k) = 1.
\end{equation}

As in the previous section, consider how each of these distributions $\Azz$ rescales under a coarse-graining by a factor $b$.  The $zz$ denote our indices for the various windowed distributions (00, 10, 01, 11).  Each $\Azz ds$, being a geometrical quantity (a fractional area), must be invariant under rescaling (with two invariant scaling variables):
\begin{align}
\label{eq:BzzScaling}
\Azz(s|W,L_k) \,ds &= \Azz(s/b^{1+\zeta}|W/b, L_k/b) ds/b^{1+\zeta}  \cr
		&= (1/s) \calBzz(s/L_k^{1+\zeta}, W/L_k) ds.
\end{align}
However, this is clearly not the form which makes the size and window-width dependence of the avalanche sizes manifest. We are allowed to factor powers of the invariant scaling variables $W/L_k$ and  $s/L_k^{1+\zeta}$ out of the scaling function $\calB$:
\begin{align}
\label{eq:AzzScaling}
&\Azz(s|W_k,L_k)  \cr
&= 	\frac{1}{s} \left(\frac{s}{L_k^{1+\zeta}}\right)^{2-\tau_{zz}} \left(\frac{W}{L_k}\right)^{-\upsilon_{zz}} 
	\calAzz \left(\frac{s}{L_k^{1+\zeta}}, \frac{W}{L_k}\right)
\cr
&= L_k^{(\tau_{zz}-2)(1+\zeta)+\upsilon_{zz}} W^{-\upsilon_{zz}} s^{1-\tau_{zz}}\calAzz \left(\frac{s}{L_k^{1+\zeta}}, \frac{W}{L_k} \right).
\end{align}
 The appropriate powers of $s$, $L_k$, and $W$ to pull outside depend upon which of the three distributions we are considering.  For the distributions $\Aoo$ and $\Alo$, $\Aol$, we choose $\tau_{zz}$ and $\upsilon_{zz}$ (powers of the invariant scaling variables $X=s/L_k^{1+\zeta}$ and $Y=W/L_k$)  to make the resulting scaling function go to a constant at small $X$ and/or $Y$.  This way the power laws we pull out describe the behavior of the limit of small $s$, and the way in which the avalanches are cut off by the window size (as $s$ approaches $W^{1+\zeta}$) are described by the scaling function.  On the other hand for $\All$ there are no small avalanches (they have to be large enough to span the window), and as $L_k \rightarrow \infty$, all avalanches span the window and become 11 avalanches, and in this limit the distribution will not go to zero, so for this distribution we instead pull out powers of $W$ and $s$.

\subsection{Internal Avalanches}
\label{subsec:00}

First let us consider $\Aoo(s|W,L_k) ds$, the window area spanned by avalanches of sizes in $(s,s+ds)$ that do not touch the boundaries. This can be computed explicitly from the function $A(w,S|L_k)$ (Eq.~\ref{eqn:Awsjoint}) which gives the area covered by avalanches of width $w$ and size $S$ (note that for $00$ avalanches, the segment pieces $s=S$):
\begin{widetext}
\begin{align}
\Aoo&(s|W,L_k) 
 = \int_a^W \frac{W-w}{W} A(S,w|L_k) dw \cr
 &= \int_a^W dw \frac{W-w}{W} 
   L_k^{(\tau-2)(1+\zeta)} S^{1-\tau-1/(1+\zeta)}
	\calAwSL\left(\frac{w}{S^{1/(1+\zeta)}}, \frac{w}{L_k}\right) \cr
 &=  L_k^{(\tau-2)(1+\zeta)} S^{1-\tau-1/(1+\zeta)}
    \int_a^W dw \frac{W-w}{W} 
	\calAwSL\left(\frac{w}{S^{1/(1+\zeta)}}, \frac{w}{L_k}\right)
\end{align}
\end{widetext}
where $(W-w)/W$ is the probability that an avalanche whose center lies within the window is entirely contained in the window (i.e., the avalanche center lies within $(W-w)/2$ of the center of the window).  Changing variables from $w$ to
$\Omega = w/s^{1/(1+\zeta)}$, 
\begin{align}
\label{eq:A00Scaling}
\Aoo(s|W,L_k) 
 &=  L_k^{(\tau-2)(1+\zeta)} s^{1-\tau}
    \int_{a/s^{1/(1+\zeta)}}^{W/s^{1/(1+\zeta)}} d\Omega \cr
 & ~~~~\frac{W-s^{1/(1+\zeta) \Omega}}{W} 
	\calAwSL\left(\Omega, \Omega\left(\frac{s}{L_k^{1+\zeta}}\right)^{\frac{1}{{(1+\zeta)}}}\right) \cr
 &=  L_k^{(\tau-2)(1+\zeta)} s^{1-\tau} \calAoo(s/W^{1+\zeta}, W/L_k),
\end{align}
with no explicit dependence on the window width $W$ (so $\upsilon_{00} = 0$),
and the same critical exponent $\tau_{00} = \tau$ that is given by the non-windowed distribution. Note that Eq.~(\ref{eq:A00Scaling}) is of the general form given by Eq.~(\ref{eq:BzzScaling}) and~(\ref{eq:AzzScaling}).  This scaling equation is also consistent with our numerics: the normalization of the distribution for small avalanches is independent of $W$, and $\tau_{00}= 1.26 \pm 0.02$ is consistent with the bulk $\tau = 1.24 \pm 0.04$.

Using this scaling form, we fit the $00$ data (jointly with $11$ and $10$ data) with a scaling function given by a parameterized functional form:
\begin{align}
&\Aoo(s|W,L_k) = L_k^{(\tau-2)(1+\zeta)} s^{1-\tau} \exp(-(T_{00}+U_{00} s_k^{1/2}+ \cr
&Z_{00} s_k^{\delta_{00}} + C_{00}(\frac{s_k}{W_k^{n_{00}}})^{m_{00}}) \exp((A_1^{00}/s+A_2^{00}/s^2))
\label{eq:A00function}
\end{align} 
with $s_k = s/L_k^ (1+\zeta)$. Our fit for the parameter $n_{00}$ is $1.62$, which we believe to be $1+\zeta$ (see next section).  This makes the term $s_k/W_k^{n_{00}} = S/W^{1+\zeta}$ which is another natural invariant scaling variable.

\begin{figure}[htbp]
\begin{center}
\includegraphics[scale=0.30]{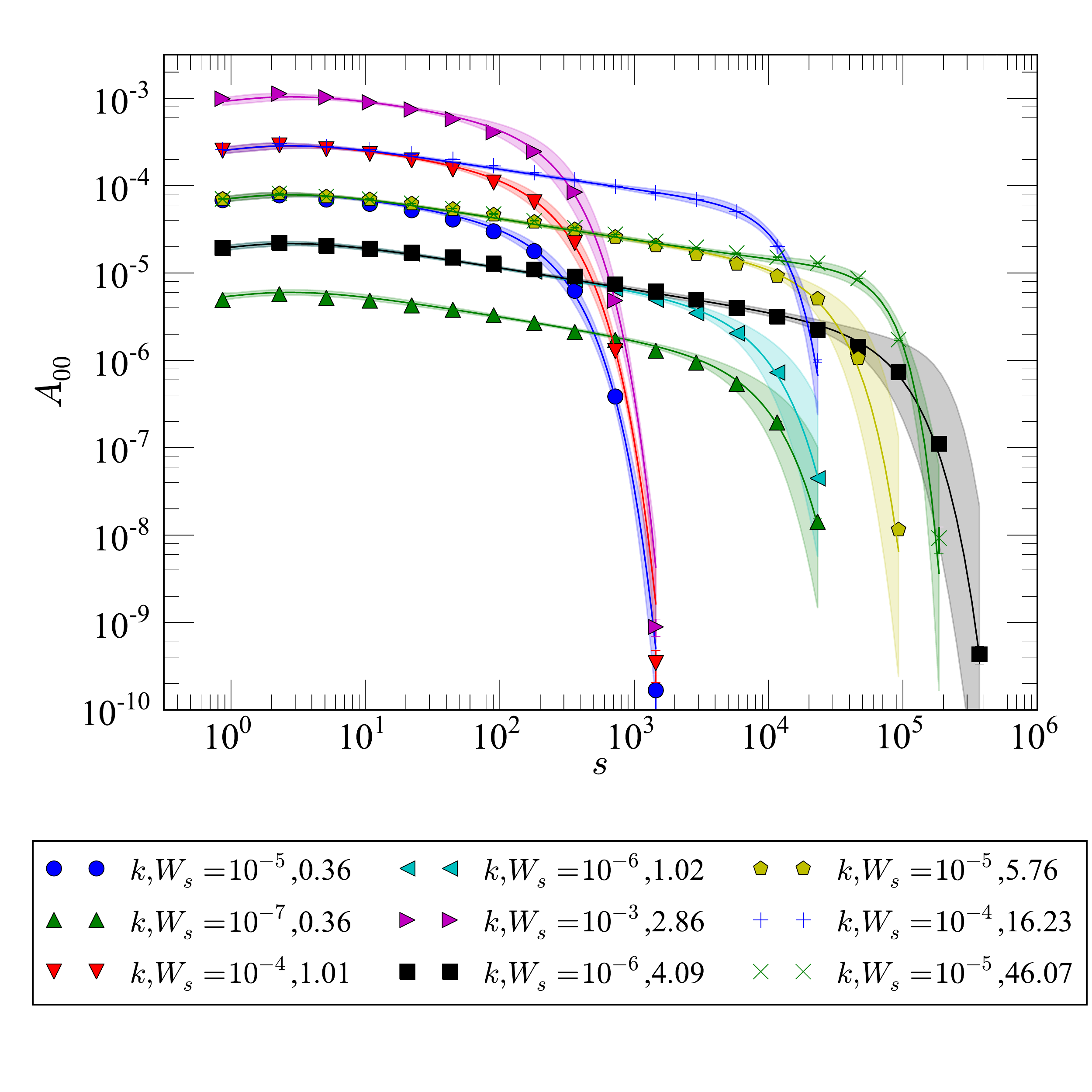}
\caption{{\bf Internal Avalanches Data and Fit} (color online) Shown here are the area-weighted size distributions for internal (00) avalanches.  The lines are the joint best fit of $A_{00}$, $A_{10}$ and $A_{11}$ to the functional forms of equations \ref{eq:A00function}, \ref{eq:A10function}, and \ref{eq:A11function}, whereas the shaded areas are the fluctuations in theory corresponding to the systematic covariant errors on our exponents and parameters (individual parameter best fit values and errors are quoted in Table{~\ref{table:fit}}).}
\label{fig:00fit}
\end{center}
\end{figure}

Figure~\ref{fig:00fit} shows the results of a nonlinear least squares fit, with shaded areas as estimations of fluctuations in the theory corresponding to systematic error bars on our parameters.  In Eq.~{\ref{eq:A00function}}, $\tau$, $\zeta$, (and $\nu_k$ which is hidden in $L_k$) are universal exponents shared amongst the three different distributions, $A_1^{00}$ and $A_2^{00}$ are (non-universal) analytic corrections to scaling reflecting lattice effects on small avalanches, and the other parameters encapsulate the shape of the universal scaling function $\calAoo$. The fitted results for the other universal and non-universal parameters are quoted in Table~\ref{table:fit}.  We describe the scaling shapes and their motivation in more detail in section~\ref{sec:shapes}.

One may note the exponent $\zeta$ in our fits is fixed to the literature value of 0.63.  If we allow for a free fit on all the parameters, it shifts to $\zeta=0.68 \pm 0.02$.  Although the free fit is $2.5 \sigma$ away from the accepted value of $\zeta$, the fit with fixed $\zeta=0.63$ has only a $50\%$ higher cost than the free fit minimum, suggesting an average of $0.5 \sigma$ drift on the parameters instead of $2.5 \sigma$ as seen in Table~\ref{table:fit}.   This suggests three cautions (1)~The estimate on our systematic error (0.02) is a lower bound estimate, and in fact the systematic error should be higher.  (2)~The fact that our systematic error should be higher also implies that the scaling functions are imperfect and may be improved upon. (3)~There could be corrections due to a crossover that depends on both $\lambda$ and $k$ which we have not accounted for, which are distorting the fit.  The subtleties and nuances in measuring the exponent $\zeta$, and the possible origins of this drift are discussed in more detail in Appendix~\ref{sec:zeta}.  

\subsection{Split Avalanches}
\label{subsec:10}

Next, consider the avalanches that are split by one side of the window, say the left side, with the distribution $\Alo(s|W,L_k)$. Physically, for small avalanches $s$ and large $L_k/W$ this is clearly proportional to $1/W$: the small avalanches extend only a small distance into the window (smaller than the window width), so the fractional area covered by them is proportional to one over the width of the window. This leaves us with a scaling form
\begin{align}
\label{eq:A10Scaling}
&\Alo(s|W,L) =  \\
&\frac{1}{W} L^{(\tau^\prime-2)(1+\zeta)} s^{1-\tau^\prime + 1/(1+\zeta)} \calAlo(s/L_k^{1+\zeta}, W/L_k). \nonumber
\end{align}
with $\tau^\prime$ to be determined.  Note again that Eq.~(\ref{eq:A10Scaling}) is of form Eq.~(\ref{eq:BzzScaling}).

We can also write  $\Alo$ in terms of the distribution of right-most pieces $A(s,w,S|x,L_k)$ from Eq.~\ref{eq:pieces}, integrating over all possible sizes $S$, all possible widths $w$, and all possible pieces $x$ (from lattice size $a$ to window width $W$):
\begin{widetext}
\begin{eqnarray}
\lefteqn{\Alo(s|W,L_k) = \int_{a^2}^\infty dS \, \int_a^\infty \, dw \int_a^W \frac{dx}{W} \, A(s,w,S|x,L_k)} \cr
&\sim& \left(\frac{s}{L_k^{1+\zeta}}\right)^{2-\tau} \frac{1}{s^2\, s^{1/(1+\zeta)}} \int_{a^2}^{L_k^{1+\zeta}} dS \, \int_a^{L_k} \, dw \int_a^W \frac{dx}{W} \,
	\calAswSxL\left(\frac{x}{s^{1/(1+\zeta)}},\frac{S}{s},
			\frac{w}{S^{1/(1+\zeta)}}, \frac{w}{L_k}\right) \cr
& = & \left(\frac{s}{L_k^{1+\zeta}}\right)^{2-\tau} \frac{1}{s} \int_{a^2/s}^{L_k^{1+\zeta}/s}  d\left(\frac{S}{s}\right) \, \int_{a/s^{1/(1+\zeta)}}^{L_k/s^{1/(1+\zeta)}} \,  d\left(\frac{w}{s^{1/(1+\zeta)}}\right) \cr
& & \times \int_{a/s^{1/(1+\zeta)}}^{W/s^{1/(1+\zeta)}} \frac{s^{1/(1+\zeta)}}{W}d\left(\frac{x}{s^{1/(1+\zeta)}}\right) \,
	\calAswSxL\left(\frac{x}{s^{1/(1+\zeta)}},\frac{S}{s},
			\frac{w}{S^{1/(1+\zeta)}}, \frac{w}{L_k}\right) \cr
&= &\frac{1}{W} L_k^{(\tau-2)(1+\zeta)} s^{1-\tau + 1/(1+\zeta)} \calAlo(s/L_k^{1+\zeta}, W/L_k)
\end{eqnarray}
\end{widetext}
$dx/W$ is the relative probability that the avalanche intersects the left-hand boundary, and we have changed the integration limits at $\infty$ to the avalanche length scale $L_k$.  (For $w<x<W$, the original distribution is naturally zero.)  After we rewrite the integration variables in terms of the invariant scaling variables, we can organize the form of the scaling function into the form of Eq.~\ref{eq:A10Scaling}.  This tells us that $\tau^\prime = \tau$.  These results are consistent with our numerical fits: $W$ has an exponent of minus one, and $\tau^\prime$ is equal to the system $\tau$.

With the correct power laws pulled out, we can now write down a function to describe the data and cutoff:
\begin{align}
&\Alo(s|W,L_k)=\frac{1}{W} L_k^{(\tau-2)(1+\zeta)} s^{1-\tau+1/(1+\zeta)} \cr
&\exp(-(T_{10}+U_{10} s_k^{1/2}+Z_{10} s_k^{\delta_{10}} + C_{10}(\frac{s_k}{W_k^{n_{10}}})^{m_{10}}) \cr
&\exp((A_1^{(10)}/s+A_2^{(10)}/s^2))
\label{eq:A10function}
\end{align} 
Again, our best fit $n_{10}$ is nearly $1+\zeta$, so $s_k/W_k^{n_{10}} \sim s/W^{1+\zeta}$.  Also note that this distribution has the same functional form as $\Aoo$ in eqn~\ref{eq:A00function} aside from a factor of $s^{1/(1+\zeta)}/W$ in front.

\begin{figure}[htbp]
\begin{center}
\includegraphics[scale=0.30]{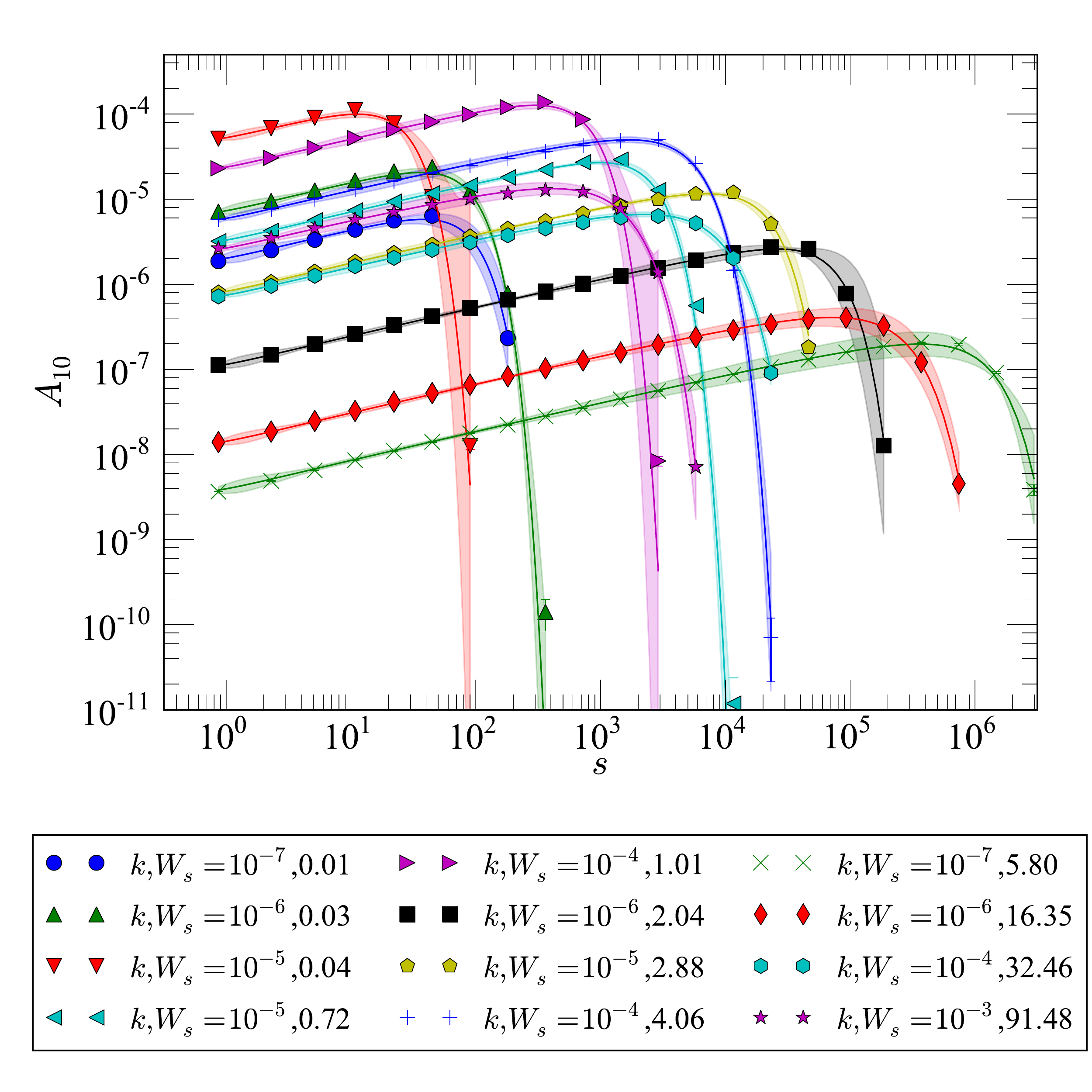}
\caption{{\bf Split Avalanches Data and Fit.} (color online) Shown here are the area-weighted size distributions for split (10) avalanches with different $k$ and window size $W$.  The lines are the joint best fit of $A_{00}$, $A_{10}$ and $A_{11}$ to equations ~{\ref{eq:A00function}}, ~{\ref{eq:A10function}}, and ~{\ref{eq:A11function}}, whereas the shaded areas are the fluctuations in theory corresponding to the covariant systematic errors on our exponents and parameters (individual parameter best fit values and errors Table{~\ref{table:fit}}).}
\label{fig:10fit}
\end{center}
\end{figure}

Figure~\ref{fig:10fit} shows the results of a joint nonlinear least squares fit with the $00$ and $11$ avalanche data, with shaded areas representing estimations of fluctuations in the theory corresponding to systematic error bars on our parameters.  Here, as in the $00$ distributions, $\tau$, $\zeta$ (and $\nu_k$ which is included in $L_k$ and the scaling variables $s_k$ and $W_k$) are universal exponents shared amongst the three different distributions, $A_1^{10}$ and $A_2^{10}$ are (non-universal) analytic corrections to scaling reflecting lattice effects on small avalanches, while the other parameters describe the shape of the universal scaling function $\calAlo(s_k, W_k)$.  Fitted results for the other universal and non-universal parameters are quoted in Table~\ref{table:fit}.  We describe the scaling shape $\cal$ and its motivation in more detail in section~{\ref{sec:shapes}}.

\subsection{Spanning Avalanches}
\label{subsec:11}
Finally, consider the spanning avalanche distribution $\All$. First, remember that most of the area in general is spanned by the largest avalanches (since $\tau < 2$). Therefore, as $L_k\to\infty$, 100\% of the area is covered by avalanches of widths much larger than $W$, and hence $\All$ must integrate to one in this limit. This makes it natural to pull out only powers of $W$ and $s$ outside the scaling form for $\All$.

\begin{figure}[tb]
\begin{center}
\includegraphics[scale=0.30]{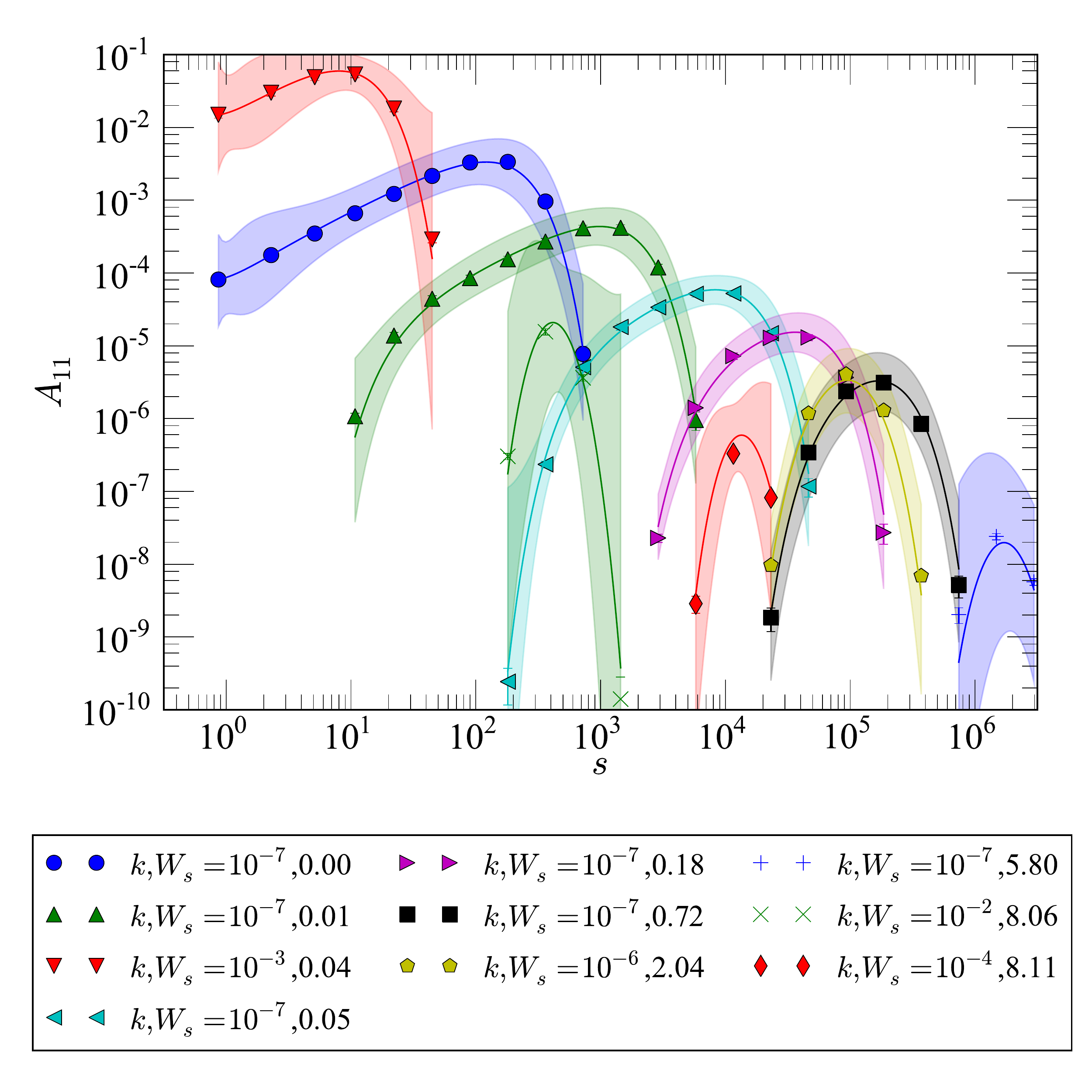}
\caption{{\bf Spanning Avalanches Data and Fit} (color online) Shown here are the area-weighted size distributions for spanning (11) avalanches with different $k$ and window size $W$.  The lines are the joint best fit value using the functional forms using equations ~\ref{eq:A00function}, ~\ref{eq:A10function}, and ~\ref{eq:A11function}, whereas the shaded areas are the fluctuations in theory corresponding to the systematic errors on our exponents and parameters (individual parameter best fit values and errors are quoted in Table{~\ref{table:fit}}). }.
\label{fig:11fit}
\end{center}
\end{figure}

Second, notice that the size of an 11 avalanche is basically given by its height. More specifically, as $W\to 0$, the avalanches have size $s = W h_x$, where $h_x$ is the height of the vertical cross section of the avalanche. Hence we can write $\All$ in the limit $W\to 0$ in terms of the distribution  $A(h_x|L_k)$ of randomly chosen vertical cross sections of avalanches (eqn~\ref{eq:AhxL}), choosing $h_x = s/W$:
\begin{align}
\All(s|W{=}0,L_k) ds 
	& \sim_{W\to0} dh_x A(h_x{=}s/W|L_k) \cr
\All(s|W{=}0,L_k)  
	& \sim_{W\to0}  \frac{1}{W} A(h_x{=}s/W|L_k)
\label{eq:A11Ahx}
\end{align}
where $dh_x/ds = 1/W$.  Remembering from Eq.~(\ref{eq:AhxL}) that 
$A(h_x|L_k) \sim 
       (h_x/L_k^{\zeta})^{(2-\tau)(1+\zeta)/\zeta} \calAhxL(h_x/L_k^\zeta)/h_x$, 
we substitute $s/W$ for $h_x$ and take the limit of $W_k \to 0$ in Eq.~\ref{eq:A11Ahx} to give: 
\begin{align}
&\All(s|W=0 ,L_k)  \sim_{W_k\to0}  \cr
& \frac{1}{W} 
 \left(\frac{s}{W\, L_k^{\zeta}}\right)^{(2-\tau)(1+\zeta)/\zeta} 
	\calAhxL\left(\frac{s}{W\, L_k^\zeta}\right)/(s/W), 
\end{align}
we cancel the two $W$'s and add the dependence on the second scaling variable $W_k = W/L_k$ to derive the scaling form for $\All$:
\begin{align}
&\All(s|W, L_k) = \cr
& \frac{1}{s} \left(\frac{s}{W L_k^\zeta}\right)^{(2-\tau)(1+\zeta)/\zeta} \calAll \left(s/(W L_k^\zeta), W/L_k\right). 
\end{align}
Here $\lim_{Y\to 0} \calAll(X,Y) = \calAhxL(X)$ and 
thus $\int \calAll(X,0)\, dX = 1$ (implied by the fact that almost 
all avalanche area touches both boundaries as $W/L_k\to 0$).  Also notice that since:
\begin{equation}
X = \frac{s}{W L_k^\zeta} = \frac{s}{L_k^{(1+\zeta)}} \frac{L_k}{W} = s_k \cdot \frac{1}{W_k},
\end{equation}
we can rewrite $\All(s| W, L_k)$ as:
\begin{align}
\All(s|W, L_k) =  \frac{1}{s} \left( \frac{s_k}{W_k} \right)^{(2-\tau)(1+\zeta)/\zeta} \calAll\left(\frac{s_k}{W_k}, W_k\right). 
\end{align}
Figure~\ref{fig:11fit} shows the results of a joint fit with the simulation data of the previous $00$ and $10$ distributions.  For the $11$ distributions we use the functional form:
\begin{align}
\All(s|W \, L_k)&=\frac{1}{s} \left(\frac{s_k}{W_k}\right)^{(2-\tau)(1+\zeta)/\zeta} \exp(-(T_{11}+U_{11} s_k^{1/2}+ \cr
&Z_{11} s_k^{\delta_{11}} +D_{11} \left(\frac{s_k}{W_k} \right)^{m1}+  C_{11} \left(\frac{s_k}{W_k^{n_{11}}}\right)^{-m2}) \cr
&\exp(A_1^{(11)}/s)
\label{eq:A11function}
\end{align}
$\tau$ (and also $\zeta$ and $\nu_k$ which are hidden in the scaling variables $s_k$ and $W_k$) are universal exponents shared amongst the three different distributions, $A^1_{11}$ is the (non-universal) analytic correction to scaling reflecting lattice effects on small avalanches, while the other parameters describe the shape of the universal scaling function $\calAll(s_k, W_k)$.  Note that we don't include the term $A^2_{11}$, as we have done in the $00$ and $10$ distributions; this term turns out to be the same as another term in the universal scaling function in the limit of $W \rightarrow 0$, and so it is redundant.  (See Appendix~\ref{subsec:corrections}).   The best fit universal and non-universal parameters are given in Table~\ref{table:fit}.  We discuss in more detail the motivation and form of the scaling function $\calAll(X, Y)$ in section~\ref{sec:shapes}.

We also test the limiting case of our scaling function with our data sets of $W=1$ in Figure~\ref{fig:AhxL} of section~\ref{subsec:joint}.  The curves drawn in Figure~\ref{fig:AhxL} are with the function given in equation~\ref{eq:A11function}, using the best fit values of the joint fit of $\All$, $\Alo$ and $\Aoo$.  Notice that the predictions of equation~\ref{eq:A11function} matches the data for $W_k=0$, indicating our function satisfies $\lim_{Y\to 0} \calAll(X,Y) = \calAhxL(X)$ as expected. 

\begin{figure}[h!]
\begin{center}
\includegraphics[scale=0.40]{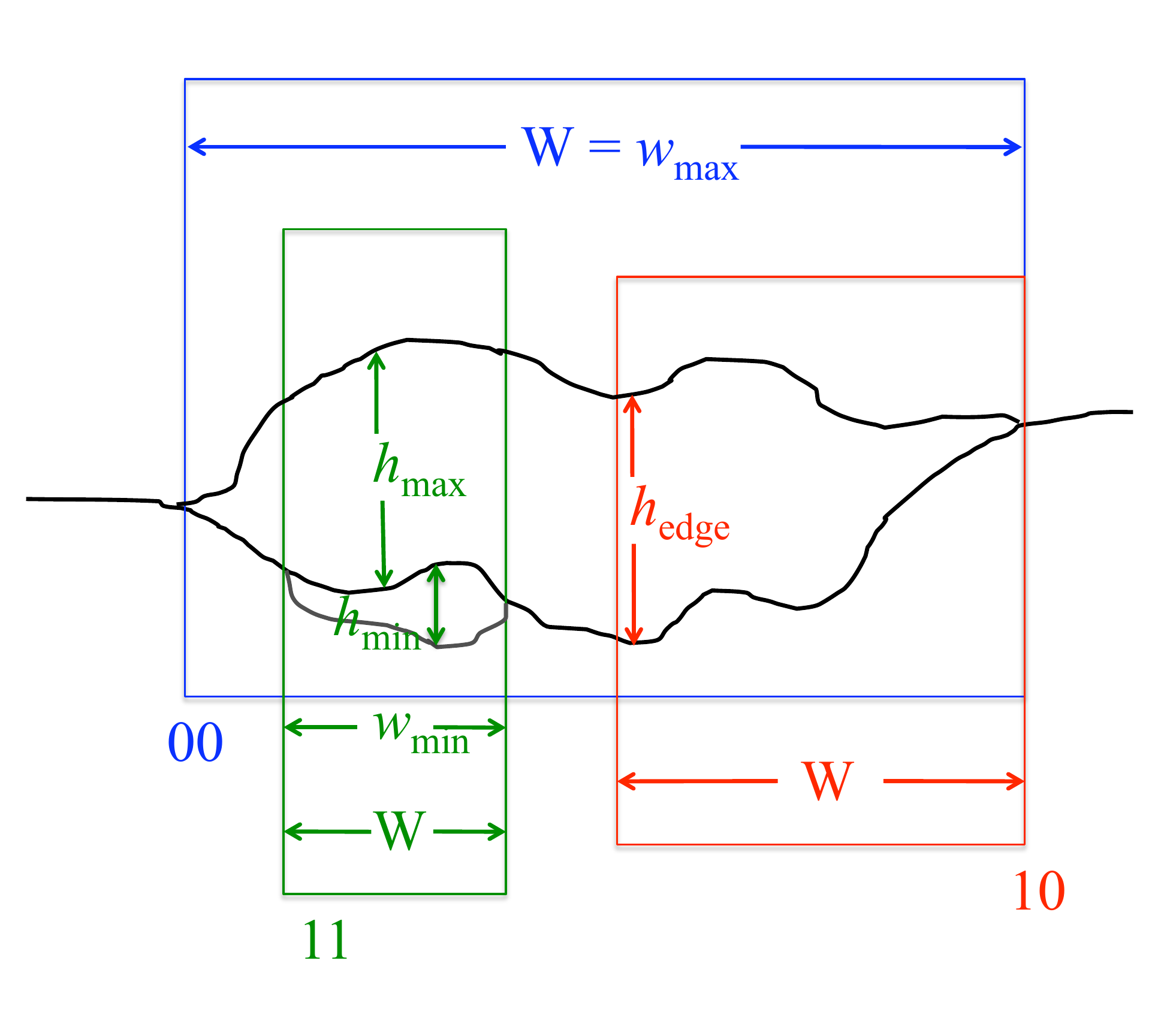}
\caption{{\bf An avalanche cut by windows in the extreme limits.} (color online)  Drawn here are two fronts separated by an avalanche event.  Here we are depicting cases where this avalanche (or its segment) is the maximum avalanche size for the $00$, $10$, and $11$ at different window widths.  Boxes of different widths and colors are used to show the cases in which this may happen.  The main avalanche may count as a $00$ avalanche for a wide window, while part of it would count as a $11$ avalanche for a smaller window; it could also count as a $10$ avalanche if it happens to cross the window boundary.  This figure illustrates our arguments for the shape of the cutoff (the exponent $n_{zz}$) given a window size $W$ for the $00$, $11$, $10$ cases.  Another small avalanche is drawn for the $11$ case to show that the minimum size to cross the $11$ window also introduces a separate cutoff.}
\label{fig:av_drawing}
\end{center}
\end{figure}

\begin{widetext}

\begin{table}[h]
\centering
\begin{tabular}{c | c |c |c|c}
\hline\hline
~~~parameter~~~&~~~~best fit~~~~&standard errors&~~~systematic~~~&drift from best fit\\
&&in linear approx.&errors&with free $\zeta$\\
\hline
\hline
\multicolumn{5}{l}{Universal Exponents} \\
\hline
$\tau$ & 1.2636 & $\pm$ 0.0006 & $\pm$ 0.02 & -0.006\\
$\nu_k$ &  0.4630 &  $\pm$ 0.0002 &$\pm$ 0.01 & -0.02\\
$\zeta$ & 0.63 (fixed) &  $\pm$ 0.0007 &$\pm$ 0.02  & +0.05\\
\hline
\multicolumn{5}{l}{$\calAoo(s_k, W_k) = \exp(-(T_{00}+U_{00} s_k^{1/2}+Z_{00} s_k^{\delta_{00}} + C_{00}(\frac{s_k}{W_k^{n_{00}}})^{m_{00}})$} \\
\hline
$T_{00}$ & 2.488 & $\pm$ 0.004 &$\pm$ 0.1 & -0.01\\
$U_{00}$ & -0.150& $\pm$ 0.005 &$\pm$ 0.1 & + 0.04\\
$Z_{00}$ & 0.0040 & $\pm$ 0.0004 &$\pm$ 0.01 & -0.0009\\
$\delta_{00}$ & 2.21& $\pm$ 0.03 &$\pm$ 0.9 &  +0.06\\
$C_{00}$ & 5.60 & $\pm$ 0.01  &$\pm$ 0.7 & +1.8\\
$m_{00}$ & 1.371 &  $\pm$ 0.003&$\pm$ 0.1 & -0.04\\
$n_{00} \, (1+\zeta)$ & 1.621 &  $\pm$ 0.004 & $\pm$ 0.7 & +0.04 \\
\hline
\multicolumn{5}{l}{$\calAlo(s_k, W_k) = \exp(-(T_{10}+U_{10} s_k^{1/2}+Z_{10} s_k^{\delta_{10}} + C_{10}(\frac{s_k}{W_k^{n_{10}}})^{m_{10}})$} \\
\hline
$T_{10}$ & 1.437 & $\pm$ 0.004 &$\pm$ 0.1 & -0.1 \\
$U_{10}$ & 0.244 & $\pm$ 0.244 &$\pm$ 0.1 & -0.03\\
$Z_{10}$ & 0.027 & $\pm$ 0.001 &$\pm$ 0.03 & +0.005\\
$\delta_{10}$ & 1.64 & $\pm$ 0.01 &$\pm$ 0.4 & -0.06\\
$C_{10}$ & 1.153 & $\pm$ 0.004 &$\pm$ 0.2 & +0.7\\
$m_{10}$ & 1.962 & $\pm$ 0.005 &$\pm$ 0.2 & +0.04\\
$n_{10} \, (1+\zeta)$ & 1.624 & $\pm$ 0.004 &$\pm$ 0.1 & +0.06\\
\hline
\multicolumn{5}{l}{$\calAll(s_k, W_k) = $}\\
\multicolumn{5}{l}{
$ \exp(-(T_{11}+U_{11} s_k^{1/2}+Z_{11} s_k^{\delta_{11}} +D_{11} (\frac{s_k}{W_k})^{m1}+  C_{11} (\frac{s_k}{W_k^{n_{11}}})^{-m2})$} \\
\hline
$T_{11}$ & 0.47 & $\pm$ 0.03 & $\pm$ 1.2 & -0.3\\
$U_{11}$ & -0.5 & $\pm$ 0.1 &$\pm$ 3.6 & -0.5\\
$Z_{11}$ & 0.21 &  $\pm$ 0.06&$\pm$ 1.7 & +0.4\\
$\delta_{11}$ & 1.102 & $\pm$ 0.03 &$\pm$ 1.0 & -0.12\\
$D_{11}$ & 0.52 & $\pm$ 0.03 &$\pm$ 1.0 & +0.1\\
$C_{11}$ & 0.83 & $\pm$ 0.05 &$\pm$ 1.6 & -0.3\\
m1 & 1.48 & $\pm$0.01 &$\pm$ 0.4 & -0.0008\\
m2 & 1.64 & $\pm$ 0.02 &$\pm$ 0.6 & -0.02\\
$n_{11} \, (1+\zeta)$ & 1.655 & $\pm$ 0.004 &$\pm$ 0.1 & +0.02\\
\hline
\multicolumn{5}{l}{Corrections to Scaling $A_1^{zz}/s+A_2^{zz}/s^2$}\\
\hline
$A^{00}_1$ & -0.94 & $\pm$ 0.02 &$\pm$ 0.5 & +0.06 \\
$A^{00}_2$ & 0.27 &$\pm$ 0.01 & $\pm$ 0.4 & -0.04 \\
$A^{10}_1$ & 0.15 &$\pm$ 0.01& $\pm$ 0.4 & -0.2 \\
$A^{10}_2$ & -0.07 & $\pm$ 0.01& $\pm$ 0.3 & +0.1 \\
$A^{11}_1$ & 0.8 &$\pm$ 0.07& $\pm$ 2.1 & -0.3 \\
\end{tabular}

\caption{{\bf Best fit exponents and parameters for windowed distributions.} Here are the results of our joint fit for the windowed $\Aoo$, $\All$, $\Alo$ distributions.  The corresponding scaling forms which were fit are quoted in the table alongside the parameter results.   Here systematic error bars which account for errors in the theory (see Appendix~\ref{subsec:systematic} for explanation) are given.  The traditional standard error bars are typically $\sim 30$ times smaller than the systematic error bars quoted.  The last column is the drift in parameters seen when allowing $\zeta$ to be a free parameter. Notice that these numbers are more or less similar or smaller than the estimated systematic error, except for $\zeta$. (The problems in measuring $\zeta$ are discussed in Appendix~\ref{sec:zeta})  As in Table~\ref{table:shw-fit}, we quote several digits more than the error bars warrant for individual parameters, because the errors are strongly correlated; truncating each parameter to its significant figures would yield a poor fit.}
\label{table:fit}
\end{table}

\end{widetext}

\section{Scaling Shapes and Results}
\label{sec:shapes}

In the previous section we wrote down scaling functions for each type of avalanche inside a window.  In principle there are many possible parameterizations we can choose that would be able to capture the behavior of the data.  In this section we explain how and why we chose each one, and also discuss the scaling function in the limit of small windows.

\subsection{Scaling shapes and functional forms}

We would like to capture the scaling behavior of both the finite size of the avalanches, and the effect of the window size on the distributions.  We choose forms inspired by a functional renormalization group expansion for static avalanche size distributions for all universality classes~\cite{DoussalWiese09}, and further motivated by heuristic arguments for the cutoff dependence on the two scaling variables, $s_k=s/L_k^{1+\zeta}$ and $W_k=W/L_k$.  

Similar to in section~\ref{sec:joint_fit}, we start with an avalanche size distribution of the form:
\begin{equation}
\label{eq:size_scaling}
\calA(s_k) = \exp(-(T+U s_k^{1/2}+Z s_k^{\delta}))
\end{equation}

We expect that as $W \rightarrow \infty$, we will not see the effect of the window size, and the scaling forms will go to the limit of our proposed avalanche size distribution in Eq.~\ref{eq:size_scaling}.  Keeping this in mind, we would like to write a function of $\calA(s_k , W_k)$, including the effect of the window size.   For the 00 and 10 distributions, we expect the avalanche sizes to be cutoff by the window size $W$ when $W/L_k < 1$, hence a cutoff dependent on $s_k/W_k^n$ should be expected, where $n>0$.  While when $W >> L_k$, their scaling forms should go to the limit of our proposed avalanche size distribution in Eq.~{\ref{eq:size_scaling}}.  Therefore for the 00 and 10 avalanches, we propose a scaling form of:
\begin{align}
&\calAzz(s_k, W_k)= \cr
&\exp(-(T_{zz}+U_{zz}s_k^{1/2}+Z_{zz}s_k^{\delta_{zz}} + C_{zz}\left(\frac{s_k}{W_k^{n_{zz}}}\right)^{m_{zz}})).
\end{align}

\begin{figure}[t]
\begin{center}
\subfigure[]{\label{fig:00shape-a}\includegraphics[scale=0.30]{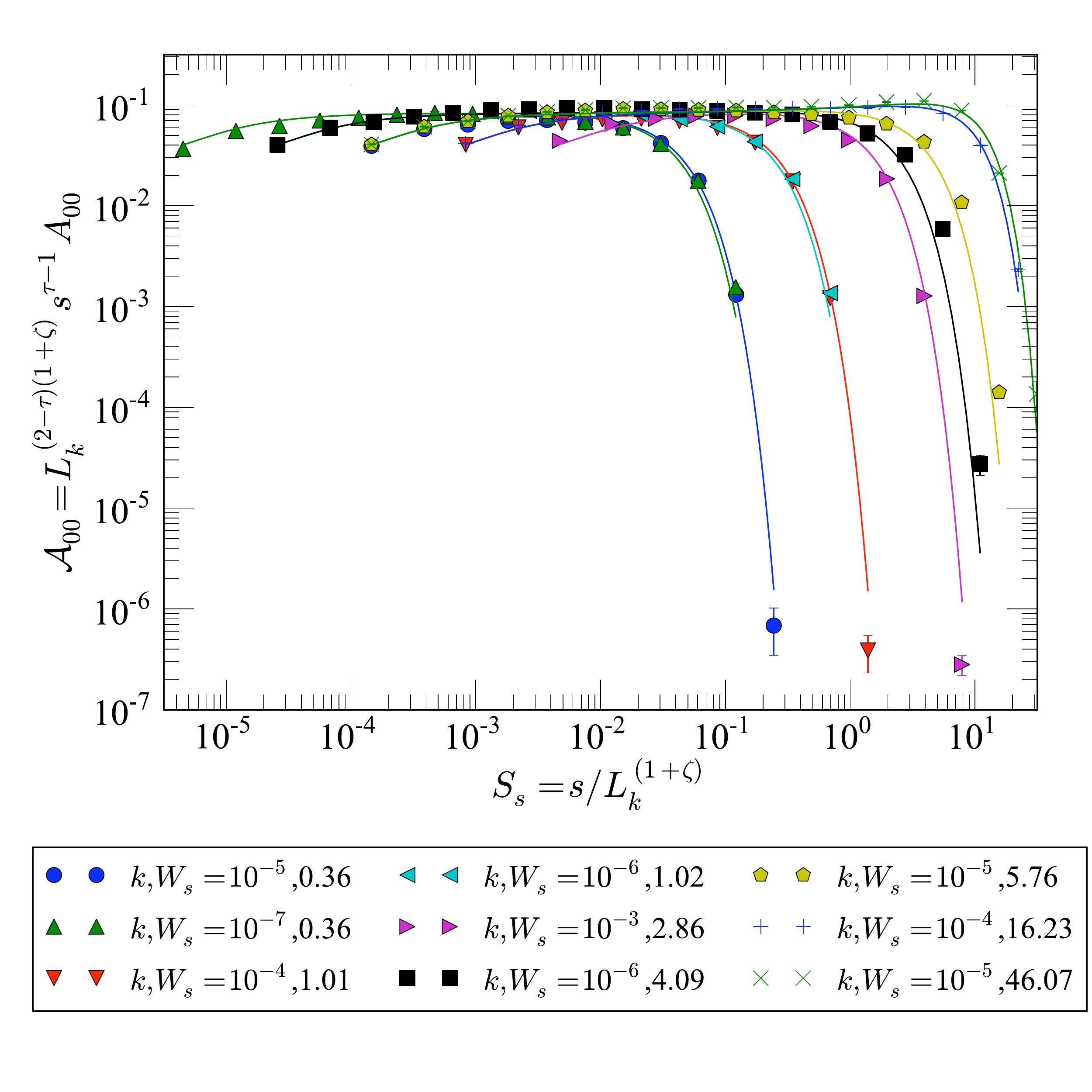}}
\subfigure[]{\label{fig:00shape-b}\includegraphics[scale=0.40]{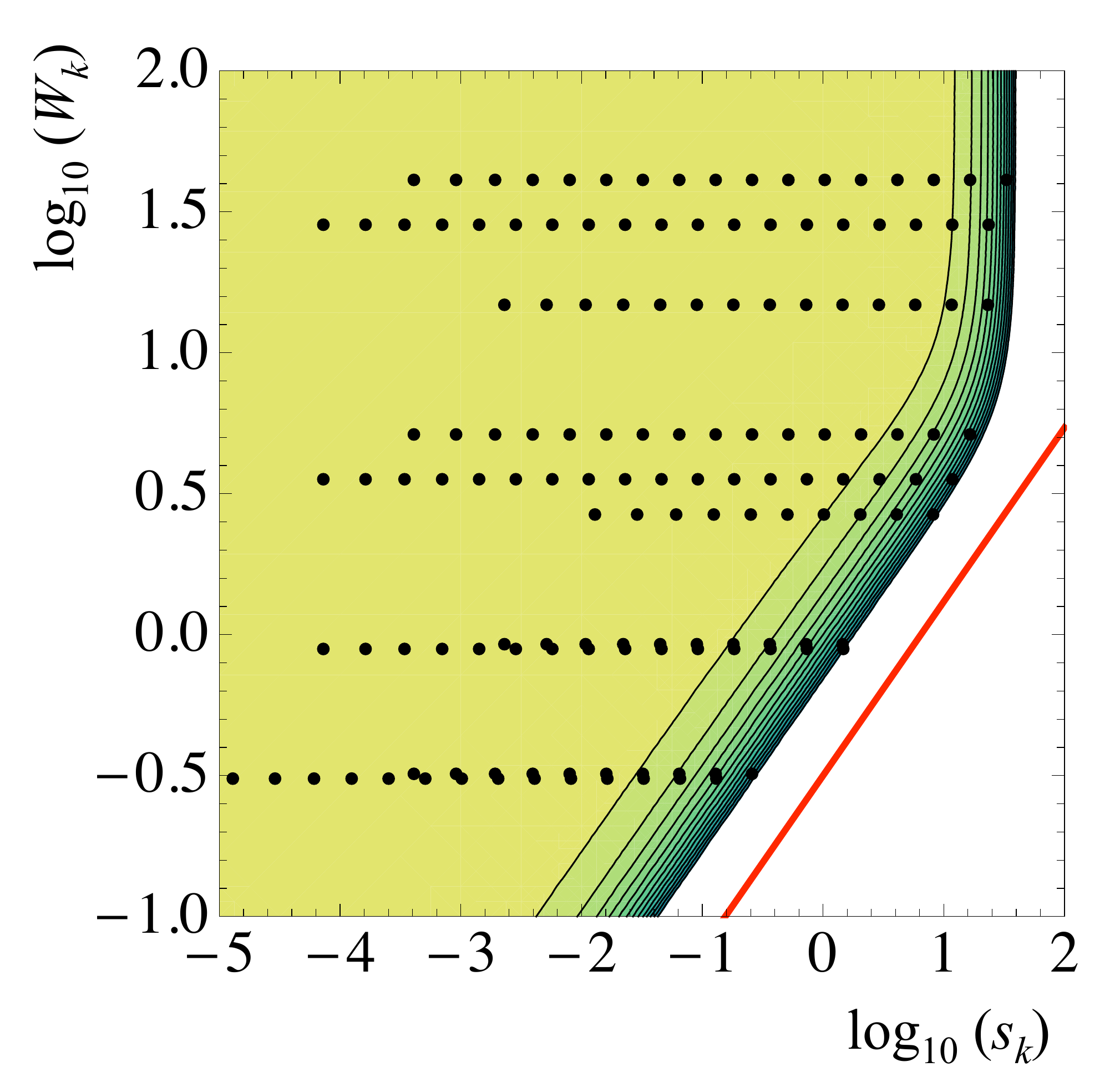}}
\end{center}
\caption{{\bf Internal Avalanches Scaling Function.} (color online) (a)~Scaling collapse showing $\calAoo(s_k,W_k)$ as a function of $s_k$.  The separate curves show the effects of the scaled window size $W_k$.  (b)~Logarithmic contour plot of best fit scaling function against both scaling variables $s_k$ and $W_k$.  Each contour reflects a drop of a factor of $e$ in the scaling function.  The black dots are at locations of the simulated data points used in the fit indicating where the fit should be a reliable prediction.  The red solid line is $\log_{10} W_k =n_{00} \log_{10} s_k$ which is the slope at the large avalanche cutoff, with $n_{00}=1.62$, the best fit value.  ($n_{00}=1+\zeta=1.63$ is the expected value from our heuristic arguments).}
\label{fig:00shape}
\end{figure}

We have heuristic arguments for what this $n_{zz}$ value should take.  Figure~\ref{fig:av_drawing}, a schematic of an avalanche cut by different windows, is meant to help illuminate our discussion. For the 00 internal avalanches, the largest avalanche contained fully within the window should have a width $w$ that is roughly $W$.  And since $s_{max} \sim w_{max} h_{max}\sim w_{max}^{1+\zeta}$, then $s_{max} \sim W^{1+\zeta}$ and it follows that $s_k \sim W_k^{1+\zeta}$, giving us an expectation value of $n_{00}=1+\zeta$.  Numerically, we find $n_{00}\sim 1.62$ when we fix $\zeta = 0.63$.  The fit plotted against one of the scaling variables $s_k$ and also the contour plot of the scaling shape is shown in Figure~{\ref{fig:00shape}}.

\begin{figure}[t]
\begin{center}
\subfigure[]{\label{fig:10shape-a}\includegraphics[scale=0.30]{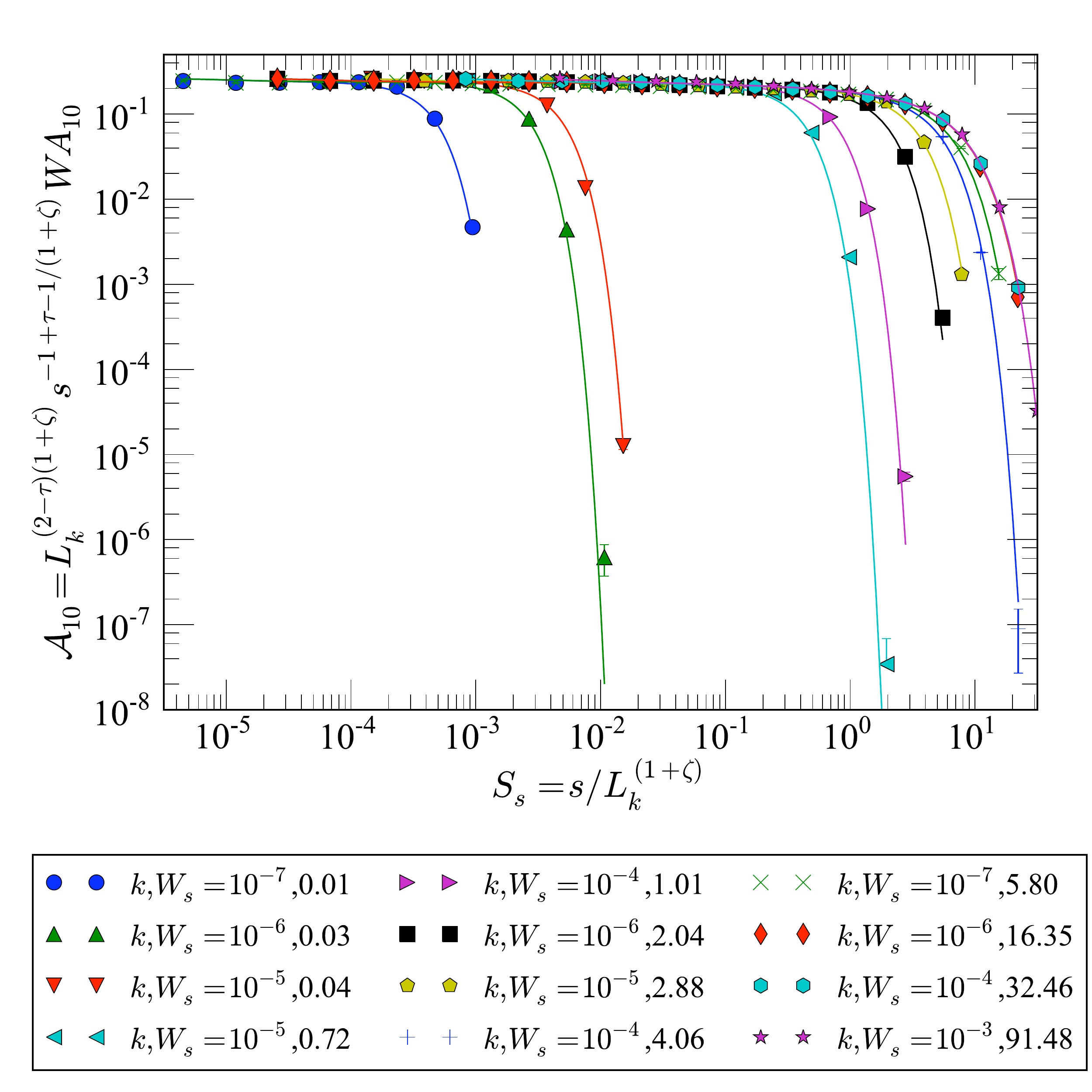}}
\subfigure[]{\label{fig:10shape-b}\includegraphics[scale=0.40]{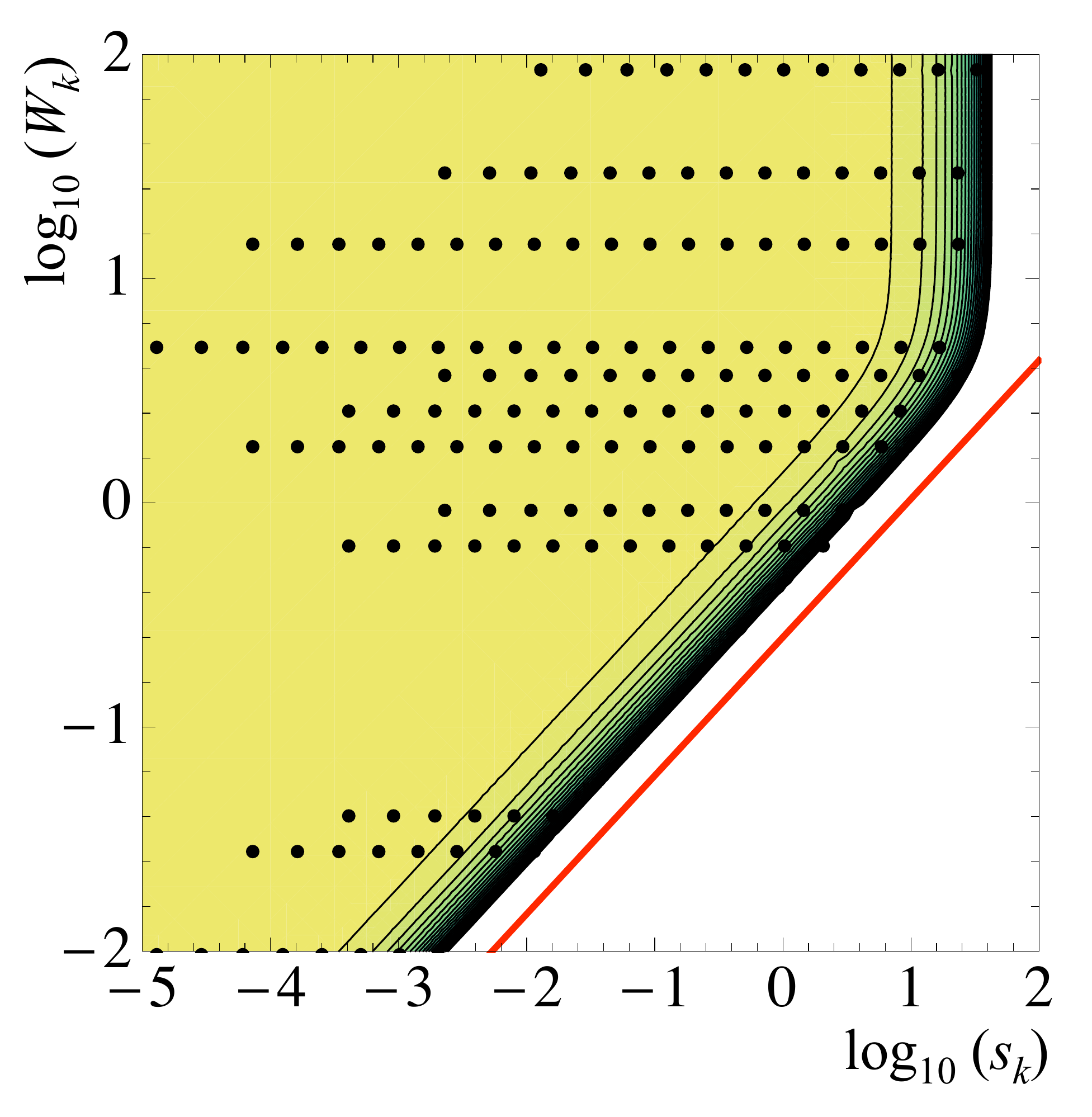}}
\end{center}
\caption{{\bf Split Avalanches Scaling Function.} (color online) (a)~Scaling collapse showing $\calAlo(s_k,W_k)$ as a function of $s_k$.  The separation between curves shows the dependence on the scaled window size $W_k$.  (b)~Logarithmic contour plot of best fit scaling function against both scaling variables $s_k$ and $W_k$.  Each contour reflects a drop of a factor of $e$ in the value of the scaling function.  The black dots are at locations of the simulated data points used in the fit.  The red solid line is $\log_{10} W_k =n_{10} \log_{10} s_k$ which is the slope at the large avalanche cutoff, with $n_{10}=1.62$, the best fit value.}
\label{fig:10shape}
\end{figure}

For the 10 or 01 spilt avalanches, since we are effectively measuring the ends of avalanches that spill into the window, $n_{10}$ depends on what the shape of the avalanche is at the edges.  The largest portion of an avalanche to spill into the window will again be limited by the size of the window $W$.  Here the shape follows the roughness of the two fronts preceding and following the avalanche, where $h(x) \sim x^\zeta$ for each, so plausibly $h_{edge} = h_{after}-h_{before} \sim W^{\zeta}$.   The size is then limited by $s_{max} \sim w_{max} h_{edge} \sim W^{1+\zeta}$, giving us an expectation value of $n_{10}=1.63$.   Numerically, we find $n_{10} \sim 1.62$  ($\zeta$ is estimated from $0.62-0.72$ in our various measures), matching our expectation.  The fit plotted against one of the scaling variables $s_k$ and also the contour plot of the scaling shape is shown in Figure~\ref{fig:10shape}.

Now we move on to discuss the 11 spanning avalanches.  Here the situation is slightly more complicated than the previous two cases, due to the distribution being strongly cut off at two length scales, as one may note from the shapes of the distributions shown in Figure~{\ref{fig:11fit}}.  First of all the avalanches need to be large enough to cross the window, implying an inner cutoff that depends on $W_k^{1+\zeta}/s_k$ (i.e. the cutoff is for $s_{min}/s \lesssim 1$ and so $s_{min}/s = W_k^{1+\zeta}/s_k$); here the argument for the minimum size 11 avalanches follows from a similar argument for the maximum size 00 avalanches.  The smallest avalanche that is able to span the window will have a width $w_{min}=W$, whereas $h_{min} \sim w_{min}^{\zeta}$, and $s_{min} \sim h_{min} w_{min} \sim W^{1+\zeta}$.   However, in this case $s_k$ should be in the denominator of the scaling variable, since for $s_{max}>>s>s_{min}$ the probability of the having a spanning avalanche grows as $s$ increases. For the outer cutoff we expect that the maximum size is given by the window size $W$ and the typical maximum height, i.e. $s_{max} \approx W h_{max}$ which implies that ${s_k}^{max} \approx {W_k}^{max}  (h^{max}/L_k^{\zeta})$.  This rescaled height  $(h^{max}/L_k^{\zeta})$ is constant.  Therefore the cutoff for the large size avalanches should depend on $s_k/W_k$.   Hence, we propose the scaling form below for the 11 spanning avalanches:

\begin{align}
&\calAll(s_k, W_k) = \exp(-(T_{11}+U_{11}s_k^{1/2}+Z_{11}s_k^{\delta_{11}} \cr 
&~~~~~~~~~+ C_{11}\left(\frac{s_k}{W_k^{n_{11}}}\right)^{-m_2}+D_{11}\left(\frac{s_k}{W_k}\right)^{m_1}))
\end{align}
where $C_{11}$controls the strength of the inner cutoff and $D_{11}$ the outer cutoff.

\begin{figure}[ht]
\begin{center}
\subfigure[]{\label{fig:11shape-a}\includegraphics[scale=0.30]{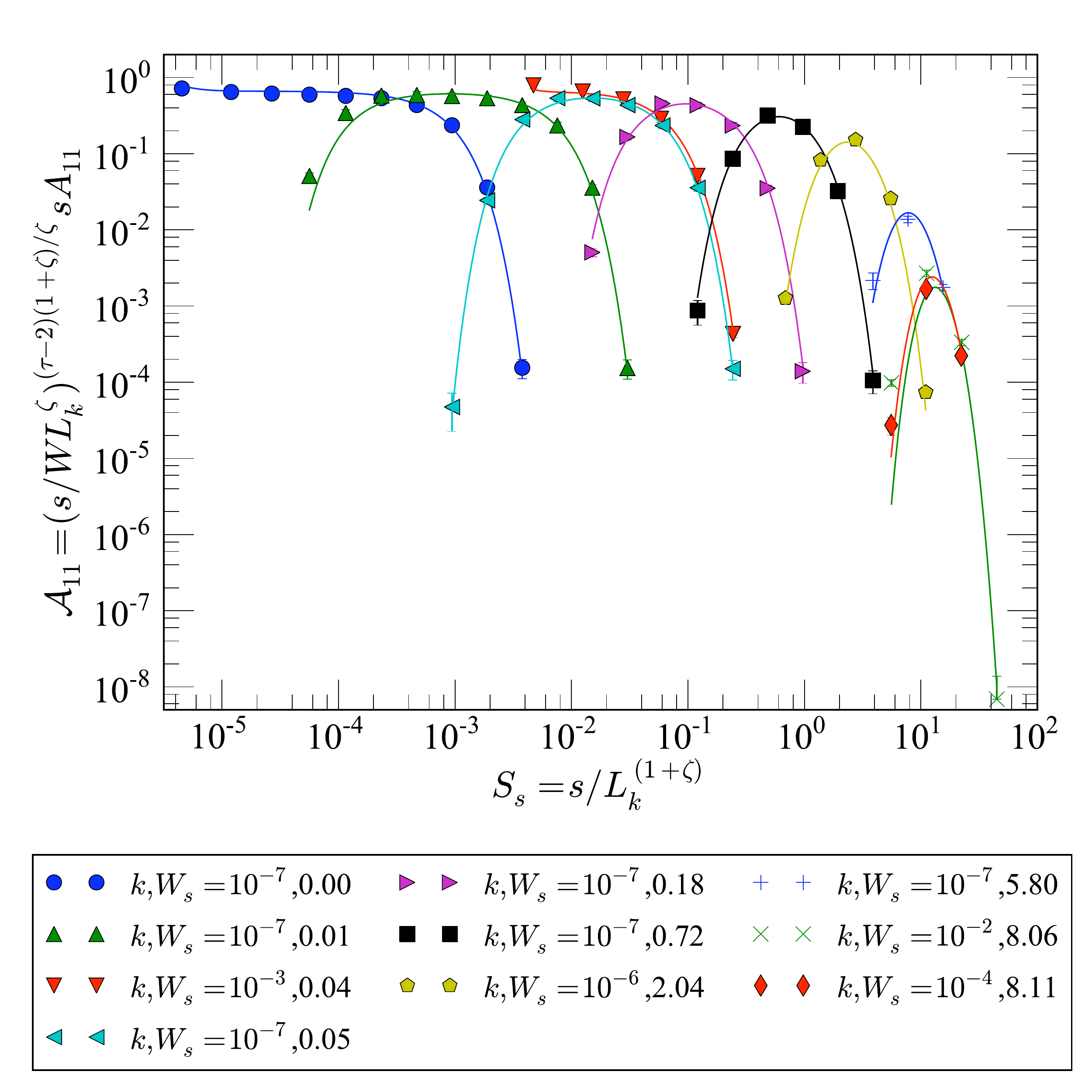}}
\subfigure[]{\label{fig:11shape-b}\includegraphics[scale=0.20]{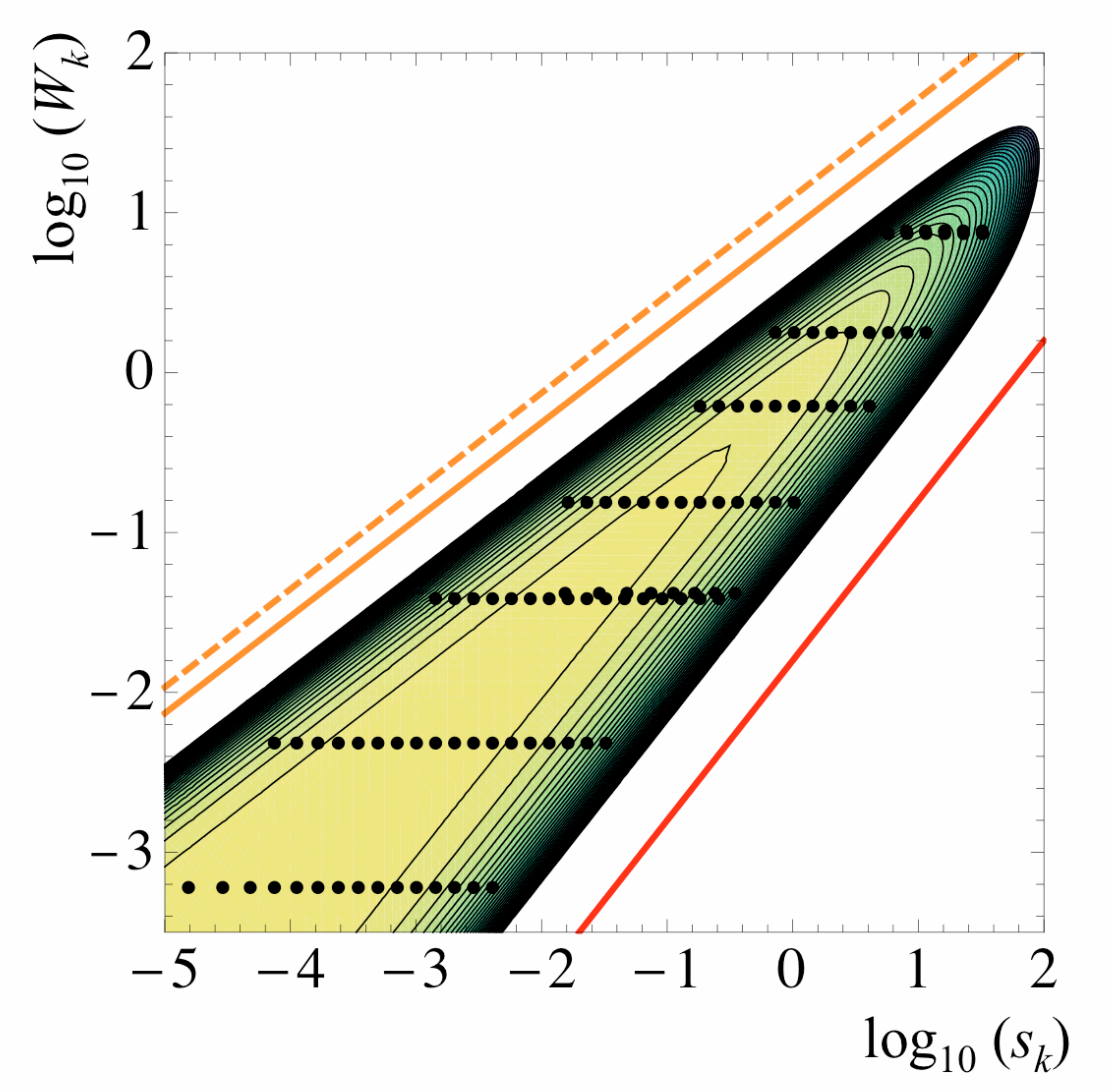}}
\end{center}
\caption{{\bf Spanning Avalanches Scaling Function.} (color online) (a)~Scaling collapse showing $\calAll(s_k,W_k)$ as a function of $s_k$.  The curves move leftward and become more sharply rounded with increasing $W_k$.  (b)~Logarithmic contour plot of best fit scaling function.  Each contour reflects a drop of a factor of $e$ in the scaling function.  The black dots are at locations of the simulated data points used in the fit.  The upper orange solid line is the slope of the contour plot at the small avalanche cutoff, and has $\log_{10} W_k = 1/1.65 \log_{10} s_k$ where $1.65$ is the best fit $n_{11}$ value.  The orange dashed  has $\log W_k = 1/1.63 \log_{10} s_k$ where $1.63$ is the $1+\zeta$ value.  The lower red solid line is $\log_{10} W_k = \log_{10} s_k$ which is the slope at the large avalanche cutoff.}
\label{fig:11shape}
\end{figure}

Figure~\ref{fig:11shape}(a) shows the shape of the scaling function plotted against one scaling variable $s_k$, and Figure~\ref{fig:11shape}(b) gives the contour plot of this function against both variables.  The best fit value of $n_{11}$ is $n_{11} \sim 1.65$, whereas the expected was $n_{11}=1+\zeta=1.63$.

Finally, one may note that in our system all the $n_{zz}$ turn out to fit to our expected $1+\zeta$ within the error bars of $\zeta$.  One may be tempted to set $n_{zz}$ as $1+\zeta$ and have fewer fit parameters in one's form; however, we recognize that our geometrical arguments do not hold for front propagation that is super-rough with $\zeta > 1$, or for models that allow overhangs, so we leave $n_{zz}$ as a free parameter to signify this geometrical constraint.

\subsection{The limit of small windows}

Although there are noticeable imperfections in the theory function, the agreement is impressive between theory and simulation as seen in figures~\ref{fig:00shape}, ~\ref{fig:10shape}, and ~\ref{fig:11shape} .

The scaling function for each of the three distributions is a competition between two types of terms: the rescaled size $s_k = S/L_k^{1+\zeta}$ and the rescaled window size $W_k = W/L_k$.  Upon examining the fit parameters of the scaling function, all three distributions share the characteristic that at small $W_k$ the terms with pure powers of $s_k^{1/2}$ and $s_k^{\delta_{zz}}$ become unimportant, leaving only $s_k/W_k^{n_{zz}}$ for the $00$ and $10$ distributions, and for the $11$ a $s_k/W_k$ term.  Notice that for the $00$ and $10$ distributions, since $n_{zz}$ is ${1+\zeta}$, the $L_k$ dependence disappears for the universal scaling function at small $W_k$.  Therefore, the shape of the scaling function is cut off mainly by the window size $W$.  In fact removing the $s_k$ terms for these functions does not affect the shape for $W_k < 5$.   For experiments that study systems in the same universality class as this one, this implies that data may be measured at large magnifications (small windows) and fit to extract exponents and scaling behavior without the extra, often unknown, scale of $L_k$.

\section{Suggestions for Experiments and Conclusions}
\label{sec:Experiments}

What does our analysis imply for current experiments?  How should one conduct the experiment and analyze the data?  Here we discuss for the particular case of magnetic avalanches in Barkhausen noise, how to take into account window effects and further enhance the collection of data.
 
There are two optical methods for detecting avalanche distributions for Barkhausen noise in 2D thin films, and both make use of the magneto-optical Kerr effect (MOKE).  When a polarized beam of light reflects off a magnetized sample, the reflected polarization is affected depending on the magnetization.  A second polarizer can be used to filter this signal, and then using either a photodiode~\cite{Puppin00}, or an optical microscope~\cite{MagniDurin09,ryu07nature}, we can collect data about the avalanches from the signal.  For experiments using a photodiode (let us call this ``laser reflectometry''), one can only measure the total magnetization change over time, and not individual avalanches.  For experiments using an optical microscope (let us call this ``avalanche visualization''), one can resolve individual avalanches and their shape.

In laser reflectometry experiments~\cite{Puppin00} we only have information for the magnitude of magnetization as a function of time, and cannot see which avalanches touch the boundary.  Furthermore, in current techniques the laser spots are Gaussian in shape, and do not have sharp boundaries.  However, there seems to be no fundamental reason why the illuminated region could not be optically generated with uniform illumination and sharp edges, up to some diffraction limit depending upon the geometry of the experiment.  (A typical avalanche of interest is a few microns in size, large compared to the wavelength of optical light which is 400-700 nm).  If one could make the edges of the laser spots sharper, one could adjust the laser spots to flicker between two sizes, one with a radius slightly larger than the other.  Events that occur with the same magnitude in both the large size measurement and the smaller size measurement would be 00 internal avalanches.  More elaborate sequences of spot shapes could be used to further distinguish 01 and 10 split avalanches from 11 spanning avalanches.

For avalanche visualization experiments~\cite{MagniDurin09,KimPRL03,ryu07nature}, it is straightforward to separate the data into 00, 10, 01, and 11 avalanches for systems which have our strip geometry and flat fronts with $\zeta <1$.   Our analysis will remain valid with minor corrections due to real-world experimental circumstances. For example, sometimes the propagation direction of the front is not parallel to the top and bottom boundaries.  In this case there would be corrections depending on the angle of the tilt $\theta$ and the size of the window.  The local heights $h_x$ would need to be adjusted with the factor of $\cos \theta$.  The 10, and 01 avalanches would be cut off at an angle, but for self-affine (short, wide) avalanches these size corrections are likely irrelevant.

For materials with dipolar interactions and zig-zag shaped fronts~\cite{ryu07nature,cerruti2006barkhausen}, we are less confident that our methods can be applied without modification.  The large vertical extent of the zig-zag front suggests that all four boundaries of the window will matter; therefore we will need to divide avalanches into more categories (0000, 1000, 0100, 0010, 1010...).  The analysis for these types of avalanches will be more complicated.  This would be an interesting problem to pursue by simultaneous analysis of simulations and the experimental data. 

Knowing how to deal with window effects can be an important tool for these visualization experiments.  By combining data at several magnifications (corresponding to different window sizes), we resolve a larger range of length scales.  Higher magnifications will show the small avalanches, while lower magnifications will allow us to both capture larger avalanches and explore more fully the 00 internal avalanches.  For example, if our CCD camera recording the images has a resolution of $1000^2$ pixels, and we have a magnification of up to 5x-50x, we can simultaneously explore a range of window sizes and extend our effective spatial resolution from $1000^2$ to $10000^2$.

More generally, this paper has provided the tools needed to extract from the experimental data for systems of similar interface dynamics the critical exponents $\tau$, $\zeta$ and the universal scaling functions $\calASL$, $\calAhL$, etc.  For these experiments, we can also measure widths and heights and the average shape, giving us an independent measure of $\zeta$, and the universal scaling function $\calAhxL$.  Our detailed analysis and comprehensive methods presented in this paper enable a more powerful interpretation of current experiments, and improved construction of future experiments.

\appendix
\section{Methods and software}
\label{sec:methods}

\subsection{Numerical simulation}
\label{subsec:numerics}
The avalanche simulations in this paper were produced using a quenched KPZ model~\cite{TangLeschhorn92,Buldyrev1992PRA}, with dynamics given by:
\begin{equation}
\label{eq:qKPZ}
\frac{\partial h (x, t)}{\partial t } = F -k \langle h \rangle + \gamma \nabla^2 h + \lambda (\nabla h)^2 + \eta(x,h)
\end{equation}
where $h(x,t)$ is the height of the front, $F$ the driving force, $k$ the ``demagnetization field'', linear and non-linear terms for the KPZ model controlled by the parameters $\gamma$ and $\lambda$ respectively, and $\eta$ gaussian random noise.  This was run for system sizes of width $L$ 4096, 8192, and 16384.  The simulations have been run in a strip geometry ($4096 \times 8192$, $8192 \times 16384$, and $16384 \times 32768$) and the bottom half of the simulations have been truncated to avoid transient effects due to the initially flat front.  The left-right boundaries have periodic boundary conditions.

 The simulations are done using a discrete cellular automaton model, in which the displacement of the string $h$, the time $t$ and the space $x$ are all discretized and take integer values \cite{leschhorn93,leschhorn96PRE}. For a configuration $\{h_i\}$, we compute the local force $F_i$ at each site $i$, leading to a discretized version of Eq.~\ref{eq:qKPZ}
\begin{eqnarray}
F_i =F -k \langle h_i \rangle + \frac{\gamma}{a^2} \sum_{nn} (h_{i+nn}-h_i) + \cr
  \frac{\lambda}{a^2} \sum_{i} (h_{i+i}-h_{i-1})^2 + \eta_i(h_i), 
\end{eqnarray}
where the sum runs over all the nearest neighbors $nn$ for the site $i$, $a$ is the discretization length that we set to $1$, and $\eta_i(h_i)$ is a random force. The automaton dynamics are as follows: (1)~increment the external field until one site is unstable ($F_i >0$); (2)~determine for each site along the interface whether it is stable ($F_i<0$) or unstable ($F_i>0$); (3)~advance all unstable sites by one step $h_i=h_i+a=h_i+1$ in parallel, generating a new value of the pinning force $\eta_i(h_i)$; (4)~repeat until no sites are unstable (the end of the avalanche); (5)~repeat (1-4) until the front passes the top of the simulated window.

\subsection{Nonlinear least squares fitting}
\label{subsec:fitting}
We use nonlinear least squares methods for fitting data to theory functions, minimizing a cost defined as:
\begin{equation}
C(\theta) = \frac{1}{2} \Sigma_i \left(\frac{y^{theory}_i - y^{data}_i(\theta)}{\sigma_i}\right)^2
\end{equation}
Here $\theta$ are the parameters, $y$ the function value, and $\sigma_i$ the error on the data points.  The weight $\sigma$ in our case is determined by fluctuations from run to run.  Namely, we bin the data (in equispaced log bins) for each of $N$ simulations, calculate the standard deviation of this value across runs, divide by $\sqrt{N-1} $ to get the fluctuation in the mean of that bin.~\footnote{This definition of fluctuations in the (simulation) data assumes that the error in each bin is uncorrelated.}

In our distributions, small avalanches occur more often (leading to small error bars), but large avalanches are more important, and the smallest avalanches suffer from non-universal lattice effects.  So during the fitting process there is a tradeoff between fitting the region where there is good data and where the variations are most important.  We use a number of methods to compensate for this imbalance.  (1)~ We set a minimum error bar ($1\%$ of the data value) on the data points, making the error bars on smaller avalanches larger, and therefore decreasing their weight.~\footnote{ We choose this 1\% empirically.  We adjust error bars to be large enough so that the theory is not distorted over the data points with small error bars, but not so large that the points don't matter.   Therefore this value depends on the size of the error bars overall in the data set.} (2)~ Analytic and singular corrections to scaling can also account for non-universal effects.  We include analytic corrections to scaling for the lattice effects in our scaling functions. These corrections appear in all distributions we discuss.   A more detailed discussion is in section E of this appendix.  (3)~One may also skip points that have non-universal behavior when fitting.   For our fits in this paper, all points are included.  

Another issue arises in regions where one has sparse data; there may be bins that do not have any observations.  For these zeros, the error bar should not be zero! We can use maximum likelihood methods to estimate theoretical errors.  Say we have $N$ experiments, and bin the data with $L_i$ sizes in each bin $i$. With the median size in the bin $S_i$, the expected probability for each size is $\rho_i = A(S_i)/S_i$, where $A(S)$ is the theoretical distribution of sizes, and so getting a zero in one of the bins will have the probability $p = (1-\rho_i)^{N L_i}$.  Using maximum likelihood, this results in a cost of $C = -\log p = -NL_i \log (1-\rho_i) \sim N L_i \rho$ given that $\rho$ is small.  The residual we add to the total cost is therefore:
\begin{equation}
r_{zero} = \sqrt{NL_i \rho_i(\theta)}.
\label{eq:ml_err}
\end{equation}  This calculation is generalizable to sparse data that is non-zero- say one has $n$ events in a bin for $N$ measurements, then the probability is $p = \frac{(NL_i)!}{n!(NL_i -n)!} (1-\rho_i)^{NL_i -n}\rho_i^n$, and the appropriate error bar we get is approximately $\sigma = \frac{1}{N L_i}$.  One could then use the larger of the error bar given by this argument or the statistical error bars from the simulation.  In practice, for our simulations, we find it sufficient to use the statistical error bars but to compensate with minimum error bars given by Eq.~\ref{eq:ml_err}.

In non-linear least squares fitting, we can also include priors in the cost if we have assumptions or information a priori about the parameters.  For example, in our problem, we put priors on the exponents inside the scaling functions ($n_{00}$, $\delta_{00}$, etc) to prevent them from going to large values and forcing their corresponding coefficients to zero.  As a result our cost function now becomes:
\begin{equation}
\label{eq:cost_prior}
C(\theta) = \frac{1}{2} \Sigma_i \left( \frac{y^{theory}_i - y^{data}_i(\theta)}{\sigma_i} \right)^2 + \Sigma n^2 + \log(\delta_{11}^2)
\end{equation}
In Eq.~\ref{eq:cost_prior}, $n$ represents all the arbitrary exponents that occur inside the scaling functions ($n_{zz}$, $m_{zz}$, $\delta_{zz}$ etc...). $\log(\delta_{11}^2)$ is included to prevent $\delta_{11}$ from going to zero.

A clear understanding of these techniques for fitting is important for increasing the reliability of our results, and acknowledging its limitations.  In a following subsection, section~\ref{subsec:systematic}, we will further discuss how to estimate the reliability of results inferred by fitting data to a theory, generating systematic error bars for fitting results.

\subsection{Software for fitting}
\label{subsec:software}
To facilitate the exploration of this problem, we have developed a software environment, {\it SloppyScaling}, in Python.  This code is downloadable at http://www.lassp.cornell.edu/
sethna/Sloppy/SloppyScaling/SloppyScaling.htm.  The main features of this code include various nonlinear-least squares fitting methods~\cite{transtrum2011geometry}, automatic plotting for visualization, and methods for generating systematic error bars on the theory.

\subsection{Systematic Error Bars}
\label{subsec:systematic}
We have quoted in our results systematic error bars instead of the more commonly used standard error bars in the parameters.   Standard errors given by the covariance matrix are expected to be erroneous for our problem, since our problem is highly nonlinear in the parameters, and also sloppy- parameter combinations in the sloppiest directions can vary an infinite amount without affecting the fit.  We use a method due to Frederiksen et. al ~\cite{FrederiksenPRL04} for Bayesian estimation of errors.  This method involves assuming that given a theory (M) which is imperfect, a spread of parameters (each corresponding to different models) may fit the data(D) in an equally acceptable matter.  We can define a probability of a certain model with
\begin{equation}
P(\theta|D, M) = \exp(-C(\theta)/T)
\end{equation}
where $C(\theta)$ is the cost at a given set of parameters $\theta$, and the effective temperature $T$ sets a scale for the fluctuations away form the best fit.  Since the cost at the best fit parameters $C_{bf}$ is a measure of how well the theory is doing, we choose $T = 2 C_{bf}/N$, where N is the number of parameters with ''equipartition'' allowing each degree of freedom a contribution of $\frac{1}{2} T $ to the total cost.

Ideally, after defining such a probability, one should sample parameter space to determine the systematic error bars on parameters.  However, in our functions sampling is non-trivial due to parameter evaporation ~\cite{TranstrumPRL10}, the ``entropy'' for parameters drifting to infinity overwhelms their cost in degrading the resulting fits.   Therefore, we make a quadratic estimate of the fluctuations in predictions, essentially using propagation of error to estimate the systematic error.  The covariance matrix gives an error $\sigma_{stat}$ that assumes the temperature of 1, corresponding to $P(\theta|D, M) = \exp(-C(\theta))$.  Using propagation of error, we calculate the systematic error according to our effective temperature $T = 2 C_{bf}/N$:
\begin{equation}
\sigma_{sys} = \sqrt{T} \sigma_{cov}
\end{equation}
The shaded plots are generated by sampling according to the Hessian at the best fit, weighting our steps in each eigendirection by the inverse square root of the eigenvalue, and scaling the steps with a low temperature ($T_L$).  Then for our ensemble of parameters we calculate the fluctuations in the theory (residuals $\delta r_{ens}$) corresponding to the ensemble.  We scale up these fluctuations according to the temperature defined by the best fit, or $ \delta r_{T}=\sqrt{\frac{T_{bf}}{T_L}} \delta r_{ens}$,
\begin{equation}
\delta r_{T} = \Sigma \frac{\partial r}{\partial \theta} \delta \theta_{T} 
\end{equation}
We have also estimated systematic errors by removing large $k$ curves (which we believe have larger corrections), and looking at the corresponding drift in exponents and parameters.  The estimate of the systematic error that this procedure gives is often similar (sometimes smaller) to the one using our temperature-scaled propagation of error.

\subsection{Corrections to scaling}
\label{subsec:corrections}

Corrections to scaling play an important role in our scaling functions, their inclusion accounts for non-universal effects, and helps increase the reliability of universal predictions.  In each of the scaling functions in this paper we have included analytic corrections to scaling that capture the lattice effects of our automata simulations.  They are of the form:
\begin{equation}
\exp(A^1/S + A^2/S^2)
\label{eq:latticecorrections}
\end{equation}  
This expansion for small $S$ corrects for the lattice effects on small size avalanches which we believe to be present in the distributions.  One can imagine that experiments may have other origins of non-universal effects, such as a nonlinear signal amplifier or distorting lens; one should always attempt to account for and include these~\cite{SethnaDahmen04}.  

One caveat is that, just as adding extra free parameters does not necessarily increase the quality of one's fits, adding corrections is not a guarantee for increasing the accuracy of one's scaling function.  One should be careful in checking that the terms included in the corrections to scaling behave as expected in the region of interest, are subdominant when taking the appropriate limits, and do not confuse the main universal scaling function, either by canceling out terms or having the same effect.  An example of this complication is seen in our studies of using the limit of the $11$ distribution at $W=0$ for the local height distributions.  

In sections~\ref{subsec:localheights} and~\ref{subsec:11}, we take the view point of using the parameterized from of $\All(s|L_k,W=0)$ as the proper limit of $A(h_x|L_k)$.  However, this data should be matched as well by $\All(s|L_k, W=1)$, since in effect this is what we are fitting for the local height distributions.  We can view the ratio between the two functions as a multiplicative correction to scaling from lattice effects (without considering the corrections to scaling for lattice effects in equation~\ref{eq:latticecorrections}):
\begin{align}
f(h_x,L_k) &= \frac{\All(s|L_k,W=1)}{\All(s|L_k, W=0)} \\
&= \exp ( -( U_{11} (\frac{h_x}{L_k^{1+\zeta}})^{1/2} + Z_{11} ( \frac{h_x}{L_k^{1+\zeta}} )^{\delta_{11}} \cr
& \, + C_{11} \left( \frac{h_x}{L_k^{1+\zeta-n_{11}}} \right)^{-m_2} )) \cr
& = \exp(-(U_{11} L_k^{-1/2} h_s^{1/2} + Z_{11}  L_k^{-\delta_{11}} h_s^{\delta_{11}} \cr
& \, + C_{11}  L_k^{(1+\zeta-n_{11}) m_2} h_x^{-m_2})) \nonumber
\label{eq:correction_hx}
\end{align}
Here we define $h_s = h_x/L_k^\zeta$.  Note however that for the term $C_{11} L_k^{(1+\zeta-n_{11})m_2} h_x^{-m_2}$, the value of the fit for $n_{11}$ equals $1.65$, which is within the range of error for $1+\zeta = 1.63 \pm 0.02$.  Therefore this term is nearly $C_{11} h_x^{-m_2}$, where $m_2 = 1.61 \pm 0.6$.  This term then has the same effect as the correction to scaling term $A_2^{11}/s^2$ at small $W$ (since $s \sim h  W$ for $11$ avalanches at small $W$).  Since $A_2^{11}/s^2$ is only significant in the range of small $s$, which only occur in the $11$ distributions at small $W$, these two parameters serve the same purpose, and it is redundant to include both for the fits.  We therefore remove the term $A_2^{11}/s^2$ for the $11$ distributions for our fits.          

One can also check the corrections and see all powers of $L_k$ are negative, and the multiplicative correction approaches unity as we get closer to the critical point.  With our current fit, the term with $L_k^{-1/2}$ dominates the corrections.  Notice that, originally, to account for lattice effects, we have added corrections in integer powers of $h_x$ ($\exp(A^{11}_1/h_x + A^{11}_2/h_x^2)$), which are subdominant compared to $L_k^{-1/2}$.  This implies that there are more dangerous corrections to scaling than originally inferred.  By this method, one might check if the corrections originally included are sufficient- in this case they are close, but could be improved upon by systematically adding similar terms.

In a continuum case, the concept of local heights $h_x$ should describe a smooth shape tracing the depinning line, in our automata, since the smallest width is naturally 1 lattice spacing, the smallest possible window width is $W=1$.  In measuring the local heights, this discreteness limits the smoothness of the shape of $h_x$,  and gives rise to the corrections we see.  Here we've seen that if we ``know'' from other measurements (in our case the $11$ spanning avalanches) the right limit the universal scaling form should take, we may find the form of the corrections.  

\subsection{Scaling collapses and their limitations}

We have argued at various points in the paper that scaling collapses are limited and may lead to questionable results.  Here we will illustrate an example of this.  Using the critical exponents given by a free fit of the windowed distributions ($\zeta$ allowed to vary), we collapse the sizes, heights and widths.  Comparing the figures included here and the ones in Section~\ref{sec:FullSystem} (Figures~\ref{fig:Ascollapse},~\ref{fig:Ahcollapse}, and ~\ref{fig:Awcollapse}.), one can see that the collapses are of similar quality.  

\begin{figure}[h]
\begin{center}
\includegraphics[scale=0.35]{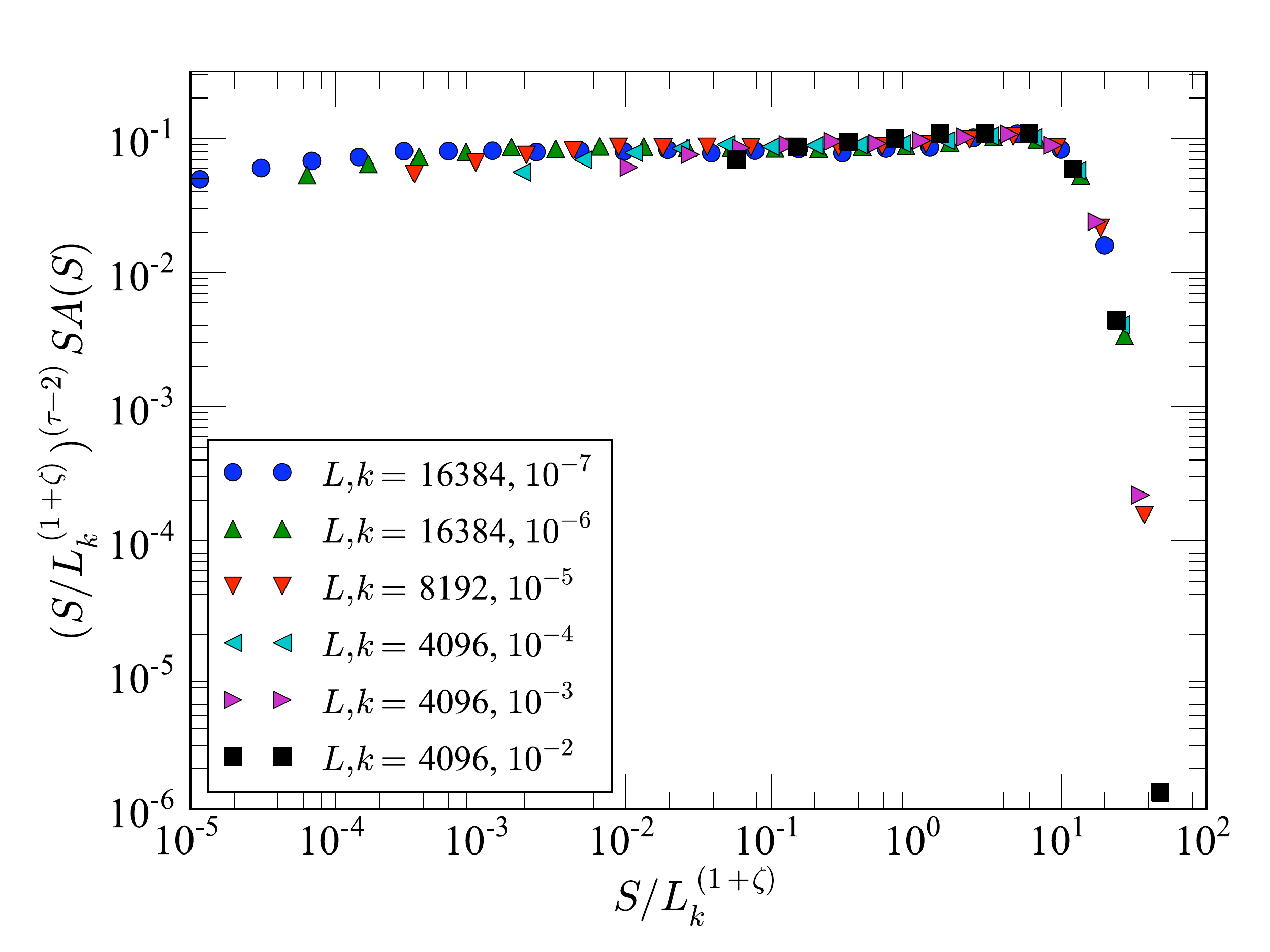}
\caption{{\bf Size distribution collapse} (color online) Here we collapse the size distributions with exponents $\tau = 1.25$, $\nu_k=0.44$, and $\zeta = 0.68$.  Notice that the collapses are similar to the ones shown in Figure~\ref{fig:Ascollapse}.   Here only the combination of $\nu_k (1+\zeta)$ affect the scaling collapse, the large shifts in $\nu_k$ and $\zeta$ mostly cancel in the product, yielding similar collapses.}
\end{center}
\label{fig:Ascollapse-068}
\end{figure}

\begin{figure}[h]
\begin{center}
\includegraphics[scale=0.35]{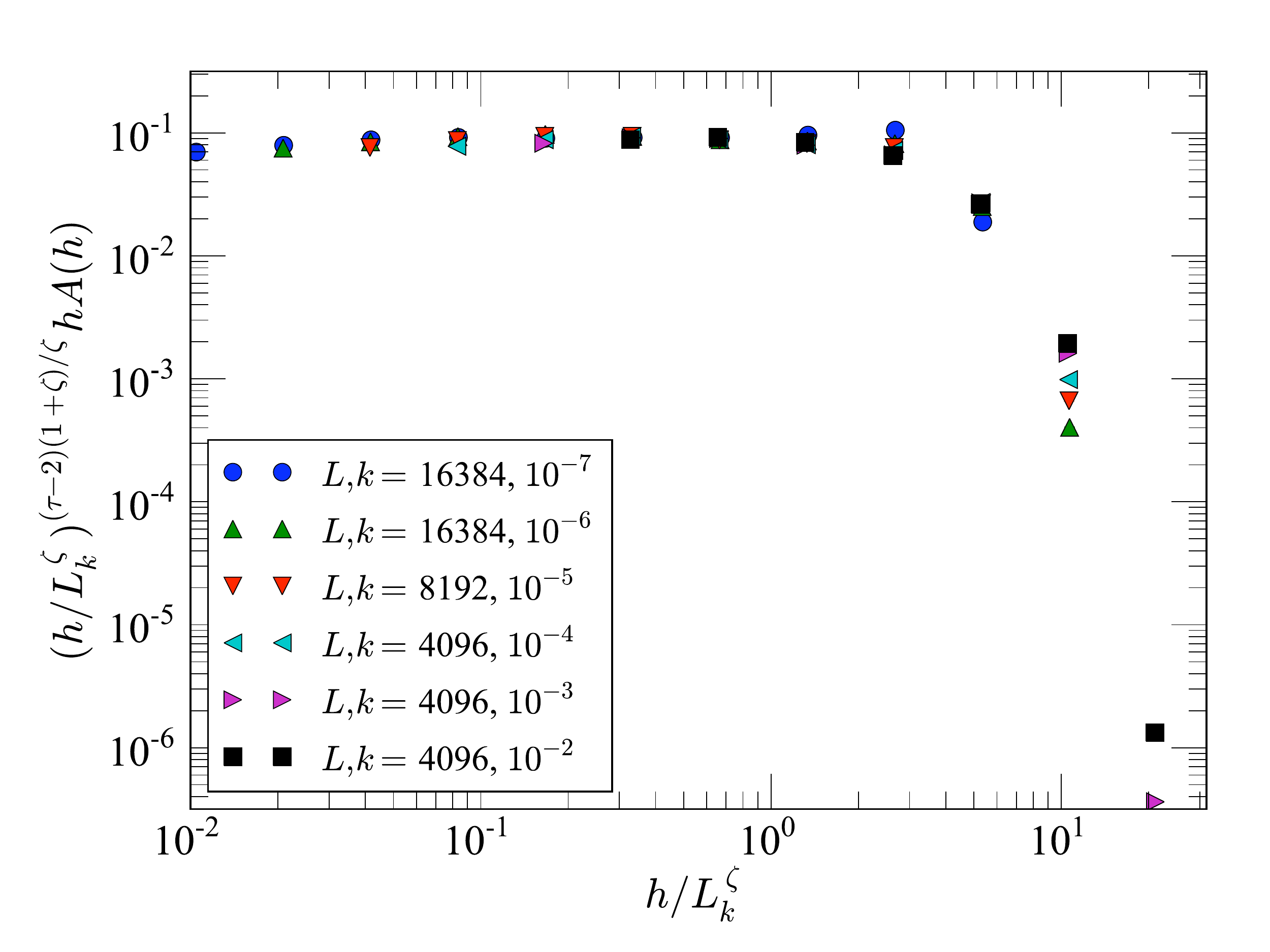}
\caption{{\bf Height distribution collapse} (color online) Here we collapse the size distributions with exponents $\tau = 1.25$, $\nu_k=0.44$, and $\zeta = 0.68$. Here only the combination of $\nu_k \zeta$ affect the scaling collapse, the large shifts in $\nu_k$ and $\zeta$ mostly cancel in the product, yielding similar collapses.  Comparing this with the collapse shown in Figure~\ref{fig:Ahcollapse}, we see that the two collapses are comparable in quality, where in Figure~\ref{fig:Ahcollapse} the large avalanche cutoff is collapsed nicely, and here the smaller avalanches are collapsed better.}
\end{center}
\label{fig:Ahcollapse-068}
\end{figure}

\begin{figure}[h]
\begin{center}
\includegraphics[scale=0.35]{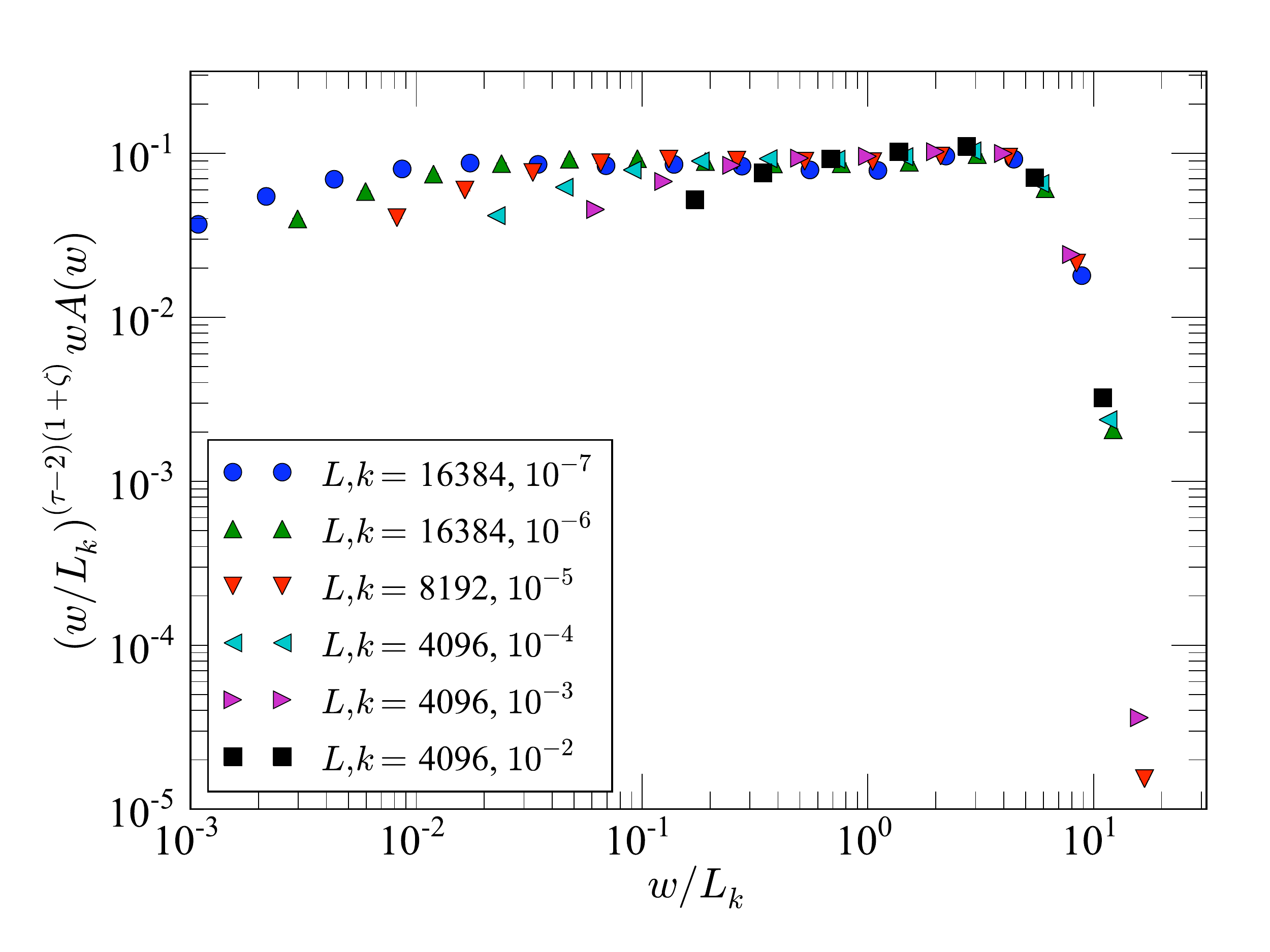}
\caption{{\bf Width distribution collapse} (color online) Here we collapse the size distributions with exponents $\tau = 1.25$, $\nu_k=0.44$, and $\zeta = 0.68$.  Notice that the collapses are similar to the ones shown in Figure~\ref{fig:Awcollapse}.   Here only the combination of $\nu_k$ affects the scaling collapse and $(\tau-2)(1+\zeta)$ affects the shape of the scaling collapse.  Since $\nu_k$ does not change significantly, the quality of the collapses are similar.}
\end{center}
\label{fig:Awcollapse-068}
\end{figure}

The fact that such distinct values of $\zeta$ can yield similar quality collapses may imply (1)~our "systematic error" bars on $\zeta$, estimated to be $\pm 0.02$, are in reality much larger, (2)~collapses do not incorporate non-universal corrections to scaling, and these may have an important effect,  (3)~collapses are not a reliable way of verifying the values of critical exponents.  In particular, we expect corrections to scaling due to large $k$ to be responsible for the drift in exponents.

\vspace{16pt}

With these software tools and analytical methods, data at critical points may be analyzed while including multiple scaling variables, allowing for the treatment of a broad range of experiments, and also allowing for a far more rigorous estimation of statistical and systematic errors.  By using automatic fits to entire scaling functions, instead of traditional collapses, and by estimating systematic error bars, we facilitate the interpretation of data with multiple scaling variables and analytic corrections to scaling.  This advance will allow for better characterization not only of noise in magnetic thin films and similar avalanche dynamics, but should be broadly applicable to all applications of critical phenomena and scaling theories to experiments and simulations.

\section{The roughness exponent $\zeta$}
\label{sec:zeta}

In the investigations presented in this manuscript, we have found that the estimates of the critical exponent $\zeta$ have been problematic.  In this appendix, we will discuss various means of measuring this exponent, the significance of the range of values we find from various measurement methods, possible origins of this range, and implications for future research. We emphasize that any value of $\zeta$ in the range we observe ($0.62\pm 0.02$ to $0.72 \pm 0.02$) can describe all of our data essentially equivalently well.

\begin{figure}[h]
\begin{center}
\includegraphics[scale=0.35]{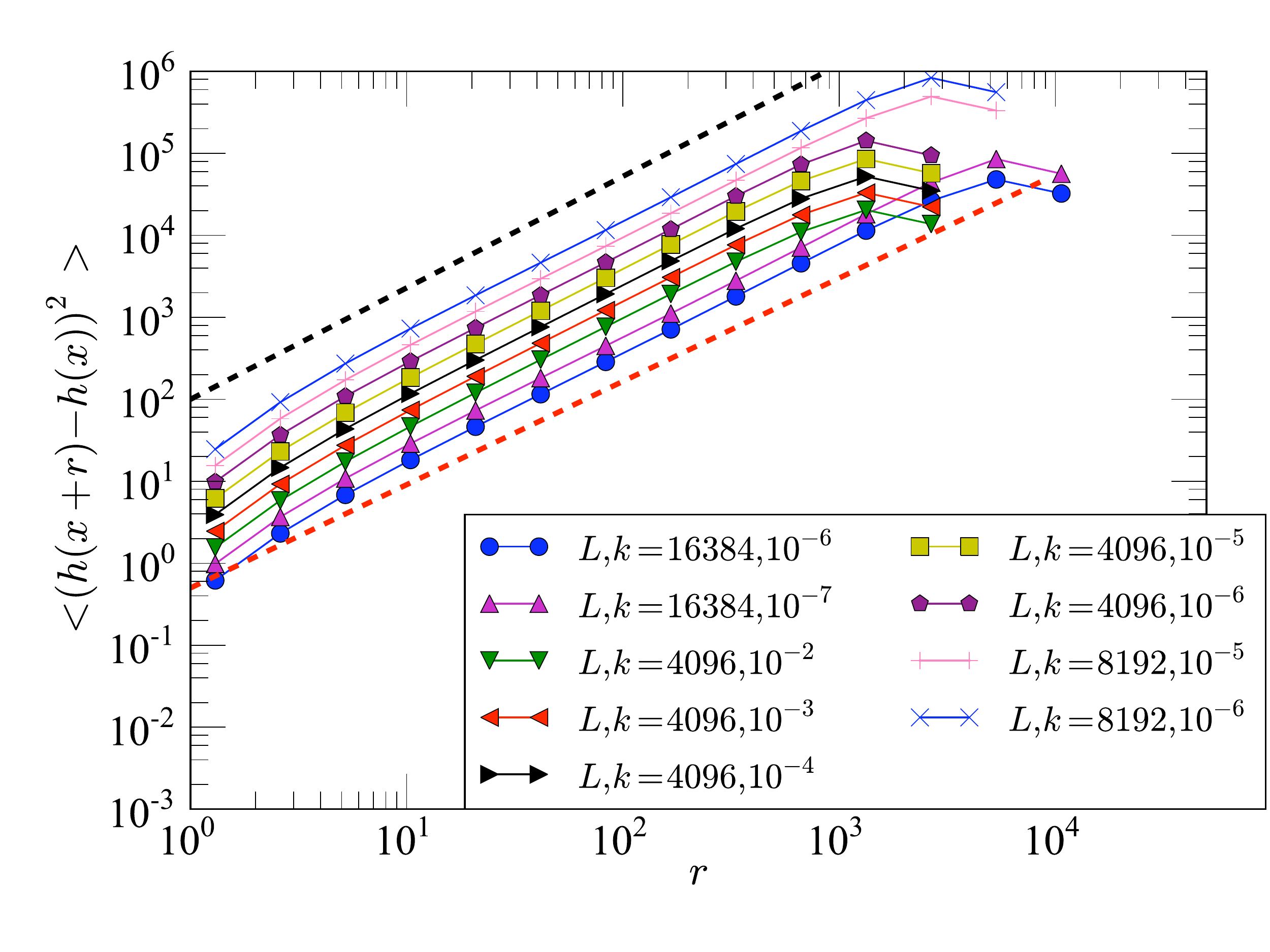}
\caption{{\bf  Height-height correlations for the qKPZ simulations} (color online) Shown here are the roughness exponents for various simulation sizes $L$ and $k$.  We measure the height-height correlation function $C(r) \sim <(h(x+r)-h(r))^2>$.  A power law fit shows $\zeta$ falls between 0.63 and 0.68.  The lower red-dashed line shows $\zeta = 0.63$ and upper black-dashed line $\zeta=0.68$.  The lines were shifted to show each individual power law.}
\label{fig:corr}
\end{center}
\end{figure}

The shape of the front has been studied as an identifying feature for front propagation models, which is usually characterized by defining a roughness exponent $\zeta$, which is measured through a height-height correlation function:
\begin{equation}
C(r) = \langle (h(x+r)-h(x))^2 \rangle \sim r^{2 \zeta}
\end{equation}

\begin{figure*}[h]
\begin{center}
\includegraphics[scale=0.65]{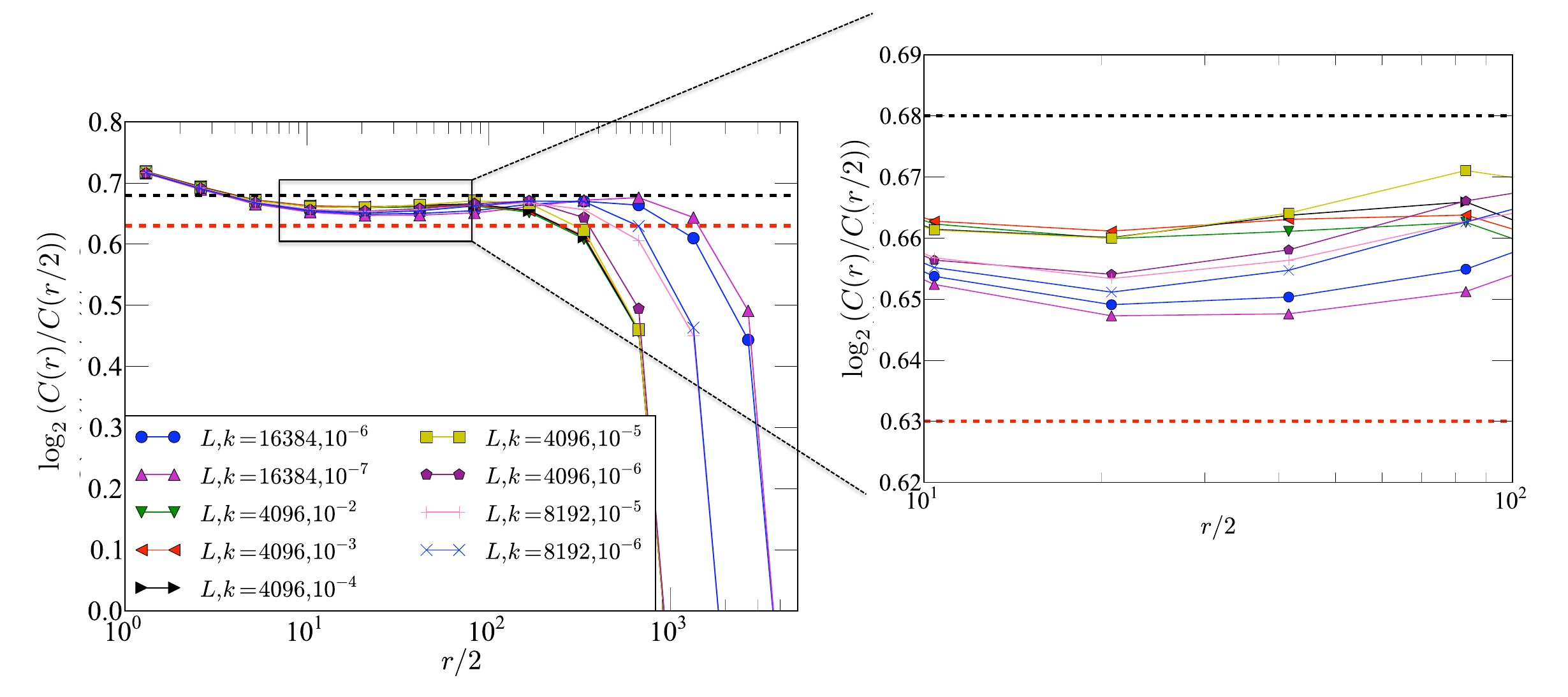}
\caption{{\bf  Roughness exponents for the qKPZ simulations} (color online) Shown here are the measurements of the local-log slope $\ln[C(r)/C(r/2)]/\ln(2)$ of the height-height correlation function;  this is a measure for the roughness exponents for various simulation sizes $L$ and $k$.  The results of the local-log slope $\ln[C(r)/C(r/2)]/\ln(2)$ is consistent with what is seen in Figure~\ref{fig:corr}.  The lower red-dashed line shows $\zeta =0.63$, corresponding to directed-percolation depinning (DPD) and suggested by literature to be the correct value for our model.  Whereas the upper black-dashed line shows $\zeta=0.68$ which is the result of our fits of windowed avalanche distributions.  In the blowup of the region of $r/2 = 10-100$, we can see there is a trend of larger $\zeta$ corresponding to larger $k$ simulations. }
\label{fig:local_log}
\end{center}
\end{figure*}

The quenched KPZ model we use is conjectured to belong to the Directed Percolation Depinning (DPD) universality class~\cite{barabasi95fractal}, which is conjectured in turn to belong to the Directed Percolation(DP) universality class.  For a pinned interface in DPD, the roughness exponent $\zeta = 0.63 \pm 0.01$~\cite{TangLeschhorn92,Buldyrev1992PRA} matches that of the ratio of correlation length exponents in DP $\nu_\bot/\nu_\parallel =0.6326 \pm0.0002$~\cite{essam1988DP}.  (Here $\xi_\bot \sim |p-p_c|^{\nu_\bot}$ and $\xi_\parallel \sim |p-p_c|^{\nu_\parallel}$, where $p$ is the branching probability, and $p_c$ is the percolation threshold.).  One may note that for a moving interface, the picture is less well known; there has been numerical study that shows $\zeta = 0.70$~\cite{buldyrev1992anomalous,maske1995scaling}, but also arguments that the interface under this condition is not self-affine~\cite{leschhorn96PRE}, that the moving regions have $\zeta = 1$ and the pinned regions have $\zeta=0.63$.  One could imagine that our demagnetizing force, like the velocity in DPD, could lead to a heterogeneous mixture of different scaling regions, converging to $\zeta=0.63$ as $k=0$.    

Figure~\ref{fig:corr} shows measurements of the height-height correlations in our model.   Using finite-size scaling for a numerical fit, we see that as $k$ is tuned away from zero, $\zeta$ falls between 0.64 and 0.68, increasing with $k$.  Measuring the local log-slope, one can see clearly a drift in the measured exponent in Figure~\ref{fig:local_log}.  

The range of our estimates (varying from $\zeta = 0.63\pm 0.02$ to $\zeta = 0.72\pm 0.02$) is large compared to our error estimates; however, best fits with $\zeta$ fixed within this range had costs within 1.5 times that of the best fit value, indicating both that our quadratic estimates for the systematic errors are too small and that it may be challenging to definitively measure $\zeta$ in either simulation or experiment.

In our fits we find that $\zeta = 0.62 \pm 0.02$ for the size, width and height distributions joint fit, and $\zeta = 0.68 \pm 0.02$ for the $00$, $10$, $11$ joint distributions.  Notice that although the direction of front propagation is in general along the y-axis in our problem, portions of the front will be at various angles to the y-axis.  Since the local direction of the front propagation is not fixed, we can also choose a rotationally invariant definition of height and width: defining the height and width of an avalanche along the axes of the moment of inertia tensor.  We define root-mean-square heights and widths as the square root of its eigenvalues.  So if we fit {\it rms} heights and widths jointly, we get $\zeta = 0.72 \pm 0.02$ much higher than $0.62$ measured along the global axes.  The difference in these two exponents seems to indicate that the local avalanche shape has a different geometry than the global avalanche front.  Many front propagation models spontaneously break rotational symmetry through the orientation of the front.  (Envision a circular front growing from a point, with differing front orientations.)  Note, however, that the qKPZ dynamics is anisotropic, breaking rotational invariance.  

We have also looked at fits with the windows scaling functions involving subsets of simulations with different $k$.  Using smaller values of $k$ generally lead to fits of $\zeta$ closer to 0.63.  This may point to imperfections in our theory function (are we missing some of the scaling behavior dealing with $k$?), or corrections to scaling (analytic or singular).  One possibility is that there is a residual crossover effect having to do with another relevant variable.  In the anisotropic form of the qKPZ model, the nonlinear term (Eq.~\ref{eqn:qKPZ}) $\lambda$ is a relevant variable, and is non-zero under renormalization ~\cite{barabasi95fractal,kardar97interfaces,ledoussal03anisotropic}, and although we simulate the model at fixed $\lambda$, we have observed there is a crossover effect following a direction having to do with both $k$ and $\lambda$.  We note that without the nonlinear KPZ term the qKPZ model becomes the quenched Edwards-Wilkinson (qEW) model, and that for the qEW $\zeta > 1$.  Literature suggests there also may be a crossover effect due to a runaway fixed point ~\cite{ledoussal03anisotropic}.  We think these last two possibilities can be explored with more simulations done on different $\lambda$ and $k$, examining a crossover to the linear version of the qKPZ model (the quenched Edwards-Wilkinson), to make a more complete picture of the phase space.

\section*{Acknowledgements}

We would like to thank A. Rosso for a useful discussion on the proper definition of universal scaling forms, and M. Transtrum for his insights on non-linear least square fitting techniques and Monte-Carlo sampling.  Y-J. C. and J.P.S. were supported by NSF DMR-1005479.  S. P.  would like to acknowledge support from DOE Basic Energy Sciences DE-FG02-07ER46393.

\bibliographystyle{apsrev}

\begin{thebibliography}{40}
\expandafter\ifx\csname natexlab\endcsname\relax\def\natexlab#1{#1}\fi
\expandafter\ifx\csname bibnamefont\endcsname\relax
  \def\bibnamefont#1{#1}\fi
\expandafter\ifx\csname bibfnamefont\endcsname\relax
  \def\bibfnamefont#1{#1}\fi
\expandafter\ifx\csname citenamefont\endcsname\relax
  \def\citenamefont#1{#1}\fi
\expandafter\ifx\csname url\endcsname\relax
  \def\url#1{\texttt{#1}}\fi
\expandafter\ifx\csname urlprefix\endcsname\relax\def\urlprefix{URL }\fi
\providecommand{\bibinfo}[2]{#2}
\providecommand{\eprint}[2][]{\url{#2}}

\bibitem[{\citenamefont{Sethna et~al.}(2001)\citenamefont{Sethna, Dahmen, and
  Myers}}]{sethna2001crackling}
\bibinfo{author}{\bibfnamefont{J.~P.} \bibnamefont{Sethna}},
  \bibinfo{author}{\bibfnamefont{K.~A.} \bibnamefont{Dahmen}},
  \bibnamefont{and} \bibinfo{author}{\bibfnamefont{C.~R.} \bibnamefont{Myers}},
  \bibinfo{journal}{Nature} \textbf{\bibinfo{volume}{410}},
  \bibinfo{pages}{242} (\bibinfo{year}{2001}).

\bibitem[{\citenamefont{{Le Doussal} and {Wiese}}(2011)}]{le2011distribution}
\bibinfo{author}{\bibfnamefont{P.}~\bibnamefont{{Le Doussal}}}
  \bibnamefont{and} \bibinfo{author}{\bibfnamefont{K.~J.}
  \bibnamefont{{Wiese}}}, \bibinfo{journal}{ArXiv e-prints}
  (\bibinfo{year}{2011}), \eprint{1104.2629}.

\bibitem[{\citenamefont{{Kardar}}(1998)}]{kardar97interfaces}
\bibinfo{author}{\bibfnamefont{M.}~\bibnamefont{{Kardar}}},
  \bibinfo{journal}{Phys. Rep.} \textbf{\bibinfo{volume}{301}},
  \bibinfo{pages}{85} (\bibinfo{year}{1998}), \eprint{arXiv:cond-mat/9704172}.

\bibitem[{\citenamefont{{Durin} and {Zapperi}}(2004)}]{DurinZapperi04}
\bibinfo{author}{\bibfnamefont{G.}~\bibnamefont{{Durin}}} \bibnamefont{and}
  \bibinfo{author}{\bibfnamefont{S.}~\bibnamefont{{Zapperi}}}, in
  \emph{\bibinfo{booktitle}{The Science of Hysteresis}}, edited by
  \bibinfo{editor}{\bibfnamefont{G.}~\bibnamefont{{Bertotti}}}
  \bibnamefont{and}
  \bibinfo{editor}{\bibfnamefont{I.}~\bibnamefont{{Mayergoyz}}}
  (\bibinfo{year}{2004}), vol.~\bibinfo{volume}{II},
  \eprint{arXiv:cond-mat/0404512}.

\bibitem[{\citenamefont{Ryu et~al.}(2007)\citenamefont{Ryu, Akinaga, and
  Shin}}]{ryu07nature}
\bibinfo{author}{\bibfnamefont{K.~S.} \bibnamefont{Ryu}},
  \bibinfo{author}{\bibfnamefont{H.}~\bibnamefont{Akinaga}}, \bibnamefont{and}
  \bibinfo{author}{\bibfnamefont{S.~C.} \bibnamefont{Shin}},
  \bibinfo{journal}{Nat. Phys.} \textbf{\bibinfo{volume}{3}},
  \bibinfo{pages}{547} (\bibinfo{year}{2007}).

\bibitem[{\citenamefont{Magni et~al.}(2009)\citenamefont{Magni, Durin, Zapperi,
  and Sethna}}]{MagniDurin09}
\bibinfo{author}{\bibfnamefont{A.}~\bibnamefont{Magni}},
  \bibinfo{author}{\bibfnamefont{G.}~\bibnamefont{Durin}},
  \bibinfo{author}{\bibfnamefont{S.}~\bibnamefont{Zapperi}}, \bibnamefont{and}
  \bibinfo{author}{\bibfnamefont{J.~P.} \bibnamefont{Sethna}},
  \bibinfo{journal}{J. Stat. Mech: Theory Exp.}
  \textbf{\bibinfo{volume}{2009}}, \bibinfo{pages}{P01020}
  (\bibinfo{year}{2009}).

\bibitem[{\citenamefont{Schwarz et~al.}(2004)\citenamefont{Schwarz, Liebmann,
  Kaiser, Wiesendanger, Noh, and Kim}}]{schwarz04PRL}
\bibinfo{author}{\bibfnamefont{A.}~\bibnamefont{Schwarz}},
  \bibinfo{author}{\bibfnamefont{M.}~\bibnamefont{Liebmann}},
  \bibinfo{author}{\bibfnamefont{U.}~\bibnamefont{Kaiser}},
  \bibinfo{author}{\bibfnamefont{R.}~\bibnamefont{Wiesendanger}},
  \bibinfo{author}{\bibfnamefont{T.~W.} \bibnamefont{Noh}}, \bibnamefont{and}
  \bibinfo{author}{\bibfnamefont{D.~W.} \bibnamefont{Kim}},
  \bibinfo{journal}{Phys. Rev. Lett.} \textbf{\bibinfo{volume}{92}},
  \bibinfo{pages}{77206} (\bibinfo{year}{2004}).

\bibitem[{\citenamefont{Christian et~al.}(2006)\citenamefont{Christian,
  Novoselov, and Geim}}]{Christian06PRB}
\bibinfo{author}{\bibfnamefont{D.~A.} \bibnamefont{Christian}},
  \bibinfo{author}{\bibfnamefont{K.~S.} \bibnamefont{Novoselov}},
  \bibnamefont{and} \bibinfo{author}{\bibfnamefont{A.~K.} \bibnamefont{Geim}},
  \bibinfo{journal}{Phys. Rev. B} \textbf{\bibinfo{volume}{74}},
  \bibinfo{pages}{064403} (\bibinfo{year}{2006}).

\bibitem[{\citenamefont{Barab{\'a}si and Stanley}(1995)}]{barabasi95fractal}
\bibinfo{author}{\bibfnamefont{A.~L.} \bibnamefont{Barab{\'a}si}}
  \bibnamefont{and} \bibinfo{author}{\bibfnamefont{H.~E.}
  \bibnamefont{Stanley}}, \emph{\bibinfo{title}{{Fractal concepts in surface
  growth}}} (\bibinfo{publisher}{Cambridge Univ Pr}, \bibinfo{year}{1995}),
  ISBN \bibinfo{isbn}{0521483182}.

\bibitem[{\citenamefont{Tang and
  Leschhorn}(1992{\natexlab{a}})}]{TangLeschhorn92}
\bibinfo{author}{\bibfnamefont{L.~H.} \bibnamefont{Tang}} \bibnamefont{and}
  \bibinfo{author}{\bibfnamefont{H.}~\bibnamefont{Leschhorn}},
  \bibinfo{journal}{Phys. Rev. A} \textbf{\bibinfo{volume}{45}},
  \bibinfo{pages}{8309} (\bibinfo{year}{1992}{\natexlab{a}}).

\bibitem[{\citenamefont{Buldyrev
  et~al.}(1992{\natexlab{a}})\citenamefont{Buldyrev, Barab\'asi, Caserta,
  Havlin, Stanley, and Vicsek}}]{Buldyrev1992PRA}
\bibinfo{author}{\bibfnamefont{S.~V.} \bibnamefont{Buldyrev}},
  \bibinfo{author}{\bibfnamefont{A.-L.} \bibnamefont{Barab\'asi}},
  \bibinfo{author}{\bibfnamefont{F.}~\bibnamefont{Caserta}},
  \bibinfo{author}{\bibfnamefont{S.}~\bibnamefont{Havlin}},
  \bibinfo{author}{\bibfnamefont{H.~E.} \bibnamefont{Stanley}},
  \bibnamefont{and} \bibinfo{author}{\bibfnamefont{T.}~\bibnamefont{Vicsek}},
  \bibinfo{journal}{Phys. Rev. A} \textbf{\bibinfo{volume}{45}},
  \bibinfo{pages}{R8313} (\bibinfo{year}{1992}{\natexlab{a}}).

\bibitem[{\citenamefont{Buldyrev
  et~al.}(1992{\natexlab{b}})\citenamefont{Buldyrev, Barab{\'a}si, Havlin,
  Kertesz, Stanley, and Xenias}}]{buldyrev1992anomalous}
\bibinfo{author}{\bibfnamefont{S.~V.} \bibnamefont{Buldyrev}},
  \bibinfo{author}{\bibfnamefont{A.~L.} \bibnamefont{Barab{\'a}si}},
  \bibinfo{author}{\bibfnamefont{S.}~\bibnamefont{Havlin}},
  \bibinfo{author}{\bibfnamefont{J.}~\bibnamefont{Kertesz}},
  \bibinfo{author}{\bibfnamefont{H.~E.} \bibnamefont{Stanley}},
  \bibnamefont{and} \bibinfo{author}{\bibfnamefont{H.~S.}
  \bibnamefont{Xenias}}, \bibinfo{journal}{Physica A}
  \textbf{\bibinfo{volume}{191}}, \bibinfo{pages}{220}
  (\bibinfo{year}{1992}{\natexlab{b}}).

\bibitem[{\citenamefont{Leschhorn}(1996)}]{leschhorn96PRE}
\bibinfo{author}{\bibfnamefont{H.}~\bibnamefont{Leschhorn}},
  \bibinfo{journal}{Phys. Rev. E} \textbf{\bibinfo{volume}{54}},
  \bibinfo{pages}{1313} (\bibinfo{year}{1996}).

\bibitem[{\citenamefont{Vives et~al.}(1994)\citenamefont{Vives, Ort{\'\i}n,
  Ma{\~n}osa, R{\`a}fols, P{\'e}rez-Magran{\'e}, and Planes}}]{martensite1994}
\bibinfo{author}{\bibfnamefont{E.}~\bibnamefont{Vives}},
  \bibinfo{author}{\bibfnamefont{J.}~\bibnamefont{Ort{\'\i}n}},
  \bibinfo{author}{\bibfnamefont{L.}~\bibnamefont{Ma{\~n}osa}},
  \bibinfo{author}{\bibfnamefont{I.}~\bibnamefont{R{\`a}fols}},
  \bibinfo{author}{\bibfnamefont{R.}~\bibnamefont{P{\'e}rez-Magran{\'e}}},
  \bibnamefont{and} \bibinfo{author}{\bibfnamefont{A.}~\bibnamefont{Planes}},
  \bibinfo{journal}{Phys. Rev. Lett.} \textbf{\bibinfo{volume}{72}},
  \bibinfo{pages}{1694} (\bibinfo{year}{1994}).

\bibitem[{\citenamefont{P{\'e}rez-Reche
  et~al.}(2004)\citenamefont{P{\'e}rez-Reche, Tadi{\'c}, Ma{\~n}osa, Planes,
  and Vives}}]{martensite2004}
\bibinfo{author}{\bibfnamefont{F.}~\bibnamefont{P{\'e}rez-Reche}},
  \bibinfo{author}{\bibfnamefont{B.}~\bibnamefont{Tadi{\'c}}},
  \bibinfo{author}{\bibfnamefont{L.}~\bibnamefont{Ma{\~n}osa}},
  \bibinfo{author}{\bibfnamefont{A.}~\bibnamefont{Planes}}, \bibnamefont{and}
  \bibinfo{author}{\bibfnamefont{E.}~\bibnamefont{Vives}},
  \bibinfo{journal}{Physical review letters} \textbf{\bibinfo{volume}{93}},
  \bibinfo{pages}{195701} (\bibinfo{year}{2004}).

\bibitem[{\citenamefont{Puppin}(2000)}]{Puppin00}
\bibinfo{author}{\bibfnamefont{E.}~\bibnamefont{Puppin}},
  \bibinfo{journal}{Phys. Rev. Lett.} \textbf{\bibinfo{volume}{84}},
  \bibinfo{pages}{5415} (\bibinfo{year}{2000}).

\bibitem[{\citenamefont{Welling et~al.}(2005)\citenamefont{Welling, Aegerter,
  and Wijngaarden}}]{welling2005fluxavalanches}
\bibinfo{author}{\bibfnamefont{M.~S.} \bibnamefont{Welling}},
  \bibinfo{author}{\bibfnamefont{C.~M.} \bibnamefont{Aegerter}},
  \bibnamefont{and} \bibinfo{author}{\bibfnamefont{R.~J.}
  \bibnamefont{Wijngaarden}}, \bibinfo{journal}{Phys. Rev. B}
  \textbf{\bibinfo{volume}{71}}, \bibinfo{pages}{104515}
  (\bibinfo{year}{2005}), ISSN \bibinfo{issn}{1550-235X}.

\bibitem[{\citenamefont{M{\aa}l{\o}y and Schmittbuhl}(2001)}]{2001slowcrack}
\bibinfo{author}{\bibfnamefont{K.~J.} \bibnamefont{M{\aa}l{\o}y}}
  \bibnamefont{and}
  \bibinfo{author}{\bibfnamefont{J.}~\bibnamefont{Schmittbuhl}},
  \bibinfo{journal}{Phys. Rev. Lett.} \textbf{\bibinfo{volume}{87}},
  \bibinfo{pages}{105502} (\bibinfo{year}{2001}), ISSN
  \bibinfo{issn}{1079-7114}.

\bibitem[{\citenamefont{Planet et~al.}(2009)\citenamefont{Planet, Santucci, and
  Ort{\'\i}n}}]{planet2009imbibition}
\bibinfo{author}{\bibfnamefont{R.}~\bibnamefont{Planet}},
  \bibinfo{author}{\bibfnamefont{S.}~\bibnamefont{Santucci}}, \bibnamefont{and}
  \bibinfo{author}{\bibfnamefont{J.}~\bibnamefont{Ort{\'\i}n}},
  \bibinfo{journal}{Phys. Rev. Lett.} \textbf{\bibinfo{volume}{102}},
  \bibinfo{pages}{94502} (\bibinfo{year}{2009}), ISSN
  \bibinfo{issn}{1079-7114}.

\bibitem[{\citenamefont{Aegerter et~al.}(2003)\citenamefont{Aegerter,
  G{\\"u}nther, and Wijngaarden}}]{aegerter2003rice}
\bibinfo{author}{\bibfnamefont{C.~M.} \bibnamefont{Aegerter}},
  \bibinfo{author}{\bibfnamefont{R.}~\bibnamefont{G{\\"u}nther}},
  \bibnamefont{and} \bibinfo{author}{\bibfnamefont{R.~J.}
  \bibnamefont{Wijngaarden}}, \bibinfo{journal}{Phys. Rev. E}
  \textbf{\bibinfo{volume}{67}}, \bibinfo{pages}{51306} (\bibinfo{year}{2003}),
  ISSN \bibinfo{issn}{1550-2376}.

\bibitem[{\citenamefont{Aegerter et~al.}(2004)\citenamefont{Aegerter,
  L{\H{o}}rincz, Welling, and Wijngaarden}}]{aegerter2004rice}
\bibinfo{author}{\bibfnamefont{C.~M.} \bibnamefont{Aegerter}},
  \bibinfo{author}{\bibfnamefont{K.~A.} \bibnamefont{L{\H{o}}rincz}},
  \bibinfo{author}{\bibfnamefont{M.~S.} \bibnamefont{Welling}},
  \bibnamefont{and} \bibinfo{author}{\bibfnamefont{R.~J.}
  \bibnamefont{Wijngaarden}}, \bibinfo{journal}{Phys. Rev. Lett.}
  \textbf{\bibinfo{volume}{92}}, \bibinfo{pages}{58702} (\bibinfo{year}{2004}),
  ISSN \bibinfo{issn}{1079-7114}.

\bibitem[{\citenamefont{Kim et~al.}(2003)\citenamefont{Kim, Choe, and
  Shin}}]{KimPRL03}
\bibinfo{author}{\bibfnamefont{D.-H.} \bibnamefont{Kim}},
  \bibinfo{author}{\bibfnamefont{S.-B.} \bibnamefont{Choe}}, \bibnamefont{and}
  \bibinfo{author}{\bibfnamefont{S.-C.} \bibnamefont{Shin}},
  \bibinfo{journal}{Phys. Rev. Lett.} \textbf{\bibinfo{volume}{90}},
  \bibinfo{pages}{087203} (\bibinfo{year}{2003}).

\bibitem[{\citenamefont{Papanikolaou et~al.}(2011)\citenamefont{Papanikolaou,
  Bohn, Sommer, Durin, Zapperi, and Sethna}}]{BeyondScaling}
\bibinfo{author}{\bibfnamefont{S.}~\bibnamefont{Papanikolaou}},
  \bibinfo{author}{\bibfnamefont{F.}~\bibnamefont{Bohn}},
  \bibinfo{author}{\bibfnamefont{R.~L.} \bibnamefont{Sommer}},
  \bibinfo{author}{\bibfnamefont{G.}~\bibnamefont{Durin}},
  \bibinfo{author}{\bibfnamefont{S.}~\bibnamefont{Zapperi}}, \bibnamefont{and}
  \bibinfo{author}{\bibfnamefont{J.~P.} \bibnamefont{Sethna}},
  \bibinfo{journal}{Nat. Phys.}  (\bibinfo{year}{2011}), ISSN
  \bibinfo{issn}{1745-2473}.

\bibitem[{\citenamefont{Durin et~al.}(2010)\citenamefont{Durin, Chen, and
  Sethna}}]{SloppyScaling}
\bibinfo{author}{\bibfnamefont{G.}~\bibnamefont{Durin}},
  \bibinfo{author}{\bibfnamefont{Y.-J.} \bibnamefont{Chen}}, \bibnamefont{and}
  \bibinfo{author}{\bibfnamefont{J.~P.} \bibnamefont{Sethna}},
  \emph{\bibinfo{title}{Sloppy scaling software}},
  \bibinfo{howpublished}{\url{https://github.com/gdurin/SloppyScaling}} (\bibinfo{year}{2010}).

\bibitem[{\citenamefont{Perkovi{\'c} et~al.}(1999)\citenamefont{Perkovi{\'c},
  Dahmen, and Sethna}}]{olga1999disorder}
\bibinfo{author}{\bibfnamefont{O.}~\bibnamefont{Perkovi{\'c}}},
  \bibinfo{author}{\bibfnamefont{K.~A.} \bibnamefont{Dahmen}},
  \bibnamefont{and} \bibinfo{author}{\bibfnamefont{J.~P.}
  \bibnamefont{Sethna}}, \bibinfo{journal}{Phys. Rev. B}
  \textbf{\bibinfo{volume}{59}}, \bibinfo{pages}{6106} (\bibinfo{year}{1999}).

\bibitem[{\citenamefont{Colaiori}(2008)}]{colaiori2008exactly}
\bibinfo{author}{\bibfnamefont{F.}~\bibnamefont{Colaiori}},
  \bibinfo{journal}{Adv. Phys.} \textbf{\bibinfo{volume}{57}},
  \bibinfo{pages}{287} (\bibinfo{year}{2008}), ISSN \bibinfo{issn}{0001-8732}.

\bibitem[{\citenamefont{Leschhorn and Tang}(1994)}]{leschhorn94PRE}
\bibinfo{author}{\bibfnamefont{H.}~\bibnamefont{Leschhorn}} \bibnamefont{and}
  \bibinfo{author}{\bibfnamefont{L.~H.} \bibnamefont{Tang}},
  \bibinfo{journal}{Phys. Rev. E} \textbf{\bibinfo{volume}{49}},
  \bibinfo{pages}{1238} (\bibinfo{year}{1994}).

\bibitem[{\citenamefont{Rosso and Krauth}(2001)}]{RossoKrauth01}
\bibinfo{author}{\bibfnamefont{A.}~\bibnamefont{Rosso}} \bibnamefont{and}
  \bibinfo{author}{\bibfnamefont{W.}~\bibnamefont{Krauth}},
  \bibinfo{journal}{Phys. Rev. Lett.} \textbf{\bibinfo{volume}{87}},
  \bibinfo{pages}{187002} (\bibinfo{year}{2001}).

\bibitem[{\citenamefont{Rosso et~al.}(2007)\citenamefont{Rosso, Le~Doussal, and
  Wiese}}]{Rosso07}
\bibinfo{author}{\bibfnamefont{A.}~\bibnamefont{Rosso}},
  \bibinfo{author}{\bibfnamefont{P.}~\bibnamefont{Le~Doussal}},
  \bibnamefont{and} \bibinfo{author}{\bibfnamefont{K.~J.} \bibnamefont{Wiese}},
  \bibinfo{journal}{Phys. Rev. B} \textbf{\bibinfo{volume}{75}},
  \bibinfo{pages}{220201} (\bibinfo{year}{2007}).

\bibitem[{\citenamefont{{Sethna} et~al.}(2004)\citenamefont{{Sethna}, {Dahmen},
  and {Perkovic}}}]{SethnaDahmen04}
\bibinfo{author}{\bibfnamefont{J.~P.} \bibnamefont{{Sethna}}},
  \bibinfo{author}{\bibfnamefont{K.~A.} \bibnamefont{{Dahmen}}},
  \bibnamefont{and}
  \bibinfo{author}{\bibfnamefont{O.}~\bibnamefont{{Perkovic}}}, in
  \emph{\bibinfo{booktitle}{The Science of Hysteresis}}, edited by
  \bibinfo{editor}{\bibfnamefont{G.}~\bibnamefont{{Bertotti}}}
  \bibnamefont{and}
  \bibinfo{editor}{\bibfnamefont{I.}~\bibnamefont{{Mayergoyz}}}
  (\bibinfo{year}{2004}), vol.~\bibinfo{volume}{II},
  \eprint{arXiv:cond-mat/0406320}.

\bibitem[{\citenamefont{Le~Doussal and Wiese}(2009)}]{DoussalWiese09}
\bibinfo{author}{\bibfnamefont{P.}~\bibnamefont{Le~Doussal}} \bibnamefont{and}
  \bibinfo{author}{\bibfnamefont{K.~J.} \bibnamefont{Wiese}},
  \bibinfo{journal}{Phys. Rev. E} \textbf{\bibinfo{volume}{79}},
  \bibinfo{pages}{051106} (\bibinfo{year}{2009}).

\bibitem[{\citenamefont{Tang and
  Leschhorn}(1992{\natexlab{b}})}]{tang1992pinning}
\bibinfo{author}{\bibfnamefont{L.~H.} \bibnamefont{Tang}} \bibnamefont{and}
  \bibinfo{author}{\bibfnamefont{H.}~\bibnamefont{Leschhorn}},
  \bibinfo{journal}{Phys. Rev. A} \textbf{\bibinfo{volume}{45}},
  \bibinfo{pages}{8309} (\bibinfo{year}{1992}{\natexlab{b}}).

\bibitem[{\citenamefont{Sneppen}(1992)}]{sneppen1992PRL}
\bibinfo{author}{\bibfnamefont{K.}~\bibnamefont{Sneppen}},
  \bibinfo{journal}{Phys. Rev. Lett.} \textbf{\bibinfo{volume}{69}},
  \bibinfo{pages}{3539} (\bibinfo{year}{1992}).

\bibitem[{\citenamefont{Amaral et~al.}(1995)\citenamefont{Amaral, Barab\'asi,
  Buldyrev, Harrington, Havlin, Sadr-Lahijany, and Stanley}}]{Amaral1995PRE}
\bibinfo{author}{\bibfnamefont{L.~A.~N.} \bibnamefont{Amaral}},
  \bibinfo{author}{\bibfnamefont{A.-L.} \bibnamefont{Barab\'asi}},
  \bibinfo{author}{\bibfnamefont{S.~V.} \bibnamefont{Buldyrev}},
  \bibinfo{author}{\bibfnamefont{S.~T.} \bibnamefont{Harrington}},
  \bibinfo{author}{\bibfnamefont{S.}~\bibnamefont{Havlin}},
  \bibinfo{author}{\bibfnamefont{R.}~\bibnamefont{Sadr-Lahijany}},
  \bibnamefont{and} \bibinfo{author}{\bibfnamefont{H.~E.}
  \bibnamefont{Stanley}}, \bibinfo{journal}{Phys. Rev. E}
  \textbf{\bibinfo{volume}{51}}, \bibinfo{pages}{4655} (\bibinfo{year}{1995}).

\bibitem[{\citenamefont{Frederiksen et~al.}(2004)\citenamefont{Frederiksen,
  Jacobsen, Brown, and Sethna}}]{FrederiksenPRL04}
\bibinfo{author}{\bibfnamefont{S.~L.} \bibnamefont{Frederiksen}},
  \bibinfo{author}{\bibfnamefont{K.~W.} \bibnamefont{Jacobsen}},
  \bibinfo{author}{\bibfnamefont{K.~S.} \bibnamefont{Brown}}, \bibnamefont{and}
  \bibinfo{author}{\bibfnamefont{J.~P.} \bibnamefont{Sethna}},
  \bibinfo{journal}{Phys. Rev. Lett.} \textbf{\bibinfo{volume}{93}},
  \bibinfo{pages}{165501} (\bibinfo{year}{2004}).

\bibitem[{\citenamefont{Cerruti and Zapperi}(2006)}]{cerruti2006barkhausen}
\bibinfo{author}{\bibfnamefont{B.}~\bibnamefont{Cerruti}} \bibnamefont{and}
  \bibinfo{author}{\bibfnamefont{S.}~\bibnamefont{Zapperi}},
  \bibinfo{journal}{J. Stat. Mech: Theory Exp.}
  \textbf{\bibinfo{volume}{2006}}, \bibinfo{pages}{P08020}
  (\bibinfo{year}{2006}).

\bibitem[{\citenamefont{Leschhorn}(1993)}]{leschhorn93}
\bibinfo{author}{\bibfnamefont{H.}~\bibnamefont{Leschhorn}},
  \bibinfo{journal}{Physica A} \textbf{\bibinfo{volume}{195}},
  \bibinfo{pages}{324} (\bibinfo{year}{1993}).

\bibitem[{\citenamefont{Transtrum et~al.}(2011)\citenamefont{Transtrum, Machta,
  and Sethna}}]{transtrum2011geometry}
\bibinfo{author}{\bibfnamefont{M.~K.} \bibnamefont{Transtrum}},
  \bibinfo{author}{\bibfnamefont{B.~B.} \bibnamefont{Machta}},
  \bibnamefont{and} \bibinfo{author}{\bibfnamefont{J.~P.}
  \bibnamefont{Sethna}}, \bibinfo{journal}{Phys. Rev. E}
  \textbf{\bibinfo{volume}{83}}, \bibinfo{pages}{036701}
  (\bibinfo{year}{2011}).

\bibitem[{\citenamefont{Transtrum et~al.}(2010)\citenamefont{Transtrum, Machta,
  and Sethna}}]{TranstrumPRL10}
\bibinfo{author}{\bibfnamefont{M.~K.} \bibnamefont{Transtrum}},
  \bibinfo{author}{\bibfnamefont{B.~B.} \bibnamefont{Machta}},
  \bibnamefont{and} \bibinfo{author}{\bibfnamefont{J.~P.}
  \bibnamefont{Sethna}}, \bibinfo{journal}{Phys. Rev. Lett.}
  \textbf{\bibinfo{volume}{104}}, \bibinfo{pages}{60201}
  (\bibinfo{year}{2010}).

\bibitem[{\citenamefont{Essam et~al.}(1988)\citenamefont{Essam, Guttmann, and
  De'Bell}}]{essam1988DP}
\bibinfo{author}{\bibfnamefont{J.~W.} \bibnamefont{Essam}},
  \bibinfo{author}{\bibfnamefont{A.~J.} \bibnamefont{Guttmann}},
  \bibnamefont{and} \bibinfo{author}{\bibfnamefont{K.}~\bibnamefont{De'Bell}},
  \bibinfo{journal}{Journal of Physics A: Mathematical and General}
  \textbf{\bibinfo{volume}{21}}, \bibinfo{pages}{3815} (\bibinfo{year}{1988}).

\bibitem[{\citenamefont{Makse and Amaral}(1995)}]{maske1995scaling}
\bibinfo{author}{\bibfnamefont{H.~A.} \bibnamefont{Makse}} \bibnamefont{and}
  \bibinfo{author}{\bibfnamefont{L.~A.~N.} \bibnamefont{Amaral}},
  \bibinfo{journal}{EPL} \textbf{\bibinfo{volume}{31}}, \bibinfo{pages}{379}
  (\bibinfo{year}{1995}).

\bibitem[{\citenamefont{Le~Doussal and Wiese}(2003)}]{ledoussal03anisotropic}
\bibinfo{author}{\bibfnamefont{P.}~\bibnamefont{Le~Doussal}} \bibnamefont{and}
  \bibinfo{author}{\bibfnamefont{K.~J.} \bibnamefont{Wiese}},
  \bibinfo{journal}{Phys. Rev. E} \textbf{\bibinfo{volume}{67}},
  \bibinfo{pages}{16121} (\bibinfo{year}{2003}), ISSN
  \bibinfo{issn}{1550-2376}.
  
\end{thebibliography}

\end{document}